\documentclass[a4paper,11pt]{article}
\pdfoutput=1 

\usepackage{jcappub} 

\usepackage[T1]{fontenc} 

\usepackage{graphicx}
\usepackage{float}
\usepackage{hyperref}
\usepackage{xcolor, xstring}
\usepackage{amssymb, amsmath, mathtools}
\usepackage{listings}
\usepackage{lscape}
\usepackage{lipsum}
\usepackage{multicol}
\usepackage{physics}
\usepackage{natbib}
\usepackage{txfonts}
\usepackage[small]{caption}
\usepackage{xspace}
\usepackage{comment}
\usepackage[normalem]{ulem} 
\usepackage{booktabs} 
\usepackage{pdflscape}
\usepackage{rotating}
\usepackage{subcaption}
\usepackage{bm}
\usepackage{soul}

\newcommand{\boostO}{\hat{\mathcal{O}}_x}

\newcommand{\DiffO}{\hat{\mathcal{D}}_x}

\newcommand{\diff}{\mathop{}\!\rm{d}}

\newcommand{\Gspec}{{G}}
\newcommand{\Mspec}{{M}}

\usepackage[dvipsnames]{xcolor}

\newcommand{\MM}{\texttt{Mathematica}\xspace}
\newcommand{\CT}{\texttt{CosmoTherm}\xspace}
\newcommand{\COBEF}{\textit{COBE/FIRAS}\xspace}
\newcommand{\Planck}{\textit{Planck}\xspace}

\newcommand{\Litebird}{{\it Litebird}\xspace}
\newcommand{\PICO}{{\it PICO}\xspace}

\newcommand{\Mpc}{{\rm Mpc}}

\newcommand{\Tz}{{T_{z}}}

\newcommand{\TCMB}{T_{\rm CMB}}

\newcommand{\xc}{x_{\rm c}}

\newcommand{\Yspec}{{{Y}}}
\newcommand{\Ynspec}[1]{{Y}_{#1}}

\newcommand{\Xe}{X_{\rm e}} 

\newcommand{\Ne}{N_{\rm e}}

\newcommand{\Drr}{\frac{\Delta \rho_{\gamma}}{\rho_{\gamma}}} 
\newcommand{\sigT}{\sigma_{\rm T}}
\newcommand{\Drrtext}{\Delta \rho_{\gamma}/\rho_{\gamma}} 

\newcommand{\Thz}{\theta_{z}}

\newcommand{\nbb}{{n_{\rm bb}}}

\newcommand{\expf}[1]{{{\rm e}^{#1}}} 
\newcommand{\id}{{\,\rm d}} 
\newcommand{\pot}[2]{#1 \times 10^{#2}}

\newcommand{\me}{m_{\rm e}} 

\newcommand{\beqa}{\begin{eqnarray}}   %
\newcommand{\eeqa}{\end{eqnarray}}   %
\newcommand{\beal}{\begin{align}}
\newcommand{\enal}{\end{align}}
\newcommand{\bealf}[1]{\begin{align} #1 \end{align}}
\newcommand{\bspl}{\begin{split}}
\newcommand{\espl}{\end{split}}
\newcommand{\bsub}{\begin{subequations}}
\newcommand{\esub}{\end{subequations}}
\newcommand{\bmulti}{\begin{multline}}   %
\newcommand{\beqm}{\begin{mathletters}}   %
\newcommand{\eeqm}{\end{mathletters}}   %

\newcommand{\vgh}{{\hat{\boldsymbol\gamma}}}

\newcommand{\beq}{\begin{equation}}   %

\newcommand{\eeq}{\end{equation}}   %

\newcommand{\vek} [1]{\mbox{\boldmath${#1}$\unboldmath}}

\newcommand{\Tin}{T_{\rm in}}

\newcommand{\zcon}{z_{\rm con}}
\newcommand{\xin}{x_{\rm in}}

\newcommand{\mdp}{m_{\rm d}}

\newcommand{\gammacon}{\gamma_{\rm con}}

\newcommand{\Xeb}{\bar{X}_{\rm e}}
\newcommand{\NHb}{\bar{N}_{\rm H}}

\newcommand{\oOx}[1]{{x^{#1}\partial_x^{#1}}}
\newcommand{\vbetah}{{\hat{{\boldsymbol\beta}}}}

\usepackage{bbm}
\def\i{\mathbbm{i}}

\title{Spectral distortion anisotropies from photon to dark photon conversions}

\author[a]{Sara Evangelista,}
\author[a]{Jens Chluba}
\author[b]{and Bryce Cyr}

\affiliation[a]{Jodrell Bank Centre for Astrophysics, School of Physics and Astronomy, The University of Manchester, Oxford Road, Manchester, M13 9PL, U.K.}

\affiliation[b]{Center for Theoretical Physics - A Leinweber Institute, Massachusetts Institute of Technology, Cambridge, MA 02139, USA}

\emailAdd{sara.evangelista@postgrad.manchester.ac.uk}
\emailAdd{jens.chluba@manchester.ac.uk}
\emailAdd{brycecyr@mit.edu}

\renewcommand{\trfont}{\small\ttfamily}
\subheader{\protect\hfill MIT-CTP-6049}
\date{June 2026}

\begin{document}

\abstract{
Dark photons are a gauge boson of a hypothetical dark sector, representing one of the most-studied minimal extensions of the Standard Model, with wide-ranging theoretical and observational implications. Here, we consider scenarios in which an initially unpopulated dark photon sector is populated via resonant photon to dark photon conversions. This process leads to observable spectral distortions in the cosmic microwave background (CMB), that can be used to constrain these models. We extend previous spectral distortion studies of the monopole spectrum to anisotropic spectral distortions, using the newly developed Frequency Hierarchy (FH) framework of {\tt CosmoTherm}. We illustrate the physics by presenting detailed computations of the photon transfer functions and distortion cross power spectra throughout the dark photon parameter space. We find that the dark photon mass explicitly controls the shape (i.e., multipole-dependence) of the signal power spectra, while the overall amplitude of the signal is determined by the kinetic mixing parameter of the model. Using these results, we place complementary limits on the minimal dark photon model using data from \Planck, finding that the constraints are only marginally weaker than those obtained with \COBEF data for the average (monopole) distortion.  
In addition, we compute the corrections to the standard temperature field, arguing that conversions at $z \gtrsim 2 \times 10^6$ may add iso-curvature type perturbations, which could lead to novel constraints in regimes where distortion anisotropies thermalize.
%
}

\maketitle
\flushbottom

\newpage

\section{Introduction}
Positing the existence of a dark $U(1)_{\rm d}$ gauge field is perhaps one of the most simple extensions to the standard model of particle physics, yet doing so can lead to a rich set of phenomenological implications. Limits on such a new gauge boson (originally called the \textit{paraphoton}) were first placed by Okun \cite{Okun1982} using precision QED data. Subsequent studies of this new gauge boson \cite{Galison1983, Holdom1985} unveiled a host of interesting consequences, such as the possibility that electrons could possess a small dark charge under this new symmetry, and that photons could oscillate into and out of the dark sector \cite{Georgi1983}. Among these seminal papers, this new gauge boson was rebranded as the \textit{dark photon}, a name that has persisted ever since. 

Dark photons have been shown to possess vast utility in model building beyond the standard model (BSM). Some examples include being used as vector mediators for more complex dark sectors (such as in inelastic \cite{TuckerSmith2001, Fitzpatrick2021, Berlin2023, Roy2026}, millicharged \cite{Davidson1993, Davidson2000, Berlin2022}, and atomic dark matter \cite{Kaplan2009, Ackerman2008, CyrRacine2012, Cline2012, Bansal2022, Adams2026} scenarios), dark matter candidates in their own right, or simply as generic gauge extensions just waiting to be constrained (or discovered!). Because of this, dark photon searches have been the subject of intense study in every conceivable experimental domain, from collider searches, to tabletop/tunable cavity experiments, and in astrophysical and cosmological processes\footnote{A conglomeration of constraints is presented in various review articles \cite{Caputo2021, Antel2023, Caputo2025, Caputo2026}, while the most up-to-date limits (at least for $m_{\rm d} \lesssim 2 m_{\rm e}$) can typically found on Ciaran O'Hare's \href{https://cajohare.github.io/AxionLimits/docs/dp.html}{Github page}.}. 

In the most minimal dark photon setups, the parameter space is fully specified by a mass ($m_{\rm d}$) and kinetic mixing parameter ($\epsilon$), which sets the strength of the coupling between the dark and standard sectors. The cosmic microwave background (CMB) has been used to set constraints on this parameter space in many different ways, though in this work we will be most interested in resonant conversions between the dark and standard model photons. In a cosmological background of charged particles (most importantly, the electrons), the standard model photon can develop a small plasma mass ($\omega_{\rm pl}$). This acts as an effective mass term for the photon which, for the redshifts we consider ($10^3 \lesssim z \lesssim 2\times 10^6$), tracks the free electron density $\omega_{\rm pl} \propto N_{\rm e}^{1/2}$. In essence, this means that the photon mass evolves over many orders of magnitude during our cosmological history. 

This fact is particularly relevant when one realizes that resonant oscillations (conversions) between the dark and standard model photon can occur if the tuning condition $\omega_{\rm pl} \simeq m_{\rm d}$ is met. In particular, for dark photons with $10^{-9} \, \lesssim m_{\rm d}/{\rm eV} \lesssim \, 10^{-4}$, this condition is met during an epoch in which primordial distortions to the frequency spectrum of the CMB can be generated ($z_{\rm rec} \lesssim z \lesssim 2 \times 10^6$). CMB spectral distortions are produced by the nonthermal injection or extraction of energy and/or entropy (direct photon number) at a time when processes that mediate thermalization are no longer efficient \citep{Sunyaev1970mu, Burigana1991, Hu1993, Chluba2011therm}. The direction of energy transfer between the dark and visible sectors during a resonant conversion depends on their relative occupation. In the extreme cases where the dark photon is initially {\it all} of the dark matter ($\Omega_{\rm dp} = \Omega_{\rm dm}$), or {\it none} of it ($\Omega_{\rm dp} \simeq 0$), leading constraints have been set on the parameter space using data from the \COBEF satellite \cite{Mirizzi:2009iz,Arias2012,McDermott2020DP,Caputo2020, Chluba2024DP, Arsenadze2024}, though the validity of the resonant conversion process in the former scenario has recently been called into question \cite{Hook2025}. 

In this work, we will be primarily concerned with the latter case in which initially $\Omega_{\rm dp} \simeq 0$, which implies that at a resonance point, CMB photons are {\it extracted} from the spectrum as they ``go dark". This scenario can be robustly constrained by CMB spectral distortions, and planned CMB spectrometer concepts such as TMS \citep{Jose2020TMS}, COSMO \citep{Masi2021} and {\it BISOU} \citep{BISOU} promise improvements over the limits from \COBEF within the next ten years, while on the longer term space-based concepts such as {\it PIXIE} \citep{Kogut2011PIXIE, Kogut2016SPIE} and {\it FOSSIL} could revolutionize CMB distortion measurements.

In addition to work on sky-averaged (global) CMB spectral distortions, resonant conversion processes of $\gamma_{\rm CMB}\rightarrow \gamma_{\rm d}$ have been considered recently in the patchy screening context \cite{Pirvu2023,McCarthy2024, Aramburo-Garcia2024}. In this mechanism, excess dimming of the CMB via resonant conversions is correlated with large scale structure, providing leading constraints on light dark photon masses ($10^{-13} \lesssim m_{\rm d}/{\rm eV} \lesssim 10^{-11}$) when cross-correlating the \Planck temperature maps with the unWISE galaxy survey \cite{Planck2018, Krolewski2019, Planck2018params}. While this is a powerful mechanism to constrain anisotropic dark photon signatures, one can also make use of the expected cross-correlations of anisotropies in both the temperature and spectral distortion fields (commonly referred to as $\mu T$ or $y T$ signals), as is the subject of this work. 
As we show here this opens a new route for independently constraining dark photon models with conversions in the pre-recombination era (i.e., masses $10^{-9} \, \lesssim m_{\rm d}/{\rm eV} \lesssim \, 10^{-4}$).

Only recently has a formalism been developed that allows one to consistently treat the sourcing and subsequent evolution of anisotropic CMB spectral distortions \cite{chluba_spectro-spatial_2023-I, chluba_spectro-spatial_2023-II, kite_spectro-spatial_2023-III, Chluba2026}. This framework generalizes the usual computation of CMB temperature and polarization anisotropies to allow for perturbations in the frequency spectrum, providing a spectro-spatial treatment of the thermal plasma. With this, it is possible to compute auto- and cross-spectra of the canonical $\mu$ and $y$ fields to provide additional constraining power on models that perturb the thermal plasma, both before and after recombination. While the $\mu\mu$ and $yy$ auto-power spectra are expected to be extremely small \citep{kite_spectro-spatial_2023-III}, there are significant benefits to looking for a $\mu T$ signal. From the observational side, this cross-correlation sidesteps the need to absolutely calibrate the CMB temperature monopole. Absolute calibration of the monopole is a major challenge faced by upcoming spectral distortion missions that aim to constrain deviations of the sky-averaged CMB blackbody at the $\Delta I /I \lesssim 10^{-6}$ level \citep[e.g.,][]{Kogut2023}.
Measurements of the $\mu T$ cross power spectra are furthermore already possible with existing data from \Planck \citep[e.g.,][]{Rotti2022muT}, ACT and SPT, using their latest data releases \citep{ACTDR62025, SPT2026}. In the future one can expect significant improvements from The Simons Observatory, \Litebird and possibly \PICO \citep{Remazeilles2018muT, Zegeye2023S4, Zegeye2024}, promising much activity in combination with planned CMB monopole measurements.

In this work, we explain how the resonant dark photon conversion process can give rise to an anisotropic spectral distortion signal and cause $\mu T$ and $y T$ correlation signals. Not only are dark photons well-motivated from the theory standpoint, but as we shall see, dark photon conversions are particularly compelling in the spectro-spatial treatment because they provide a multipole ($\ell$) dependent source term which tracks the CMB anisotropies, delivering a rich phenomenology. 

The remainder of the paper is organized as follows. First, we review the frequency hierarchy treatment in Sec.~\ref{sec:FH-treatment}, describing the spectral basis we consider and providing the backbone of the calculation. Next, in Sec.~\ref{sec:Dark_photon_source_terms}, we show how the dark photon distortion can be embedded in the frequency hierarchy, highlighting in particular its multipole dependence. We present our main numerical results in Sec.~\ref{sec:Results}, which primarily consists of a detailed illustration of the transfer functions and cross-correlation spectra in the dark photon model. In Sec.~\ref{sec:constraints} we provide constraints derived from $\mu T$ and $\mu E$ cross-correlations using \Planck data (at $\ell \lesssim 1000$) and compare against monopole distortion constraints before concluding in Sec.~\ref{sec:conclusions}.

\section{The Frequency Hierarchy Treatment}
\label{sec:FH-treatment}
Constraining anisotropic spectral distortions (SDs) of the CMB spectrum can offer valuable insights into various physical processes occurring throughout cosmic history. Until recently, the treatment of SDs from various sources was restricted to a computation of the sky-averaged (monopole) signal. This allowed for a more simplistic mathematical treatment and reduced computational cost, when compared with a treatment that considers all possible $k$-modes. However, since the introduction and subsequent development an anisotropic formalism (with various useful analytical approximations), detailed studies of distortion anisotropies have become increasingly appealing. In a series of three papers \cite{chluba_spectro-spatial_2023-I, chluba_spectro-spatial_2023-II, kite_spectro-spatial_2023-III}, it has been shown that using a discretised basis which accurately represents all possible SDs shapes, one can derive a new set of equations, referred to as the Frequency Hierarchy (FH) system, which allows one to quickly and accurately compute SD anisotropies. The FH provides an approach to evolving the SDs both in time and frequency in the presence of the usual thermalisation processes [i.e., Compton scattering (CS), Bremsstrahlung (BR) and double Compton (DC) emission], generalizing the standard photon hierarchy usually applied to CMB temperature and polarization anisotropies \citep[e.g.][]{Ma1995, CMBFAST}.

The newly defined spectral basis consists of:
\bealf{
\label{eq:basis}
&\Gspec(x)=\boostO \nbb
=\frac{x\,\expf{x}}{(\expf{x}-1)^2},
\quad 
\Mspec(x)=\Gspec(x)\left[\frac{1}{\beta_M}-\frac{1}{x}\right],
\nonumber \\[1mm]
& \Yspec(x)
=\Gspec(x)\left[x\frac{\expf{x}+1}{\expf{x}-1}-4\right],
\quad 
Y_k=(1/4)^k \,\boostO^k Y,
\nonumber \\[2mm]
&\Delta n= \vek{B} \cdot \vek{y}, 
\quad
\vek{B}=(\Gspec, Y_0, Y_1, \cdots, Y_N, M)^T, 
\quad
\vek{y}=(\Theta, y_0, y_1, \cdots, y_N, \mu)^T, 
}
where $\nbb=1/[\expf{x}-1]$ is the blackbody spectrum at dimensionless frequency $x=h\nu/k_{\rm B} T_z$, where $T_z$ is a convenient reference temperature that scales as $\propto (1+z)$ \citep[e.g.,][for explanation]{Chluba2011therm}, $\Mspec(x)$ and $\Yspec(x)$ are the standard spectral shapes for the $\mu$ and $y$ distortion respectively, and $\Gspec(x)$ is the spectrum of a temperature shift. Finally, $\Ynspec{k}(x)$ is a new set of functions obtained by repeatedly applying the boost operator $\boostO=-x\partial_x$ to $Y_0\equiv Y$. This set is truncated at some $N$, which we usually set to $N=15$ for best convergence \citep{chluba_spectro-spatial_2023-I}. 
We also highlight that the vector of spectral amplitudes, $\vek{y} \left( \eta, \vek{r}, \Hat{\gamma} \right)$, depends on the conformal time $\eta$, the position $\vek{r}$ and the observation direction $\Hat{\gamma}$. This will be evaluated either at the background level, $\vek{y}^{(0)}$, where the spatial and angular dependence is lost, or at first order in perturbations, $\vek{y}^{(1)}$. Here, we also introduce the unit vectors of the spectral basis $\vek{e}_G=(1, 0, \ldots, 0)^T$, $\vek{e}_Y=(0, 1, 0, \ldots, 0)^T$, $\vek{e}_{Y_1}=(0, 0, 1, 0, \ldots, 0)^T$, $\vek{e}_{M}=(0, \ldots, 0, 1)^T$.

The final set of equations, first derived in \cite{chluba_spectro-spatial_2023-II}, was recently improved by \cite{Chluba2026} to include a previously-omitted stimulated scattering term (i.e., $2\Theta^{(1)}_0\,\DiffO^* \Delta n^{(0)}_0 \boostO \nbb$) of the Comptonisation equation. With these changes, the FH system looks a bit different from the original one, with some extra terms and some cancellations:
\bsub
\label{eq:evol_yn_final}
\bealf{
&\frac{\partial \vek{y}^{(0)}_0}{\partial \eta}
= 
{\rm C}^{(0)}_{\rm d}[\vek{y}]+{\rm C}^{(0)}_{\rm h}[\vek{y}] +{\rm C}^{(0)}_{\rm S}[\vek{y}] \, ,
\\[2mm]
\label{eq:evol_y1_final}
&\frac{\partial \vek{y}^{(1)}}{\partial \eta}+\vgh\cdot \nabla \vek{y}^{(1)}+
\vek{b}^{(0)}_0 \left(\frac{\partial \Phi^{(1)}}{\partial \eta}+ \vgh \cdot \nabla\Psi^{(1)} \right)
= {\rm C}^{(1)}_{\rm T}[\vek{y}]+{\rm C}^{(1)}_{\rm d}[\vek{y}]+{\rm C}^{(1)}_{\rm h}[\vek{y}]+{\rm C}^{(1)}_{\rm S}[\vek{y}] \, ,
\\[2mm]
&{\rm C}^{(1)}_{\rm T}[\vek{y}]=\tau'\left[\vek{y}^{(1)}_0+\frac{1}{10}\,\vek{y}^{(1)}_2 - \vek{y}^{(1)} + \vek{b}^{(0)}_0 \beta^{(1)}\chi\,\right], 
\qquad
\vek{b}^{(0)}_0 = \vek{e}_G + M_{\rm B} \vek{y}^{(0)}_0,
}
\esub
where $\vgh$ is the photon direction, $\chi=\vgh\cdot \vbetah $ is the direction cosine with the baryon velocity vector $\vbetah$ and $\eta=\int c\id t/a$ is the conformal time with scale factor $a$. We furthermore introduce the Newtonian potential perturbations $\Phi^{(1)}$ and $\Psi^{(1)}$, and the Thomson scattering rate  $\tau'=\sigT \Ne a$. Finally, $\vek{b}^{(0)}_0$ is the matrix representation of $\boostO n^{(0)} \simeq \vek{b}_0^{(0)} \cdot \vek{B}$, where we made the monopolar dependence explicit and use the {\it Boost matrix}, $M_{\rm B}$, to evaluate $\vek{b}_0^{(0)}$ \citep[see][for details on how to compute $M_B$]{chluba_spectro-spatial_2023-II}. 

In this system of equations, we can observe many different collision terms. At both background and perturbed level, we have ${\rm C}^{(i)}_{\rm d}$ accounting for thermalisation effects, while ${\rm C}^{(i)}_{\rm h}$ and ${\rm C}^{(i)}_{\rm S}$ denote the heating and source terms, respectively. The latter refers to any physical process that injects or absorbs photons in the CMB, simultaneously affecting the photon number and energy densities. This is usually characterised by a spectral shape depending on the specific process. An example is the photon-to-dark-photon conversion case studied in this paper, but many others are possible. Conversely, we define heating as the scenario where the electrons are heated first and then transfer energy to the photons via Compton scattering, only sourcing a $y$ component. 
At first order in perturbations, we also find the Liouville operator on the left-hand side and the Thomson collision term, ${\rm C}^{(1)}_{\rm T}$ from the standard Boltzmann hierarchy. We have made use of the standard angular decomposition and defined $\vek{y}_\ell(t, \vek{r}, \vgh)=\sum_{m} \vek{y}_{\ell m}(t, \vek{r}) Y_{\ell m}(\vgh)$ using spherical harmonic coefficients. 
At the background level, the collision terms are defined as:
\bsub
\label{eq:evol_yn_final_colls}
\bealf{
{\rm C}^{(0)}_{\rm d}[\vek{y}] & \equiv {\rm C}^{(0)}_{\rm th}[\vek{y}]=\tau'\Thz M_{\rm T} \vek{y}^{(0)}_0 \, ,
\\[2mm]
{\rm C}^{(0)}_{\rm h}[\vek{y}] & = \frac{{\vek{Q}'}^{(0)}}{4}, \qquad {\vek{Q}'}^{(0)}
= \vek{e}_Y \,\frac{{Q'_{\rm c}}^{(0)}}{\rho_z}
= \left(0,\frac{{Q'_{\rm c}}^{(0)}}{\rho_z},\ldots, 0\right)^T \, ,
\\[2mm]
{\rm C}^{(0)}_{\rm S}[\vek{y}] & = \vek{S}'^{(0)}
= \left({S'_G}^{(0)},{S'_Y}^{(0)},\ldots, {S'_{Y_N}}^{(0)}, {S'_M}^{(0)}\right)^T.
}
\esub
Here, we introduced the {\it thermalization matrix}, $M_{\rm T}$, which captures both scattering and emission/absorption effects. It is computed by replacing the $\mu$ column of the \textit{Kompaneets matrix}\footnote{The Kompaneets matrix was first introduced in \citep{chluba_spectro-spatial_2023-II}. Its detailed form depends on the size of the FH basis.}, $M_{\rm K}$, with the emission vector $\vek{D}_0=\xc \left(\gamma_T, 0, \ldots, 0, -\gamma_N\right)^T$, which models the conversion $\mu \rightarrow \Theta$ \citep[see][for details]{Chluba2026}. The vectors ${\vek{Q}'}^{(0)}$ and $\vek{S}'^{(0)}$, describe the sourcing of distortions by heat and photon sources, respectively. Comparing the two, we can clearly observe the difference between the heating terms (only sourcing $y$) and the photon source (with a general injection spectrum)
We also notice that the thermalisation terms are weighted by $\tau ' \Thz$, with the dimensionless temperature variable, $\Thz=k\Tz/\me c^2$, presenting a distinct timescale with respect to the Thomson terms, that are only weighted by $\tau '$. 

For the perturbed terms, we must now include spatial variations. These are given by variations in the local clocks, through $ \Psi^{(1)}$, and fluctuations in the baryon density, $\delta^{(1)}_{\rm b}$, and CMB temperature, $\Theta^{(1)}_
\ell$. The collision terms now look as follows:
\bsub
\label{eq:evol_yn_final_colls_II}
\bealf{
{\rm C}^{(1)}_{\rm d}[\vek{y}] & = {\rm C}^{(1)}_{\rm th}[\vek{y}]-\beta^{(1)} \chi\, (1-M_{\rm B})\, {\rm C}^{(0)}_{\rm th}[\vek{y}] \, ,
\\[2mm]
{\rm C}^{(1)}_{\rm h}[\vek{y}] & = \frac{{\vek{Q}'}^{(1)}}{4}+ \Theta^{(1)}_0\,\vek{C}^{(0)}_0 -\beta^{(1)}\chi\,(1-M_{\rm B})\, \frac{{\vek{Q}'}^{(0)}}{4}\, ,
\\[2mm]
{\rm C}^{(1)}_{\rm S}[\vek{y}] & = {\vek{S}'}^{(1)}-\beta_S^{(1)}\chi_S\,(1-M_{\rm B})\,\vek{S}'^{(0)} \, ,
\\[2mm]
{\rm C}^{(1)}_{\rm th}[\vek{y}] & = \tau' \Thz 
M_{\rm T} \vek{y}^{(1)}_0 
+\tau' \Thz \left[(\delta^{(1)}_{\rm b} +\Theta^{(1)}_0+ \Psi^{(1)})\, M_{\rm T} +\Theta^{(1)}_0 M_{\rm OK}\right]\vek{y}^{(0)}_0  \, ,
\\[2mm]
{\vek{Q}'}^{(1)}&= \left(0,\frac{{Q'_{\rm c}}^{(1)}}{\rho_z}+\Psi^{(1)}\frac{{Q'_{\rm c}}^{(0)}}{\rho_z},\ldots, 0\right)^T,
\quad 
\vek{C}^{(0)}_0 = \frac{{Q'_{\rm c}}^{(0)}}{ \rho_z}\left(0,-1, 1, 0,\ldots, 0\right)^T, 
\\
{\vek{S}'}^{(1)} 
&= \left({S'_G}^{(1)},{S'_Y}^{(1)},\ldots, {S'_{Y_N}}^{(1)}, {S'_M}^{(1)}\right)^T+\Psi^{(1)}\,\vek{S}'^{(0)}.
}
\esub
Here, we introduced the operator $M_{\rm OK}=M_{\rm B} M_{\rm K} - M_{\rm K} M_{\rm B}$. Comptonisation terms, due to the scattering of photons and electrons with non-zero energy exchange, are only taken into account for the monopole, since for higher multipoles the Thomson scattering is dominant \citep{chluba_spectro-spatial_2023-II}. However, if the electrons have a bulk velocity $\beta$, it is necessary to apply a boost transformation to the Kompaneets equation. This provides a transformation from the thermal electron frame to the observer frame, giving rise to \textit{kinematic correction} terms, as derived in \cite{Chluba2026,chluba_spectro-spatial_2023-II}. These are the terms multiplied by $\beta^{(1)}\chi$ in the collision terms.  
To keep things general, we have assumed that the velocity of the baryons/electrons and of the source term particles can differ, meaning generally that $\chi \neq \chi_S$. However, for the specific case of photon-to-dark-photon conversion, they actually coincide, since the process is mediated by the electrons in the plasma, which implies $\chi = \chi_S$.

\subsection{Multipole hierarchy}
To obtain the actual transfer functions we need to go to Fourier space and decompose the functions in terms of the Legendre polynomials, leading to the full spectro-spatial FH \citep{chluba_spectro-spatial_2023-II}. We will now follow all the steps, in analogy to what is done for the standard perturbation equations \cite{CAMB,CLASSCODE} for temperature anisotropies. As mentioned above, all the variables in Eq. (\ref{eq:evol_yn_final}) are of the type $X(\eta,\chi,\vek{r})$. Going to Fourier space, $X(\eta,\chi,\vek{r}) \rightarrow X(\eta, \chi, k)$, and the gradient becomes $ \vgh\cdot \nabla X \rightarrow \i k \chi X$. Here, $\chi$ is the cosine angle between the photon direction and the wavenumber of the mode, which for scalar perturbations, coincides with our definition from above\footnote{The baryon velocity field is parallel to the wavenumber for scalar perturbations, which we consider here.}. At this point, it is natural to expand with respect to this angle using $X(\eta, \chi, k)=\sum_\ell (2\ell+1)\,(-\i)^\ell\tilde{X}_\ell(\eta, k)\,P_\ell(\chi)$, where $P_\ell(\chi)$ are the Legendre polynomials. Not having spatial or angular dependence, the background equation remains unchanged under these transformations. 
For notational simplicity, we drop the tilde in what follows, but stress that we work with the transformed quantities. One has to remember that subscript number in this case always represents the multipole $\ell$ of the decomposition with $m=0$, implying scalar perturbations. With this, the system becomes \citep[compare][]{Chluba2026}:
\bsub
\label{eq:evol_1_final_hierarchy}
\bealf{
\label{eq:evol_1_final_hierarchy_bg}
\frac{\partial \vek{y}^{(0)}_0}{\partial \eta}
&=\tau'\Thz M_{\rm T} \vek{y}^{(0)}_0+\frac{{\vek{Q}'}^{(0)}}{4}+\vek{S}'^{(0)}_0,
\\[1mm]
\label{eq:evol_1_final_hierarchy_0}
\frac{\partial \vek{y}^{(1)}_0}{\partial \eta}
&=-k\,\vek{y}^{(1)}_1\!-\!
\frac{\partial \Phi^{(1)}}{\partial \eta}
\vek{b}^{(0)}_0
+\frac{{\vek{Q}'^{(1)}}}{4}+ \Theta^{(1)}_0\,\vek{C}^{(0)}_0+{\vek{S}'^{(1)}_0}
\\
\nonumber
&\qquad
+\tau' \Thz \Bigg\{M_{\rm T} \vek{y}^{(1)}_0+\left[(\delta^{(1)}_{\rm b} +\Theta^{(1)}_0+ \Psi^{(1)})\, M_{\rm T} +\Theta^{(1)}_0 M_{\rm OK}\right]\vek{y}^{(0)}_0 \Bigg\},
\\
\label{eq:evol_1_final_hierarchy_1}
\frac{\partial \vek{y}^{(1)}_1}{\partial \eta}
&=k \,
\left(\frac{1}{3}
\vek{y}^{(1)}_{0}-\frac{2}{3}
\vek{y}^{(1)}_{2}\right)
+\frac{k}{3}\Psi^{(1)} \,\vek{b}^{(0)}_0
-\tau'\left[
\vek{y}^{(1)}_1-\frac{\beta^{(1)}}{3}\vek{b}^{(0)}_0\right] + {\vek{S}'^{(1)}_1}
\\
\nonumber
&\qquad
- \frac{\beta^{(1)}}{3}\,(1-M_{\rm B})\, \left[\tau'\Thz M_{\rm T} \vek{y}^{(0)}_0 + \frac{{\vek{Q}'}^{(0)}}{4} + \vek{S}'^{(0)}_0 \right] ,
\\
\frac{\partial \vek{y}^{(1)}_2}{\partial \eta}
&=
k \,
\left(\frac{2}{5}\vek{y}^{(1)}_1-\frac{3}{5}
\vek{y}^{(1)}_3
\right)
-\frac{9}{10}\,\tau'\,\vek{y}^{(1)}_2 +{\vek{S}'^{(1)}_2},
\\
\frac{\partial \vek{y}^{(1)}_{\ell\geq 3}}{\partial \eta}
&=
k \,
\left(\frac{\ell}{2\ell+1}
\vek{y}^{(1)}_{\ell-1}-
\frac{\ell+1}{2\ell+1}
\vek{y}^{(1)}_{\ell+1}\right)
-\tau'\vek{y}^{(1)}_\ell + {\vek{S}'^{(1)}_\ell}.
}
\esub
Here, we have assumed that the source term ($\vek{S}'_{\ell}$) contributes to all multipoles. We will see in the following sections that this turns out to be the case for the photon-to-dark-photon conversion problem. Note that we typically do not specify the multipole number for isotropic quantities (i.e., variables that only have $\ell = 0$ projections). One can observe that the kinematic corrections are included in Eq.~\eqref{eq:evol_1_final_hierarchy_1} as part of the dipole equation, where the baryon velocity in Fourier space is defined as $\tilde{\beta}^{(1)} \equiv \i \beta^{(1)}$. Finally, no corrections due to polarisation have been included in this system. Although for the standard temperature variables we do include these effects, we leave their imprint on spectral distortion anisotropies to future work.

\subsubsection{Corrections to the standard perturbation variables}
The treatment above did not explicitly isolate how the {\it small} distortion-induced temperature corrections and other corrections to the non-distortion variables (e.g., $\beta$, $\Phi$, $\Psi$) evolve. As explained in \citep{Chluba2026}, one can rewrite all non-distortion variables as $X^{\rm full} \approx X+\delta X$ to explicitly calculate the corrections, $\delta X$, with respect to the solutions {\it without} distortions. This then defines an additional hierarchy for $\delta X$, as detailed in Sect.~7.2 of \citep{Chluba2026}.

We stress this point since in the computations presented here, we will indeed calculate the correction $\delta \Theta$ to the standard temperature variables. This allows us to argue that for dark photon conversions in the temperature-era ($z \gtrsim \pot{2}{6}$) iso-curvature type perturbations could be produced (see section~\ref{sec:delta_Theta_terms}).
However, here we do not treat the sourcing of new perturbations in the corresponding dark photon fluid, which means that the results presented here for the $T \delta \Theta $ cross power spectra are mainly for illustration, with more work needed in the future.

\subsection{Line of sight solution} \label{sec:line-of-sight}
Using the solutions to Eq. (\ref{eq:evol_1_final_hierarchy}), we can compute the power spectra and cross-power spectra as: 
\begin{align}
C_\ell^{XY}(\eta)
&=
    \frac{2}{\pi}
    \int {\diff} k \, k^2  \, P(k)\, \hat{X}_\ell(\eta, k) \, \hat{Y}_\ell(\eta, k)
\end{align}
where $\hat{Y}$ and $\hat{X}$ are the transfer functions of two generic observables (e.g., $\Theta$ and $\mu$), and $P(k)$ is the initial power spectrum of the scalar perturbations. However, solving this integral directly is numerically challenging. Therefore \cite{chluba_spectro-spatial_2023-II} generalized the well-known line of sign approach \citep{CMBFAST} to include the effects of spectral distortions. To proceed, one first takes the Fourier transform of Eq.~\eqref{eq:evol_y1_final}:
\bsub
\label{eq:evol_Yi_1st_ord_Fourier_I}
\begin{align}
&\frac{\partial \vek{y}^{(1)}}{\partial \eta}
    +\i k \chi \,\vek{y}^{(1)}
    +\tau' \vek{y}^{(1)}
    = \,
    \vek{S}^{\rm T}_\text{LOS} + \vek{S}^{\rm d}_\text{LOS}, \, 
    \\
\label{eq:S_T_LOS}
&\vek{S}^{\rm T}_\text{LOS} \equiv 
    -\left(\frac{\partial \Phi^{(1)}}{\partial \eta}+ \i k \chi\Psi^{(1)} \right)\vek{b}^{(0)}_0
+\tau'\left[\vek{y}_0^{(1)}+\frac{1}{10}\,\vek{y}_2^{(1)}+\beta^{(1)}\chi\,\vek{b}^{(0)}_0\right]
\\
&\vek{S}^{\rm d}_\text{LOS} \equiv 
    {\rm C}^{(1)}_{\rm d}[\vek{y}]+{\rm C}^{(1)}_{\rm h}[\vek{y}]+{\rm C}^{(1)}_{\rm S}[\vek{y}],
\end{align}
\esub
where we isolated the Thomson terms and used the collision terms as introduced in Eq.~\eqref{eq:evol_yn_final_colls_II}. Following the same procedure as for the temperature-only case \cite[e.g.,][]{CMBFAST},  while ensuring we deal carefully with the vectors, we can then obtain integral solutions for the required $\vek{y}^{(1)}_\ell$. 
Dropping the tilde on the Legendre transformed terms, we find $\vek{y}^{(1)}_\ell(\eta_f, k)$ computed at the final conformal time, $\eta_f$:
\bsub
\label{eq:formal_sol_Leg_fin}
\begin{align}
\vek{y}^{(1)}_\ell(\eta_f, k) &= \int_0^{\eta_f} \id \eta \,g(\eta)\, \mathcal{\vek{S}}_\ell(\eta, \eta_f, k), 
\\[2mm]    
\mathcal{\vek{S}}_\ell(\eta, \eta_f, k)&=\mathcal{\vek{S}}^{\rm T}_\ell(\eta, \eta_f, k)+\mathcal{\vek{S}}^{\rm th}_\ell(\eta, \eta_f, k)
+\mathcal{\vek{S}}^{\rm ex}_\ell(\eta, \eta_f, k).
\end{align}
\esub
Here $g(\eta)=\tau'\,\expf{-\tau_{\rm b}}=\partial_\eta \expf{-\tau_{\rm b}}$ is the Thomson visibility function and where $\tau_{\rm b}=\tau(\eta_f)-\tau(\eta)$ is the Thomson optical depth between $\eta$ and $\eta_{ f}$. The sources were split into those from Thomson scattering ($\mathcal{\vek{S}}^{\rm T}_\ell$), thermalization effects ($\mathcal{\vek{S}}^{\rm th}_\ell$), and external sources from heating and photon injection ($\mathcal{\vek{S}}^{\rm ex}_\ell$). Explicitly, these source terms take the form \citep[see][]{Chluba2026}
\bsub
\label{eq:formal_sol_Leg_fin_sources}
\begin{align}
\mathcal{\vek{S}}^{\rm T}_\ell(\eta, \eta_f, k)&=
\left[
 \vek{y}_0^{(1)}+\Psi^{(1)} \vek{b}^{(0)}_0
 +
 \left(\frac{\partial \Psi^{(1)}}{\partial \eta}
 -\frac{\partial \Phi^{(1)}}{\partial \eta}
 \right)\frac{\vek{b}^{(0)}_0}{\tau'}
 \right]\,j_\ell(k\Delta\eta)
\nonumber\\
&\qquad\qquad
+\beta^{(1)} \vek{b}^{(0)}_0\,j^{(1,0)}_\ell(k\Delta\eta) 
+ \frac{\vek{y}_2^{(1)}}{2} \,j^{(2,0)}_\ell(k\Delta\eta)
\\[1mm]
\mathcal{\vek{S}}^{\rm th}_\ell(\eta, \eta_f, k)&=
\Thz \Bigg\{M_{\rm T} \vek{y}^{(1)}_0+\left[(\delta^{(1)}_{\rm b} +\Theta^{(1)}_0+ \Psi^{(1)})\, M_{\rm T} +\Theta^{(1)}_0 M_{\rm OK}+ \Psi^{(1)} M_{\rm B} M_{\rm T}\right]\vek{y}^{(0)}_0 \Bigg\} j_\ell(k\Delta\eta)
\nonumber\\
&\qquad\qquad
-\Thz \,\beta^{(1)} (1-M_{\rm B})\, M_{\rm T} \vek{y}^{(0)}_0  \,j^{(1,0)}_\ell(k\Delta\eta)
\\[1mm]
\mathcal{\vek{S}}^{\rm ex}_\ell(\eta, \eta_f, k)&=
\left[\Psi^{(1)}M_{\rm B}\left\{\frac{{\vek{Q}'^{(0)}}}{4\tau'}+\frac{\vek{S}'^{(0)}_0}{\tau'}\right\}+\frac{{\vek{Q}'^{(1)}}}{4\tau'}+\Theta^{(1)}_0\,\frac{\vek{C}^{(0)}_0}{\tau'}+\frac{\vek{S}'^{(1)}_0}{\tau'}\right]
j_\ell(k\Delta\eta)
\\ \nonumber
&\qquad +\Bigg[\frac{3 \vek{S}'^{(1)}_1}{\tau'} - \beta^{(1)} (1-M_{\rm B})\, \left\{\frac{{\vek{Q}'}^{(0)}}{4\tau'} + \frac{\vek{S}'^{(0)}_0}{\tau'} \right\}\Bigg]\,j^{(1,0)}_\ell(k\Delta\eta)
+\frac{5 \vek{S}'^{(1)}_2}{\tau'}\,j^{(2,0)}_\ell(k\Delta\eta)
\end{align}
\esub
with $\Delta \eta=\eta_f-\eta$. Following \citep{Hu1997}, the $j^{(a,b)}_\ell(x)$ are defined using the spherical Bessel functions $j_\ell(x)$. Specifically we have $j^{(0,0)}_\ell(x)=j_\ell(x)$, $j^{(1,0)}_\ell(x)=j'_\ell(x)=\partial_x j_\ell(x)$ and $j^{(2,0)}_\ell(x)=\frac{1}{2}\left[3 j''_\ell(x)+j_\ell(x)\right]$.
For more details we refer to the full derivation in \cite{Chluba2026} and \citep{chluba_spectro-spatial_2023-II}. These expressions allow us to compute the power spectra more easily.

We can in principle drop the contributions from $\mathcal{\vek{S}}^{\rm th}_\ell$ in the calculation of the CMB power spectra. This is because around last scattering, where most of the CMB signals originate, $\Thz \lesssim 10^{-6}$, meaning that direct thermalization corrections become negligible when the Thomson visibility is significant. Similarly, for sources that are mainly active at $z\gtrsim 2000$ we can in principle omit the $\mathcal{\vek{S}}^{\rm ex}_\ell$ terms since $g(\eta)/\tau' = \exp(-\tau_{\rm b})$ remains very small until around or after the hydrogen recombination era. However, in our computations we always include these terms and note that for late ($\zcon \lesssim 10^3$) photon conversions the contributions from $\mathcal{\vek{S}}^{\rm ex}_\ell$ actually dominate and thus have to be considered carefully.

\subsection{Analytic approximations}
\label{sec:analytics}
Below we will present numerical results for the photon transfer functions and CMB power spectra including distortion physics. As shown in \citep{Chluba2026TC}, the leading order solution to the FH with spectral distortion terms can be written as
\begin{align}
\label{eq:stationary_sol}
\vek{y}^{(1)}_\ell(\eta, k) &\approx \vek{b}_0^{(0)}(\eta)\,\Theta^{(1)}_\ell(\eta, k)
\end{align}
with a correction $\Delta \vek{y}^{(1)}_\ell(\eta, k)$ that decays after the source terms become negligible. The rate of decay depends on the time when the sources are active, with sourcing at early times ($z\gtrsim 10^4$) having very short decay times, while residual effects of the source can be expected at $z\lesssim 10^4$. 

We will refer to the approximation in Eq.~\eqref{eq:stationary_sol} as a {\it stationary solution}. For comparison, we will furthermore use this solution in our discussion of the transfer functions and power spectra. This already suggests that all the distortion cross power spectra will take the approximate form
\begin{align} \label{eq:stationary_sol_ps}
C_\ell^{XT}(\eta)
&\approx b^{(0)}_{X}(\eta_*)\,C_\ell^{TT}(\eta),
\qquad 
C_\ell^{XE}(\eta)
\approx b^{(0)}_{X,0}(\eta_*)\,C_\ell^{TE}(\eta),
\end{align}
where $b^{(0)}_{X}(\eta_*)$ is the corresponding boosted background distortion variable at the time of last scattering, $\eta_*$. Given that due to thermalization physics $b^{(0)}_{X}(\eta_*)$ depends on when the distortion was initially sourced, this modifies the constraining power of the distortion anisotropies as we discuss in section~\ref{sec:constraints}.

\section{Dark photon distortion source terms}
\label{sec:Dark_photon_source_terms}
In this section, we develop an understanding of how to treat the dark photon conversion problem using the frequency-hierarchy approach of \citep{chluba_spectro-spatial_2023-I, chluba_spectro-spatial_2023-II, kite_spectro-spatial_2023-III} with recent improvements \citep{Chluba2026} as discussed above. We first explain how in the presence of photon conversion processes the initial conditions for the blackbody field are determined. After that, we consider the average (i.e., monopole) dark photon conversion problem before including the effects of a spatially-varying source. 

\subsection{Initial conditions} \label{sec:initial_conditions}

To setup the problem with correct initial conditions, we need to account for the fact that resonant dark photon conversions perturb the usual temperature redshift relation. This follows from the fact that the photon energy density is reduced due to the conversion, as discussed in \cite{Chluba2024DP}. Defining $\Tin=\TCMB(z)(1+\bar{\Theta})$ and $\xin=h\nu/k\Tin=x/(1+\bar{\Theta})$ and comparing the photon energetics before and after the conversion yields the fixed temperature shift $\bar{\Theta}\approx \gammacon G_2/[4G_3] >0$. If we furthermore include the perturbations in the medium, the initial temperature anisotropies then exhibit effective temperature variations given by $\Delta = (1+\bar{\Theta})(1+\Theta)-1=\bar{\Theta} + \Theta (1+\bar{\Theta})$. For small $\Delta$ and $\bar{\Theta}$, this results in \citep{Chluba2026}: 
\begin{align} 
\label{eq:initial_conditions}
\Delta n^{(0)}_{\rm in}&\approx 
\bar{\Theta} \,G(x), \qquad 
\Delta n^{(1)}_{\rm in}\approx 
\Theta^{(1)}(1+3\bar{\Theta}) \,G(x) + \Theta^{(1)}\bar{\Theta} \,Y(x) = -\oOx{} n_{\rm in}^{(0)} \, \Theta^{(1)}.
\end{align}
where $n_{\rm in}=\nbb(x) + \Delta n^{(0)}_{\rm in}+\Delta n^{(1)}_{\rm in}$ is the directionally-dependent blackbody occupation prior to a conversion, and $n_{\rm in}^{(0)}=\nbb(x)+\bar{\Theta} \,G(x)$.
The expression for $\Delta n^{(0)}_{\rm in}$ is easy to understand as a simple temperature shift from one blackbody to another for $|\bar{\Theta}|\ll 1$. The second comes from the effect of changes to the directionally-dependent reference blackbody temperature up to order $\mathcal{O}(\Theta^{(1)}\bar{\Theta})$. We note that even though a term $\propto Y(x)$ is present, the initial CMB spectrum is undistorted \citep{Chluba2026}. 

\subsection{Average dark photon distortion source term}
To compute the average dark photon conversion source term, we make use of the fact that immediately after the conversion the average spectrum is given by \citep{Chluba2024DP}
\bsub
\label{eq:Dn_dp}
\begin{align}
\Delta n_{\rm d}(x)&\equiv
\left\{D(x)-\kappa_{\rm d}\,G(x)
\right\}\gammacon=S_{\rm d}(x)\,\gammacon+\bar{\Theta}\,G(x)
\\
\kappa_{\rm d}&=\frac{G_1}{3\,G_2}-\frac{G_2}{4\,G_3}\approx 0.1355, \qquad \bar{\Theta}\approx \frac{G_2}{4G_3}\,\gammacon 
\\
D(x)&=\frac{G_1}{3\,G_2}\,G(x)-\frac{n_{\rm bb}(x)}{x},
\qquad S_{\rm d}(x)=  -\frac{n_{\rm bb}(x)}{x}\equiv D(x)-\frac{G_1}{3\,G_2}\,G(x).
\end{align}
\esub
Here, we defined the distortion caused by the conversion as $S_{\rm d}(x)$, to distinguish it from the initial temperature shift contribution, $\bar{\Theta}\,G(x)$. The moments $G_k$ over the Planckian distribution are given by $G_k = \int \frac{x^k}{\expf{x}-1} \id x$. Finally,  $\gammacon \ll 1$ is the {\it conversion parameter} defined as:
\begin{align} 
\label{eq:gammacon}
\gammacon(\epsilon, \mdp) = 
\left.\frac{\pi m_{\rm d}^2 \epsilon^2}{ \TCMB(z) H(z)\, (1+z)} \left| \frac{1}{\omega_{\rm pl}^2 }\frac{\id \omega^2_{\rm pl}}{\id z} \right|^{-1}\right|_{z=\zcon}
\equiv
\left.\frac{\pi m_{\rm d}^2 \epsilon^2}{\TCMB(z)} \left| \frac{1}{\omega_{\rm pl}^2 }\frac{\id \omega^2_{\rm pl}}{\id t} \right|^{-1}\right|_{z=\zcon}.
\end{align}
This expression depends on the dark photon mass ($m_{\rm d}$), the mixing parameter ($\epsilon$), and the plasma frequency of the medium, 
$\omega^2_{\rm pl}(z) \simeq 4\pi \alpha_{\rm em} N_{\rm e}(z)/m_{\rm e}$.
where $\alpha_{\rm em}$ is the fine-structure constant, $m_{\rm e}$ is the electron mass and $N_{\rm e}$ is the free electron number density. This simple expression is valid in the pre-recombination epoch when electrons are non-relativistic, though we have neglected a subdominant frequency dependent term coming from the presence of neutral and singly ionized helium at $z\lesssim 10^4$ \citep[e.g., Fig. 1 of][]{Cyr2024Axions}. Using $N_{\rm e}=N_{\rm H}(z)\Xe(z)$, with the free electron fraction ($\Xe$) and number density of hydrogen nuclei ($N_{\rm H}$) we can also rewrite the conversion parameter as 
\bsub
\label{eq:lambda}
\begin{align}
&\gammacon=\left.\frac{\pi m_{\rm d}^2 \epsilon^2}{ \TCMB(z) H(z)\, \lambda(z)}\right|_{z=\zcon} \equiv \frac{g(z)}{\lambda(z)}, 
\\[2mm]
&\lambda=\left|\frac{1+z}{\omega_{\rm pl}^2 }\frac{\id \omega^2_{\rm pl}}{\id z}\right| = \left|\frac{\id \ln \omega^2_{\rm pl}}{\id \ln z}\right| \approx \frac{\id \ln N_{\rm H}}{\id \ln z}+\frac{\id \ln \Xe}{\id \ln z}.
\end{align}
\esub
The conversion redshift is furthermore determined by the condition $m_{\rm d} \simeq \omega_{\rm pl}(\zcon)$. In our work, we are concerned with conversions happening before last scattering, and can thus drop the absolute value in the computation of $\lambda$, as positivity is always ensured at these early times.
Bringing things together, the average distortion source term is then given as
\begin{align} \label{eq:dS0}
\frac{\id \mathcal{S}^{(0)}(z, x)}{\id z}&\equiv \gammacon^{(0)} \, \delta(z-z^{(0)}_{\rm con})\,S_{\rm d}(x),
\end{align}
where $\gammacon^{(0)}$ and $z^{(0)}_{\rm con}$ are obtained by using background values for all variables (see more details below). This leads to a spectral distortion that is similar to a $\mu$-distortion \citep{Chluba2024DP}. 

\subsection{Spatially-varying dark photon distortion source term}
How do we generalize this to include the spatial variations in the conversion process? The fluctuating source term has two main contributions. One comes from the transformation of the coordinate time into the local inertial frame, giving a factor $\Psi^{(1)}$. The second arises from the variation in local quantities such as the electron density and blackbody temperature. Together these then give the perturbed photon source term, $\id \mathcal{S}^{(1)}/\id z$ as we will explain below.

\subsubsection{Perturbed conversion parameter}
We start by looking at the effect of these variations on the conversion parameter and redshift. With regards to the former, we only have to include the spatial variations of $\lambda$ in Eq.~\eqref{eq:lambda}, which leads to small changes in the conversion redshift along different lines of sight. 
This is because the factor of $\TCMB$ in $\gammacon$ appears from scaling the frequency variable by the average CMB blackbody temperature, without any perturbations. Similarly, the factor of $1/H$ has to be considered at the background level, since variations in the local clocks are already captured by $\Psi^{(1)}$. This is also clear from the second equality of Eq.~\eqref{eq:gammacon}, where this factor is absorbed in the definition of cosmic time. 

To fully account for perturbations in local quantities, we use Eq.~\eqref{eq:lambda}, where we have to vary $\lambda$ at fixed $\zcon$ and then also perturb the conversion redshift. Defining $\delta_{\zcon}=\delta \zcon/\zcon$, at first order we can write:
\begin{align}
\gammacon & \approx \frac{g^{(0)}}{\lambda^{(0)}+\lambda^{(1)}} + \frac{\id \gammacon^{(0)}}{\id \ln z}\,\delta_{\zcon} 
\approx \frac{g^{(0)}}{\lambda^{(0)}} \left( 1 - \frac{\lambda^{(1)}}{\lambda^{(0)}} \right) + \frac{\id \gammacon^{(0)}}{\id \ln z}\,\delta_{\zcon} 
\nonumber \\[2mm]
& = \gammacon^{(0)} - \gammacon^{(0)}\frac{\lambda^{(1)}}{\lambda^{(0)}} + \frac{\id \gammacon^{(0)}}{\id \ln z}\,\delta_{\zcon} = \gammacon^{(0)}-\gammacon^{(0)}\left\{\frac{\lambda^{(1)}}{\lambda^{(0)}}-\frac{\id \ln \gammacon^{(0)}}{\id \ln z}\,\delta_{\zcon}
     \right\}.
\end{align}
Here, $\lambda^{(0)}$ and $\lambda^{(1)}$ denote $\lambda$ at the background level and first perturbed order, respectively.

Introducing $N_{\rm H}\approx \NHb(1+\delta_{\rm b})$ and $\Xe\approx \Xeb(1+\delta_{\rm e})$, where the bar denotes the background quantities [i.e., $\NHb \propto (1+z)^{3}$ and $\Xeb$, which depends on the recombination history] and defining the relative perturbation variables, $\delta_{\rm e}=\delta \Xe/\Xeb$ and $\delta_{\rm b}=\delta N_{\rm H}/\NHb$, we then obtain:
\bsub
\label{eq:lamba_0_1}
\begin{align}
\lambda^{(0)}&=3+\frac{\id \ln \Xeb}{\id \ln z} \equiv 3 \xi_{\rm e} 
&\rightarrow & &\gammacon^{(0)}&=
\left.\frac{\pi m_{\rm d}^2 \epsilon^2}{3\TCMB(z) H(z)\,\xi_{\rm e}(z)} 
\right|_{z=\zcon^{(0)}}
\\[2mm]
\lambda^{(1)}&=\frac{\id \delta_{\rm b}}{\id \ln z}+\frac{\id \delta_{\rm e}}{\id \ln z}
&\rightarrow &&\gammacon^{(1)}&=-\left\{\frac{\gammacon^{(0)}}{3\xi_{\rm e}(z)}\left[\frac{\id \delta_{\rm b}}{\id \ln z}+\frac{\id \delta_{\rm e}}{\id \ln z}\right]
-\frac{\id \gammacon^{(0)}}{\id \ln z}\,\delta_{\zcon}\right\}_{z=\zcon^{(0)}}.
\end{align}
\esub
Here, we introduced the function $\xi_{\rm e}=1+\frac{1}{3}\frac{\id \ln \Xeb}{\id \ln z}$ for convenience (see section~\ref{sec:xi_e_gamma} for details).

To compute the variation of the conversion redshift, we use that $\zcon$ is a solution of the implicit equation $f(z)=m^2_{\rm d}-\omega^2_{\rm p}(z)\equiv 0$. This can be solved at background order to yield $\zcon^{(0)}$ used above. We obtain the perturbation of $\zcon$ due to variations in the electron number density, knowing that the variational derivative of $f$ should vanish, since $\zcon$ remains a solution:
\begin{align}
0 = \delta f = \frac{\partial f}{\partial \delta_{\zcon}} \,\delta_{\zcon} + \frac{\partial f}{\partial \delta_{\rm b}}\,\delta_{\rm b}+\frac{\partial f}{\partial \delta_{\rm e}}\,\delta_{\rm e}\qquad \rightarrow \qquad 
\delta_{\zcon}= \frac{\delta \zcon}{\zcon}\approx - \frac{\delta_{\rm b}+\delta_{\rm e}}{3\xi_{\rm e}}
\end{align}
where we used $\frac{\partial f}{\partial \delta_{\zcon}}\approx - 3[\omega_{\rm p}^{(0)}]^2\xi_{\rm e}$ and $\frac{\partial f}{\partial \delta_{\rm b}}=\frac{\partial f}{\partial \delta_{\rm e}}=-[\omega_{\rm p}^{(0)}]^2$.
Overall, this then means
\begin{align}
\label{eq:gamma_1}
\gammacon^{(1)}&=-\frac{\gammacon^{(0)}}{3\xi_{\rm e}}\left[\frac{\id \delta_{\rm b}}{\id \ln z}+\frac{\id \delta_{\rm e}}{\id \ln z}
+\frac{\id \ln \gammacon^{(0)}}{\id \ln z}\,\left(\delta_{\rm b}+\delta_{\rm e}\right)\right]_{z=\zcon^{(0)}}.
\end{align}
Here, we will not consider the effects of perturbed recombination \citep{Novosyadlyj2006, Khatri2009, Senatore2009} and thus set $\delta_{\rm e} \approx 0$ for our numerical calculations.  

\subsubsection{Final Anisotropic source term}
To include the spatial variations in the photon source term, we have to go back to its derivation. Right after the conversion, one has
\begin{align}
n(x)&=
n_{\rm in} \, \expf{-\frac{\gammacon}{x}}
\approx 
\left[\nbb  + \Delta n^{(0)}_{\rm in} +\Delta n^{(1)}_{\rm in} \right]\left\{1-\frac{\gammacon^{(0)}+\gammacon^{(1)}}{x}\right\}
\nonumber\\
&\approx \left[\nbb  + \Delta n^{(0)}_{\rm in} +\Delta n^{(1)}_{\rm in} \right]\left\{1-\frac{\gammacon^{(0)}}{x}\right\}-\nbb \,\frac{\gammacon^{(1)}}{x}
\nonumber\\
&\approx \nbb  + \Delta n^{(0)}_{\rm in}  - \nbb \frac{\gammacon^{(0)}}{x}
+ \Delta n^{(1)}_{\rm in}  - \Theta^{(1)}\,G \frac{\gammacon^{(0)}}{x}-\nbb(x)\,\frac{\gammacon^{(1)}}{x}
\nonumber\\
&=\nbb + \Delta n^{(0)}_{\rm in} + S_{\rm d}\, \gammacon^{(0)} + \Delta n^{(1)}_{\rm in} + S_{\rm d}\, \gammacon^{(1)}+\Theta^{(1)}\,\gammacon^{(0)}\,\left[ M - \frac{G}{\beta_M} \right],
\end{align}
with $\beta_M=3G_2/[2 G_1]=2.1923$ and the usual $\mu$ distortion spectrum\footnote{We note that $\gammacon^{(1)}$ is $\mathcal{O}(\gammacon^{(0)} \, \delta)$, meaning that it only affects the blackbody part of the CMB spectrum at linear order in the distortion variables as terms $\Delta n^{(0)}_{\rm in}\gammacon^{(0)}$ are of higher order \citep{chluba_spectro-spatial_2023-II}. } $M=G[\beta_M^{-1}-x^{-1}]$.
To first order in the perturbations, we then find the dark photon induced distortion to be
\begin{align} 
\label{eq:Dn_d_def}
\Delta n_{\rm d} \approx \Delta n^{(0)}_{\rm in} + \Delta n^{(1)}_{\rm in} + S_{\rm d}\, \gammacon^{(0)}  + S_{\rm d}\, \gammacon^{(1)}+\Theta^{(1)}\,\gammacon^{(0)}\,\Delta S_{\rm d}.
\end{align}
This means that the fluctuating part of the photon source term caused by variations in the local CMB temperature has a spectrum 
\begin{align} 
\label{eq:DS_def}
\Delta S_{\rm d}(x)=M (x) - \frac{G(x)}{\beta_M}. 
\end{align}
We note that in Eq.~\eqref{eq:Dn_d_def}, $\Delta n_{\rm in}=\Delta n^{(0)}_{\rm in} + \Delta n^{(1)}_{\rm in}$ just captures the initial fluctuations of the photon background, but the conversion source is given by $S_{\rm d}\, \gammacon^{(0)}$ for the isotropic part and $S_{\rm d}\, \gammacon^{(1)}+\Theta^{(1)}\,\gammacon^{(0)}\,\Delta S_{\rm d}$ for the fluctuating part. 
Through the term $\propto \Theta^{(1)}$ this source directly depends on $\ell$. Thus this source term enters at all multipoles in the radiation field, which is in stark contrast to more typical distortion sources coming from energy injection that tend to simply excite only the local monopole spectrum.

Putting things together and including the variations in the local clocks for $\gammacon$, this then implies the final fluctuating photon monopole source term is
\bsub
\label{eq:dS1}
\begin{align} 
\frac{\id \mathcal{S}^{(1)}(z, 
\vek{r}, 
\vgh, x)}{\id z}& \equiv\gammacon^{(0)} \,\delta(z-z^{(0)}_{\rm con})
\left\{
\left[\frac{\gammacon^{(1)}}{\gammacon^{(0)}}
+\Psi^{(1)}\right]\,S_{\rm d}(x)
+\Theta^{(1)}\,\Delta S_{\rm d}(x)
\right\},\\[2mm]
\frac{\gammacon^{(1)}(z, 
\vek{r})}{\gammacon^{(0)}(z)}&=
\frac{1}{\xi_{\rm e}}\left[\frac{1}{3\mathcal{H}}\left(\partial_\eta\delta^{(1)}_{\rm b}+\partial_\eta\delta^{(1)}_{\rm e}\right)
-\frac{1}{3}\frac{\id \ln \gammacon^{(0)}}{\id \ln z}\,\left(\delta^{(1)}_{\rm b}+\delta^{(1)}_{\rm e}\right)\right],
\end{align}
\esub
where we partially switched to conformal time, $\eta$, using $\id \delta_{i}/\id \ln z =-\delta_{i}'/\mathcal{H}$ with $\delta_{i}'=\partial_\eta \delta_{i}$, such that we can directly use their differential equations. In Eq.~\eqref{eq:dS1}, we have neglected the small correction to the phase of the photon source term from the change in the conversion redshift, but did account for the redshift variation of the overall conversion parameter. 

\subsubsection{Anisotropic source term deep into the pre-recombination era}
One can further simplify the expression for Eq.~\eqref{eq:dS1} by assuming \textit{radiation domination}. A more detailed discussion on the redshift dependence will be provided in section~\ref{sec:xi_e_gamma}. Given the definitions in Eq.~\eqref{eq:lamba_0_1}, we start by writing:
\begin{align} 
\label{eq:dlngammacon_dlnz}
\frac{\id \ln \gammacon^{(0)}}{\id \ln z}=
-\frac{\id \ln  \TCMB }{\id \ln z}-\frac{\id \ln  H }{\id \ln z}-\frac{\id \ln \xi_{\rm e}}{\id \ln z}=
-1-\frac{\id \ln  H }{\id \ln z}-\frac{\frac{\id^2 \ln \Xeb}{\id \ln z^2}}{3+\frac{\id \ln \Xeb}{\id \ln z}} \approx -3,
\end{align}
where in the last step we used the relations $H\propto(1+z)^2$ and $\Xe = {\rm const}$. We also have $\xi_{\rm e}\approx 1$, $\delta^{(1)}_{\zcon}\approx -\delta^{(1)}_{\rm b}/3$ and $\id \delta_{\rm b}/\id \ln z =-\delta_{\rm b}'/\mathcal{H}$, where we switched to conformal time ($\eta$) with $\delta_{\rm b}'=\id \delta_{\rm b}/\id \eta$. This then gives the simplified fluctuating photon source term
\begin{align} \label{eq:dS1_rad}
\frac{\id \mathcal{S}^{(1)}(z, 
\vek{r}, 
\vgh, x)}{\id z}&\approx \gammacon^{(0)} \,\delta(z-z^{(0)}_{\rm con})
\left\{
\left[\delta^{(1)}_{\rm b}+\Psi^{(1)}+\frac{\partial_\eta \delta^{(1)}_{\rm b}}{3\mathcal{H}}\right] S_{\rm d}(x)
+\Theta^{(1)}\,\Delta S_{\rm d}(x)
\right\}
\end{align}
for conversions in the radiation-dominated era. We can see that for early conversions, the dominant anisotropic source contributions stem from fluctuations in $\delta^{(1)}_{\rm b}$, $\Psi^{(1)}$, and $\Theta^{(1)}$. 

\begin{figure}
\centering
\includegraphics[width=\columnwidth]{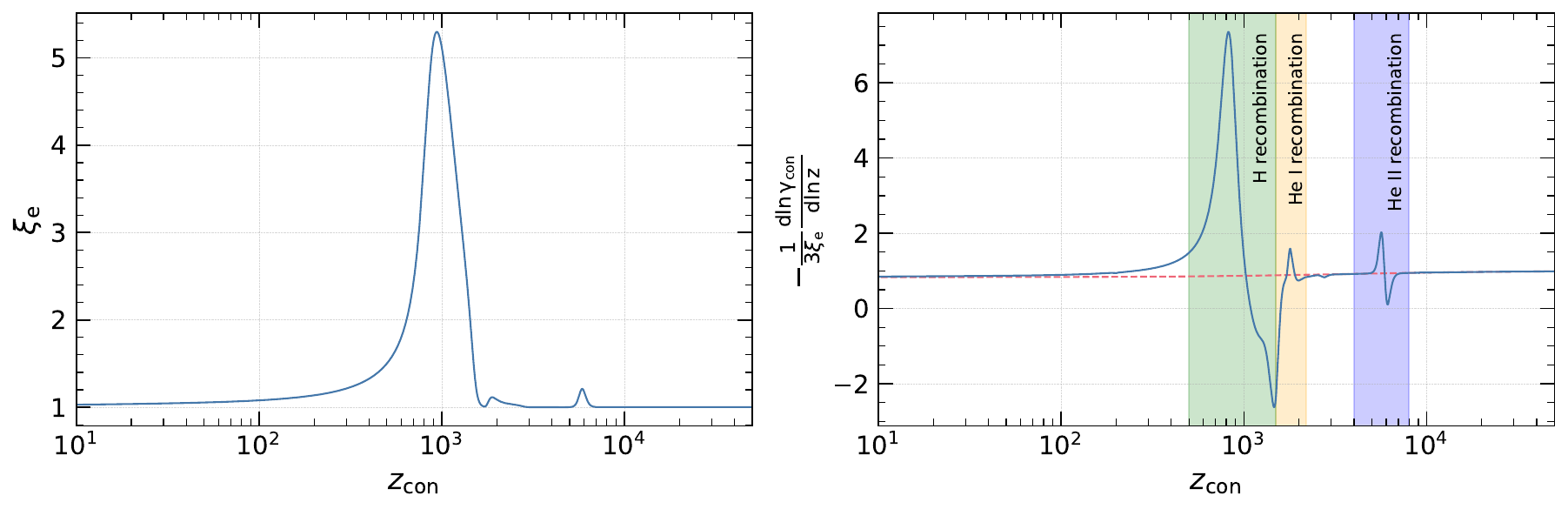}
\caption{The left panel illustrates the evolution of $\xi_{\rm e}$ with the conversion redshift. Detailed results for the free electron fraction through recombination have been included, leading to features around $z\simeq 10^3-10^4$ as highlighted by the coloured bands. In the right panel, we can observe the evolution of the conversion variable $-\frac{1}{3\xi_{\rm e}}\frac{\id\ln\gamma^{(1)}}{\id\ln z}$ [see Eq.~\eqref{eq:dS1}] compared to the simplified case where the free electron fraction does not evolve (dashed red line). The only appreciable difference is visible around the recombination redshifts.}
\label{fig:gamma_evol}
\end{figure}

\subsubsection{Effect of the free electron fraction on the conversion parameter}
\label{sec:xi_e_gamma}
If we are interested in the full evolution of the conversion factor and the source terms, we need to take into account details in the ionization history, $\Xe(z)$, which are especially important around recombination. 
Here we illustrate how well the approximation in Eq.~\eqref{eq:dlngammacon_dlnz} holds when using the full ionization history obtained from {\tt CosmoRec} \citep{Chluba2010b}.
In Fig.~\ref{fig:gamma_evol}, we show the redshift evolution of  $\xi_{\rm e}$ if we allow the free electron fraction to vary.\footnote{To obtain accurate results for the numerical derivatives, we used the central difference method for the first derivative and five-point stencil for the second derivative. We also note that although we always write $\frac{\id \ln \gammacon^{(0)}}{\id \ln z}$, we actually mean $\frac{\id \ln \gammacon^{(0)}}{\id \ln (1+z)}$, which makes a noticeable difference at low redshifts.} As one could expect, the only non-trivial modulation appears around the recombination period, presenting three distinct peaks corresponding to the first and second recombination of Helium, followed by Hydrogen recombination. Outside of these regimes, $\gammacon$ can be computed using $\xi_{\rm e} = 1$.
On the other hand, in the right panel of Fig.~\ref{fig:gamma_evol} we observe how $\frac{\id \ln \gammacon^{(0)}}{\id \ln z}$ is modified by the evolution of $\Xe$, again with features visible around $\zcon = 10^4 - 10^3$. 

In our computations, we always include the full ionization history. However, for early conversions (particularly for $z \gtrsim 10^4$), the evolving free electron fraction should not make a significant difference to the final results, a conclusion that can be helpful for analytic approximations.

\subsection{Numerical implementation of the source term}
\label{sec:num_setup}
\label{sec:decomposition}
Now that we have an analytical expression for the photon-to-dark-photon source terms, we implement them into $\CT$ to determine the effects on the various transfer functions and power spectra. To study the anisotropic problem, we use a modified \CT setup which includes a treatment of the frequency hierarchy, as introduced in section \ref{sec:FH-treatment}.
\begin{figure}[t]
    \centering
    \includegraphics[width=\linewidth]{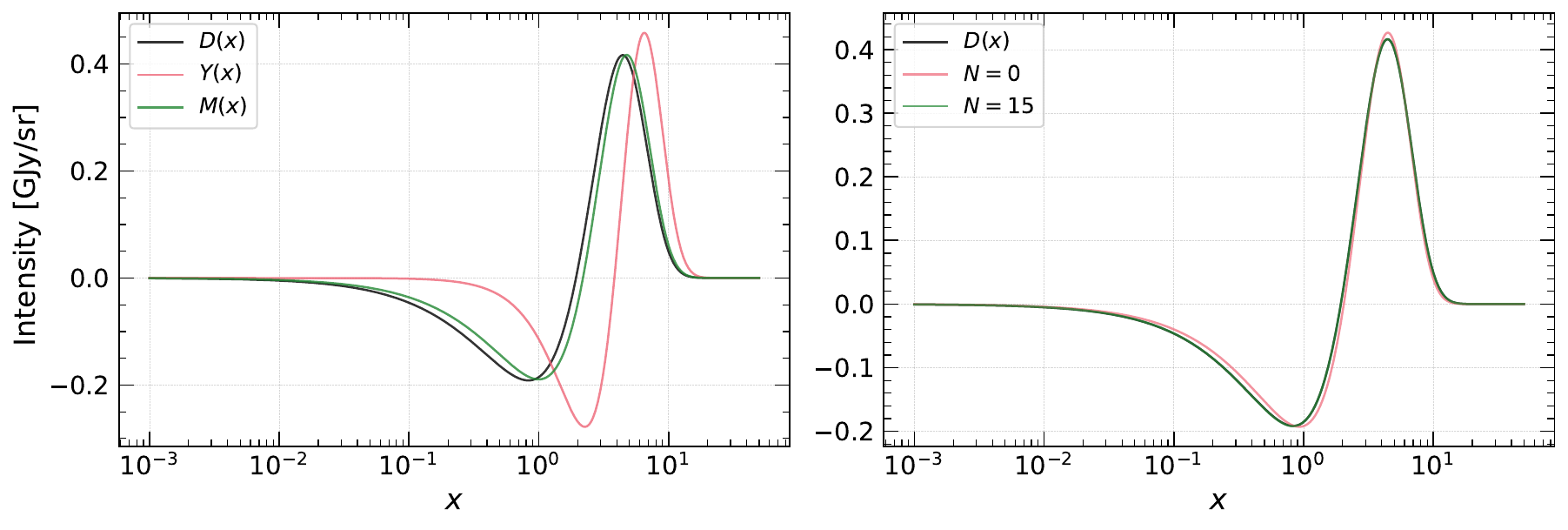}
\caption{On the left, a comparison between the analytical functions $\Yspec, M , D$, demonstrating how the source term due to photon to dark photon conversion is close to a $\mu$-distortion. On the right, the analytical spectral shape $D(x)$ is compared to the numerical approximations obtained by decomposing it in the FH spectral basis. The agreement is very good already at the lowest order, and for the full basis the match is improved.}
\label{fig:DYM}
\end{figure}

\begin{table}[t]
\centering
\resizebox{\textwidth}{!}{%
\begin{tabular}{|l|ccccccccc|}
\hline
$N$ & 0 & 1 & 3 & 5 & 7 & 9 & 11 & 13 & 15 \\
\hline
$y$       & $-2.070\times10^{-2}$ & $-2.065\times10^{-2}$ & $-2.061\times10^{-2}$ & $-2.060\times10^{-2}$ & $-2.060\times10^{-2}$ & $-2.060\times10^{-2}$ & $-2.060\times10^{-2}$ & $-2.060\times10^{-2}$ & $-2.060\times10^{-2}$ \\
$\mu$     & $8.385\times10^{-1}$ & $8.379\times10^{-1}$ & $8.376\times10^{-1}$ & $8.375\times10^{-1}$ & $8.375\times10^{-1}$ & $8.374\times10^{-1}$ & $8.374\times10^{-1}$ & $8.374\times10^{-1}$ & $8.374\times10^{-1}$ \\
$r_1$     & $0$ & $5.342\times10^{-3}$ & $6.941\times10^{-3}$ & $7.119\times10^{-3}$ & $7.153\times10^{-3}$ & $7.160\times10^{-3}$ & $7.163\times10^{-3}$ & $7.164\times10^{-3}$ & $7.164\times10^{-3}$ \\
$r_2$     & $0$ & $-4.437\times10^{-4}$ & $-9.874\times10^{-5}$ & $-4.904\times10^{-6}$ & $2.216\times10^{-5}$ & $3.116\times10^{-5}$ & $3.451\times10^{-5}$ & $3.583\times10^{-5}$ & $3.586\times10^{-5}$ \\
$r_3$     & $0$ & $9.145\times10^{-4}$ & $-6.735\times10^{-4}$ & $-4.669\times10^{-4}$ & $-3.012\times10^{-4}$ & $-2.188\times10^{-4}$ & $-1.804\times10^{-4}$ & $-1.627\times10^{-4}$ & $-1.623\times10^{-4}$ \\
$r_4$     & $0$ & $1.164\times10^{-4}$ & $-1.263\times10^{-4}$ & $1.655\times10^{-5}$ & $3.205\times10^{-5}$ & $2.235\times10^{-5}$ & $1.225\times10^{-5}$ & $5.925\times10^{-6}$ & $5.772\times10^{-6}$ \\
$r_5$     & $0$ & $5.286\times10^{-5}$ & $8.896\times10^{-5}$ & $1.041\times10^{-4}$ & $1.260\times10^{-5}$ & $-1.819\times10^{-5}$ & $-2.164\times10^{-5}$ & $-1.694\times10^{-5}$ & $-1.678\times10^{-5}$ \\
$r_6$     & $0$ & $3.209\times10^{-6}$ & $1.809\times10^{-4}$ & $-3.502\times10^{-5}$ & $-4.593\times10^{-5}$ & $-1.620\times10^{-5}$ & $-1.577\times10^{-7}$ & $5.899\times10^{-6}$ & $6.025\times10^{-6}$ \\
\hline
\end{tabular}%
}
\caption{Coefficients obtained by fitting the analytical spectral shape $D(x)$ with the FH basis, then converting to the scattering basis. The values are given with three significant figures as obtained for varying $N$, i.e., the number of $\Ynspec{N}$ components in the computational basis.}
\label{tab:coeffs}
\end{table}

Since for $S_{\rm d}$ and $\Delta S_{\rm d}$ all spectra aside from $D(x)$ are already part of the computational basis in \CT, we only need to decompose the new spectral shape $D(x)$ into the usual spectral basis $(Y_0, Y_1, \cdots, Y_N, M)$. Here, the temperature shift contribution is omitted as $D(x)$ is photon number-conserving by definition (see \cite{Chluba2024DP} for more details). The coefficients are obtained by fitting the analytical spectrum to our approximate model using \MM. Neglecting the residuals distortions ($N=0$, making it a simple $\mu+y$ decomposition) we obtain:
\begin{equation}
    D(x) \simeq 0.835 M(x) -0.0207 \Yspec(x) \,,
\end{equation}
where the coefficients have been projected in the scattering basis. This new basis, which differ from the computational basis used above, improves the representation of $D(x)$ in terms of $\mu$ and $y$ distortions, while ensuring that photon number is conserved \citep{chluba_spectro-spatial_2023-I}.
The coefficients for $N > 0$ are given in Table~\ref{tab:coeffs}. The $\mu/y$ decomposition of $D(x)$ is already a good approximation, as shown in the right panel of Fig.~\ref{fig:DYM}, but our decomposition matches the analytical function almost exactly if we go up to $N = 15$, which is what we normally use in the computations. Moreover, if we use the coefficients to estimate the total energy carried by the distortion, we find that this differs from the analytical result only at the level of $10^{-3}$ (apart from the case with $N=0$ which differs at the percent-level), further confirming the accuracy of the decomposition. The agreement between the two could be further improved at low frequencies if we take advantage of the updated numerical template for $M(x)$ developed in \cite{Evangelista2025}. For the purposes of the dark photon conversion problem, however, we find this to be an unnecessary complication. Finally, Table~\ref{tab:coeffs} shows that the projection of $D(x)$ along $\mu$ dominates, with all remaining coefficients being significantly smaller. If we look at the corresponding energy integrals, we see that the $\mu$ distortion part carries $\simeq 110 \%$ of the total energy for all considered cases, as opposed to the $\simeq -15\%$ from $y$ (and $\simeq 5\%$ in the residual distortions). This implies that $D(x)$ is indeed very similar to $M(x)$, as also highlighted by \cite{Chluba2024DP}. 

Now that we have the FH decomposition of the sources, we can import them into \CT and compute the source terms given in Eqs.~\eqref{eq:dS0} and \eqref{eq:dS1}. Numerically, we model the $\delta$-function using a narrow Gaussian with a relative width of $\sigma = 1.0 \times 10^{-2}$. The value for $\gammacon$ is set keeping in mind that $\gammacon\ll 1$. One can start by choosing $\epsilon$ and $m_{\rm d}$ to compute the conversion parameter with the expression in Eq.~\eqref{eq:gammacon}. Alternatively, one can instead set the conversion redshift, which, using $m_{\rm d} \simeq \omega_{\rm pl}(\zcon)$, corresponds to a specific mass value, and then fix the amount of the distortion energy released in the process, $\Delta \rho / \rho\big|_{\rm d}$. Being able to fix the latter is very convenient if we want to study scenarios which do not violate the constraints from \COBEF \citep{Mather1994, Fixsen1996}.
When properly accounting for the energy versus entropy change (see Appendix~\ref{sec:gammacon-to-Drr}), one obtains \citep{Chluba2024DP}
\begin{equation} 
\label{eq:gamamcon-to-Drr}
\gammacon = \left(\frac{4 G_1}{3G_2}-\frac{G_2}{G_3}
\right)^{-1}\frac{\Delta \rho}{\rho}\Bigg|_{\rm d}  \approx 1.8448\, \frac{\Delta \rho}{\rho}\Bigg|_{\rm d},
\qquad 
\bar{\Theta}=\frac{G_2}{4 G_3}\,\gammacon\approx 0.1708 \, \frac{\Delta \rho}{\rho}\Bigg|_{\rm d} 
\end{equation}
for conversion between variables. We note that in the dark photon source terms, the same amount of energy is contained in the distortion and temperature variable \citep{Chluba2024DP}.

\section{Results}
\label{sec:Results}

In this section, we highlight solutions for the distortion transfer functions and illustrate the impact of various physical parameters. We then present results for the signal power spectra, initially focusing on $\mu T$, $y T$, $\mu E$ and $y E$ but then also briefly highlighting corrections to the temperature power spectra themselves, arguing that iso-curvature type perturbations with a phase shift may be created even for conversions deep in the temperature era ($z > 10^6$).
%

\subsection{Distortion transfer functions for photon conversion} \label{sec:tf}
To gain a better understanding of how the dark photon source terms can impact spectral distortion anisotropies, we first look at the transfer functions. This shows how the distortion amplitudes vary after the conversion and how they evolve until late times. The results presented in this section are obtained by performing a rotation to the \textit{scattering basis}, as defined in \cite{chluba_spectro-spatial_2023-I}. This new basis is obtained by mapping the number conserving spectral shapes, $M(x)$, $\Yspec(x)$ and $\Ynspec{N}(x)$ into a new orthogonal basis using a Principal Component Analysis (PCA), while mapping $G(x)$ onto itself \citep{Chluba2013PCA}. The redefinition allows $\mu$ and $y$ to carry most of the information, with the remaining residual modes being negligible. 

\begin{figure}
    \centering
    \includegraphics[width=1\linewidth]{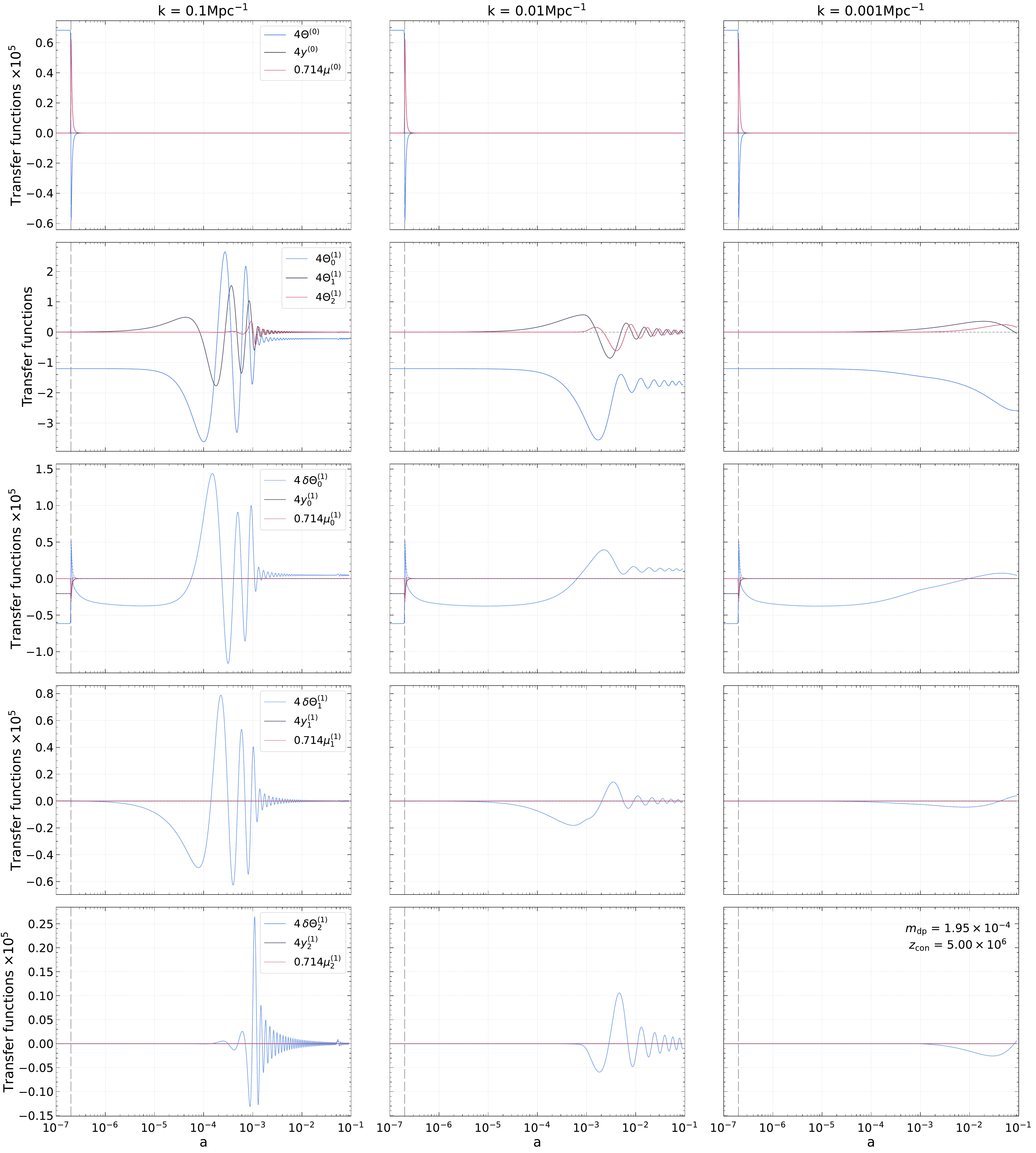}
    \caption{Distortion transfer functions for conversion redshift $\zcon = 5 \times 10^6$ with $\Drrtext = 10^{-5}$ for different wavenumbers. The gray dashed vertical line marks the conversion redshift. The dashed lines show the stationary solution $\vek{y}^{(1)}_\ell(\eta) \approx \vek{b}^{(0)}_0(\eta) \,\Theta^{(1)}_\ell(\eta)$ for reference. In this epoch, both BR and DC are still very efficient, therefore the average distortions thermalise immediately. The solutions are computed using $N = 15$ and then converted into the scattering basis. }
    \label{fig:tf6}
\end{figure}

\begin{figure}
    \centering
    \includegraphics[width=1\linewidth]{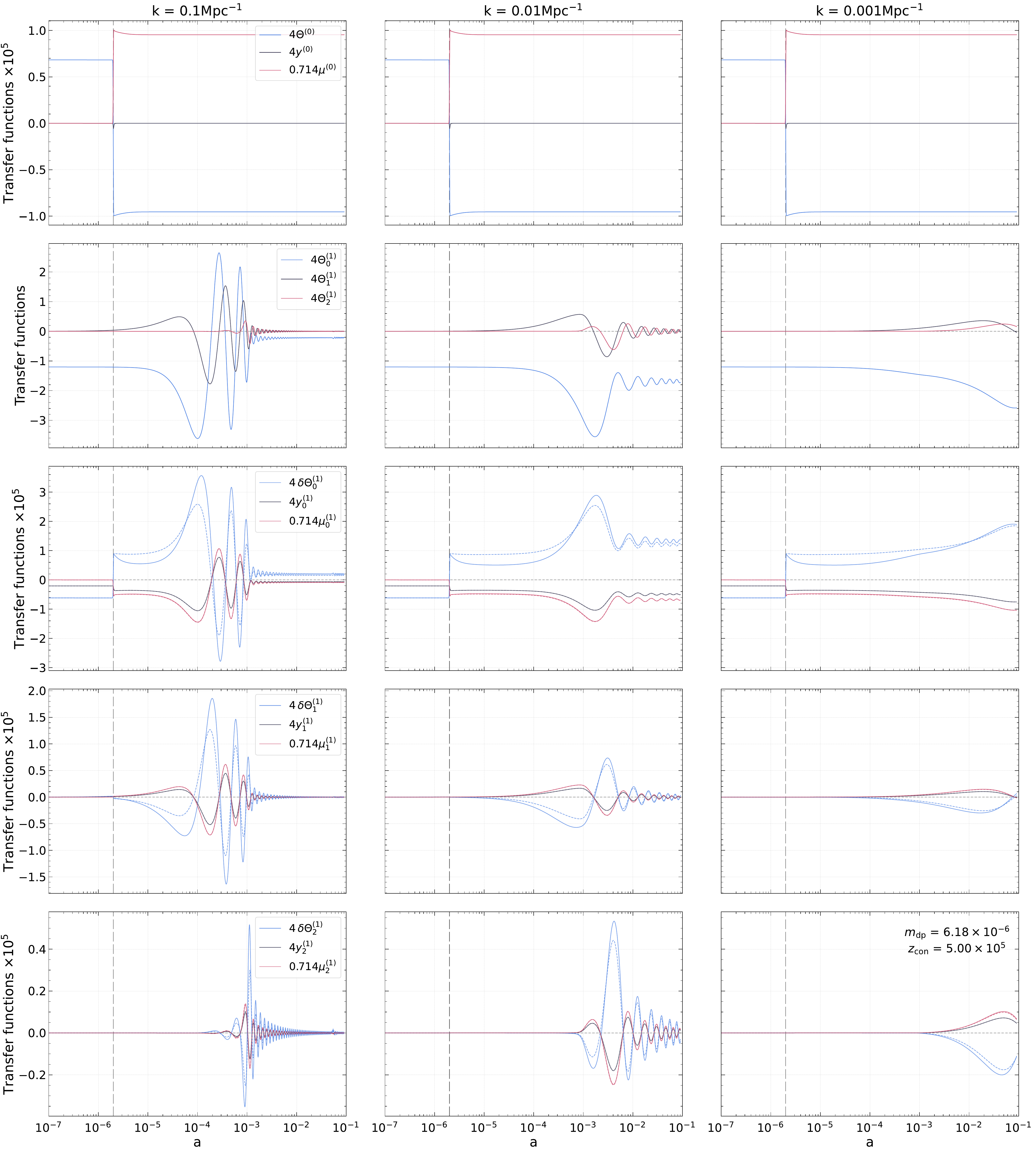}
    \caption{Same as figure~\ref{fig:tf6} but for $\zcon = 5 \times 10^5$.
    We are in the $\mu$-epoch, therefore the average $y$ distortion quickly converts into a $\mu$ distortion, which does not evolve much. }
    \label{fig:tf5}
\end{figure}

\begin{figure}
    \centering
    \includegraphics[width=1\linewidth]{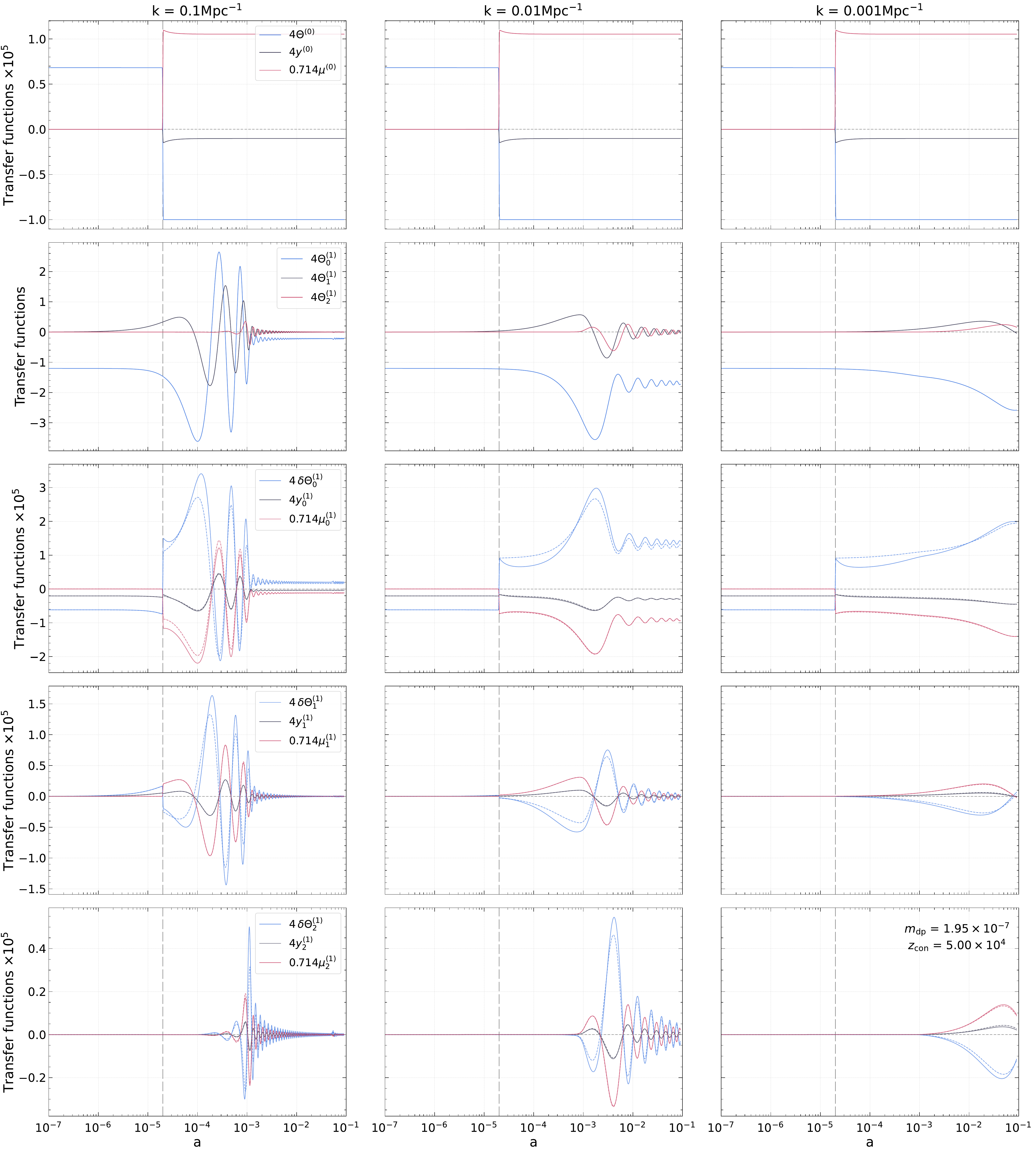}
    \caption{Same as figure~\ref{fig:tf6} but for $\zcon = 5 \times 10^4$. 
    Here, we are in the transition regime between $\mu$ and $y$, Compton scattering starts to be inefficient and the average $y$-distortion does not disappear completely. }
    \label{fig:tf4}
\end{figure}

\begin{figure}
    \centering
    \includegraphics[width=1\textwidth]{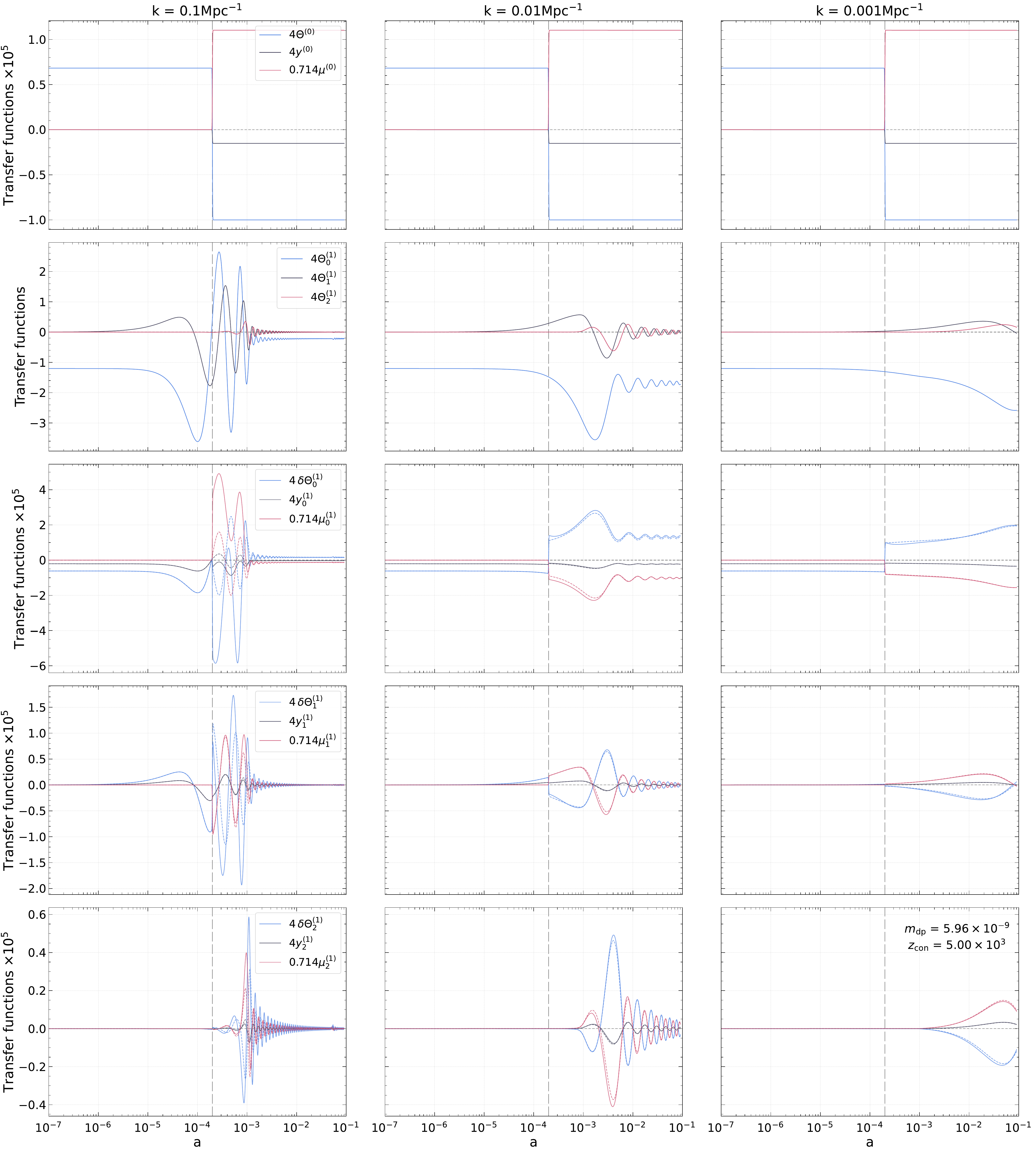}
    \caption{Same as figure~\ref{fig:tf6} but for $\zcon = 5 \times 10^3$.
    We are well within the $y$-epoch but still before recombination, the background do not evolve significantly after the sourcing.}
    \label{fig:tf3}
\end{figure}

In figures~\ref{fig:tf6}--\ref{fig:tf3}, we illustrate the solutions for several wavenumbers and dark photon masses, i.e., conversion redshifts, while keeping the total energy release constant to $\Drrtext = 10^{-5}$ and highlighting some of the physical effects. The system of coupled equations provided by the FH, derived in \cite{chluba_spectro-spatial_2023-II, Chluba2026}, has been solved with \CT using a sixth-order Gear’s method with adaptive time-stepping \citep{Chluba2010}, assuming adiabatic initial conditions. For the transfer functions, we considered $\ell_{\rm max} = 150$ to reduce truncation errors \citep{Ma1995} and obtain accurate results even at small scales and late times. We set an initial temperature shift as described in section~\ref{sec:initial_conditions}.

We start by considering the quasi-instantaneous anisotropic conversion of photons into dark photons occurring at four different redshifts before recombination, spanning both the $\mu$- and $y$-epochs, as well as the transition between the two, commonly known as the \textit{residual} distortion era. 
For each conversion redshift\footnote{We neglect corrections to the photon mass from the presence of neutral atoms, which can lead to multiple conversions especially in the post recombination era \citep{Cyr2024Axions}.}, we determine the corresponding photon mass; the approximate values for each case are reported in the figures. In each figure, the panels in each column represent a different wavenumber, allowing us to explore different scales relevant for the computation of the power spectrum. In addition, we multiplied each transfer function by their energy normalization factor (for instance, see~\citep{Chluba2025encylopedia} for more details), thus essentially illustrating the relative energy density, $\Delta \rho_{\gamma} / \rho_{\gamma}$, carried by each component of the spectral basis. 

In figures~\ref{fig:tf6}--\ref{fig:tf3}, the first row represents the background contribution for the three main distortion variables, $\Theta^{(0)}, \mu^{(0)}, y^{(0)}$ (all just monopoles). The second row illustrates the standard temperature perturbation monopole, dipole and quadrupole, while the last three rows show the monopoles through to quadrupole of $\delta \Theta^{(1)}$, $\mu^{(1)}$ and $y^{(1)}$. We also show the stationary solutions $\vek{y}^{(1)}_\ell(\eta) \approx \vek{b}^{(0)}_0(\eta) \,\Theta^{(1)}_\ell(\eta)$ (dashed lines) as expected in the limit of many scatterings for the distortion field (see section~\ref{sec:analytics}).
In all figures, we can always clearly see the initial perturbations set at $z>\zcon$ (e.g., to the left of the horizontal dashed line). {\tt CosmoTherm} correctly conserves these initial perturbations \citep{Chluba2026}, and without photon conversions we would simply recover the solutions $\Theta^{(0)}=\bar{\Theta}$, $\delta \Theta^{(1)}=3\bar{\Theta}\,\Theta^{(1)}$ and $y^{(1)}=\bar{\Theta}\,\Theta^{(1)}$ at all times, with the other components vanishing.

Focusing on the first rows in figures~\ref{fig:tf6}--\ref{fig:tf3}, for $z<\zcon$, we see the injection of an average $\mu$ distortion which is closely mirrored by a {\it negative} temperature shift and small {\it negative} $y$-distortion, as anticipated from the decomposition given in section~\ref{sec:decomposition}. Given the energy normalization used, we expect the various components to always sum to zero after the conversion, meaning that the CMB is perfectly thermalised with respect to the energy density today. We verified this condition in all the cases considered.  
However, the behaviour immediately after the conversion is drastically influenced by the cosmological epoch. For very early conversions (e.g., $\zcon=\pot{5}{6}$ in figure~\ref{fig:tf6}), when Bremsstrahlung (BR) and Double Compton (DC) are still extremely efficient, the background spectrum thermalises immediately. By construction this leaves no trace of the conversion event in the non-perturbed CMB spectrum.  Instead, for $\zcon = \pot{5}{5}$ (see figure~\ref{fig:tf5}), only the $y$-distortion disappears, being quickly converted into $\mu$, which instead persists in time together with the negative temperature shift. This is because DC and BR emission have slowed down preventing full thermalization while Compton scattering still forces the photon field into kinetic equilibrium with the free electrons. 
For even later conversion redshifts (i.e., figure~\ref{fig:tf4} and figure~\ref{fig:tf3}), there is less and less evolution, freezing the average spectrum generated by the source term. The temperature shift contribution present after the conversion comes from the energy conservation argument explained above. We note that although redundant (as the average spectrum is independent of scale), we repeated the same panel for all wavenumbers considered to allow better comparison with the spatially-varying components.

Now turning to the spatial components (rows two to four) in each figure, we can observe that once the conversion takes place, the various distortion components start to evolve. In figure~\ref{fig:tf6}, the rapid thermalization of the average spectrum makes all the distortion anisotropies disappear, leaving only a non-vanishing temperature correction, $\delta \Theta$. The details of the solution reflect the interplay between anisotropic sourcing terms, $\Delta S_{\rm d}$, rapid thermalization sourcing (i.e., $\mu$ anisotropies converting into $\delta \Theta$) and modifications of the potentials. However, since we do not accurately follow the induced dark photon perturbations, the obtained solutions are only meant for illustration. Importantly, it is clear that photon to dark photon conversions change the composition of the cosmic fluid, thus causing an {\it iso-curvature type perturbation}, which can in principle be used to constrain scenarios with large dark photon mass ($\mdp>10^{-4}\,{\rm eV}$) even when no spectral distortion signatures are present. We leave an exploration of this possibility to future work, noting that the anticipated constraints are likely not as stringent as those obtained for lower masses from distortion signals.

Moving to later conversions, we can observe that the $\mu$ and $y$ distortion solutions are very close to the stationary solution, $\vek{y}^{(1)}_\ell(\eta) \approx \vek{b}^{(0)}_0(\eta) \,\Theta^{(1)}_\ell(\eta)$. Departures increase as the conversion redshift decreases. This is because the corrections to the solutions no longer decay away rapidly as $\tau'$ drops \citep{Chluba2026TC}. In addition, the departures are generally larger at smaller scales (e.g., see figure~\ref{fig:tf4} and especially figure~\ref{fig:tf3}). This is in line with the findings of \citep{Chluba2026TC} and implies that time-dependent information will be imprinted into the signal power spectra.

On the other hand, for $\delta \Theta$, we can always observe some departures from the simple expectation $\delta \Theta^{(1)}\simeq 3 \Theta^{(0)}\Theta^{(1)}$. Given that for early conversions, one has $\Theta^{(0)}\rightarrow 0$, we see that $\delta \Theta^{(1)}$ indeed shows increasing departures with increasing conversion redshift.
This is because additional terms are present in the relevant evolution equation \citep[e.g., Eq.~7.6 of][]{Chluba2026} that cause the corrections to not decay with time \citep{Chluba2026TC}.  
As already mentioned, this reflects small iso-curvature type perturbations, with a visible phase shift, that may hold additional constraining power for large dark photon masses with conversions at $\zcon\gtrsim \pot{2}{6}$.
\begin{figure}[t]
    \centering
    \includegraphics[width=1\linewidth]{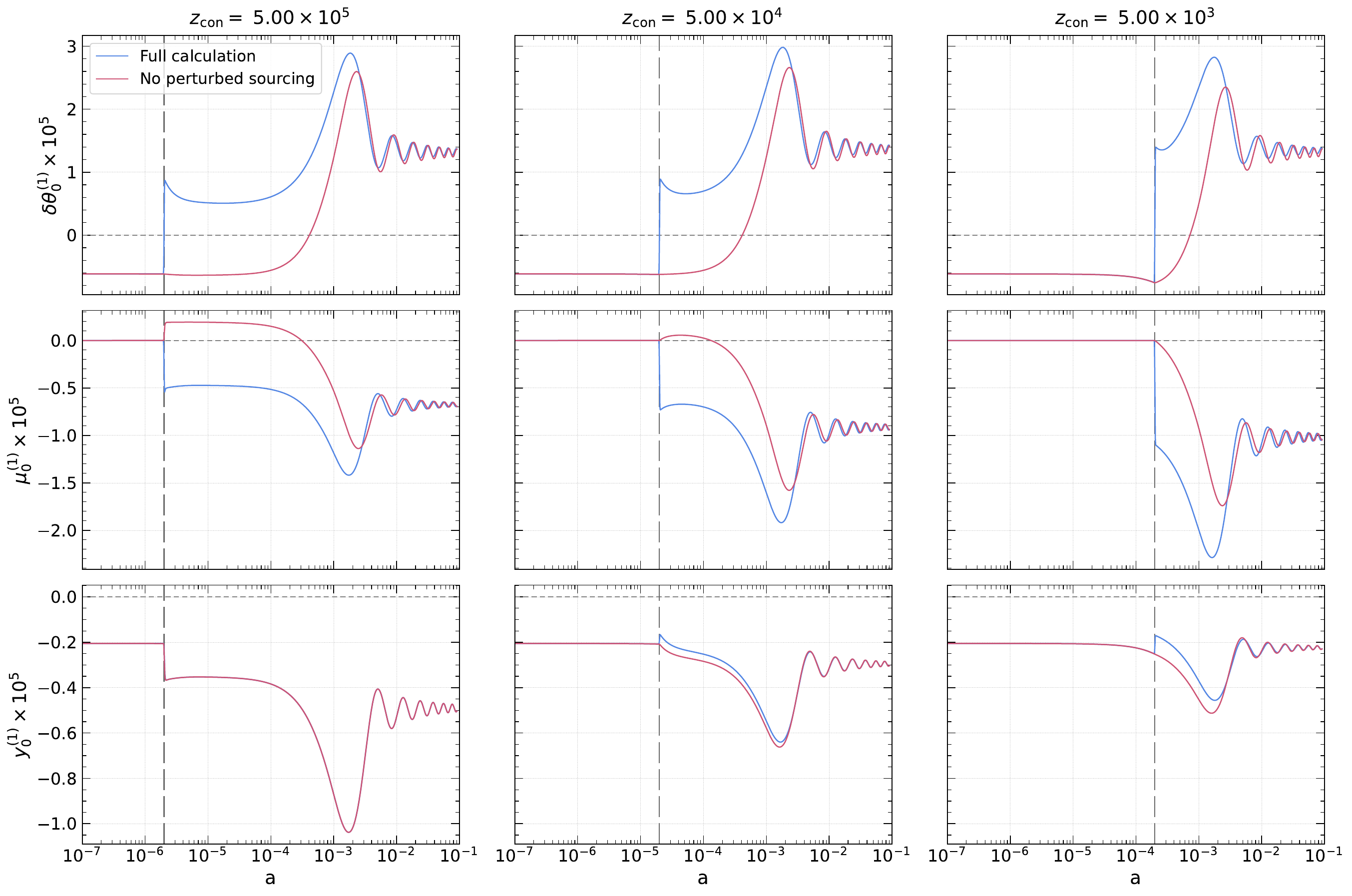}
    \caption{Distortion transfer functions for photons quasi-instantaneously converting into dark photons at various $\zcon$ with $\Drrtext = 10^{-5}$ and for $k = 0.01 \, \Mpc^{-1}$. The gray dashed vertical line marks the conversion redshift.  For each case, we illustrate what happens if we exclude the perturbed source term. We can clearly see the difference at the conversion redshifts, apart from the $y$ distortion at early times ($\zcon\geq \pot{5}{5}$), for which rapid conversion into $\mu$ prevents any noticeable anisotropic source contributions.}
    \label{fig:tf_comparison}
\end{figure}

\subsubsection{Effect of the anisotropic source term} \label{sec:tf-noS1}
In Figure~\ref{fig:tf_comparison}, we illustrate how the transfer functions for $\delta \Theta_0^{(1)}$, $\mu_0^{(1)}$ and $y_0^{(1)}$ are modified if the perturbed source term, $\id \mathcal{S}^{(1)} / \id z$ defined in Eq.~\eqref{eq:dS1} is switched off. In all cases, clear differences are visible in $\delta \Theta_0^{(1)}$ and $\mu_0^{(1)}$ right after the conversion redshift. As time passes, the differences diminish given that the anisotropic sourcing excites decaying distortion modes as discussed in section~\ref{sec:analytics} \citep{Chluba2026TC}. 

The differences are slightly less pronounced for the $y$-distortion. Specifically, for conversions at $\zcon\geq \pot{5}{5}$ we see almost no difference between the two cases. This stems from the fact that external $y$-sources are extremely rapidly converted into $\mu$, converging quickly to the stationary solution set relative to the background spectrum. However, it is clear that in general anisotropic sourcing has a significant effect on the final results for the transfer functions. 

We also note that in principle we would not expect any variation in $\mu^{(1)}$ after the conversion if we only include the background source term. However, we can clearly see a rise in the transfer functions, which is more pronounced for early conversions. This increase is not as sharp as in the case with $\id \mathcal{S}^{(1)} / \id z$, and it stems from the background term present in ${\rm C}^{(1)}_{\rm th}$ [see Eq.~\eqref{eq:evol_yn_final}], for which the importance depends on the efficiency of Compton scattering.

\subsection{Power Spectra for CMB spectral distortions}
Using the computed transfer functions and the line-of-sight (LOS) formalism introduced in section~\ref{sec:line-of-sight}, we are now able to calculate the power spectra for the photon-to-dark-photon conversion scenario. This is particularly important because it allows us to compare the predictions with observational data from existing experiments, such as \Planck, ACT or the upcoming \Litebird mission, without the need to calibrate the monopole. Applying standard CMB analysis pipelines to the cross-correlation of spectral distortions with temperature and $E$-mode polarization, we can open the path to novel constraints on early universe physical processes \citep[e.g.,][]{kite_spectro-spatial_2023-III}. As we will see, for our specific scenario, it would be possible to infer information about the mass and the coupling strength of the dark photon, distinguishing it from other possible energy injection processes. 

Unlike the transfer functions, the results for the power spectra are given in the \textit{observational basis}, to allow a more direct comparison with frequency-binned real data \citep[e.g.,][]{Rotti2022muT}. This basis compresses the information to a reduced number of spectra, using a principal component analysis (PCA) to map the full computational basis onto the standard $\Theta$, $\mu$, $y_0$ plus some residual distortion parameters, $r_i$, with decreasing observability \citep{Chluba2013PCA}. By construction, the residual distortions receive contribution only from the $y_N$ parameters, while the new $\Theta$, $\mu$, $y_0$ will be a combination of all the components. This introduces some ambiguity in the photon number conservation of the individual spectral shapes; nevertheless, the global evolution and conservation properties remain unchanged. It should also be noted that this basis describes the solutions only within a given frequency range, in analogy to frequency limitations of real experiments. For more details about the topic and the conversion between various bases, one can refer to \cite{chluba_spectro-spatial_2023-I}.

\begin{figure}
    \centering
    \includegraphics[width=0.49\linewidth]{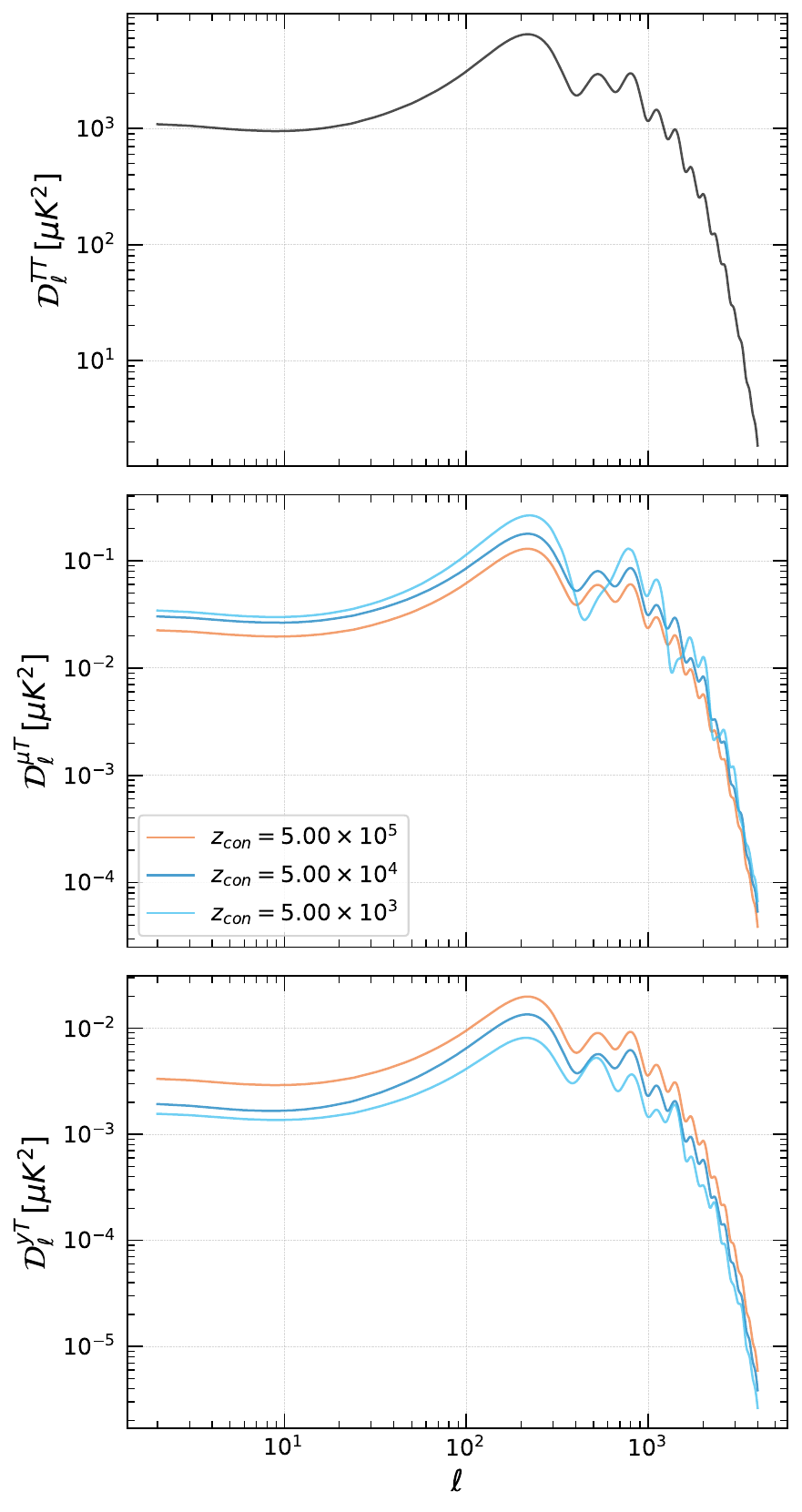}
    \hspace{2mm}\includegraphics[width=0.49\linewidth]{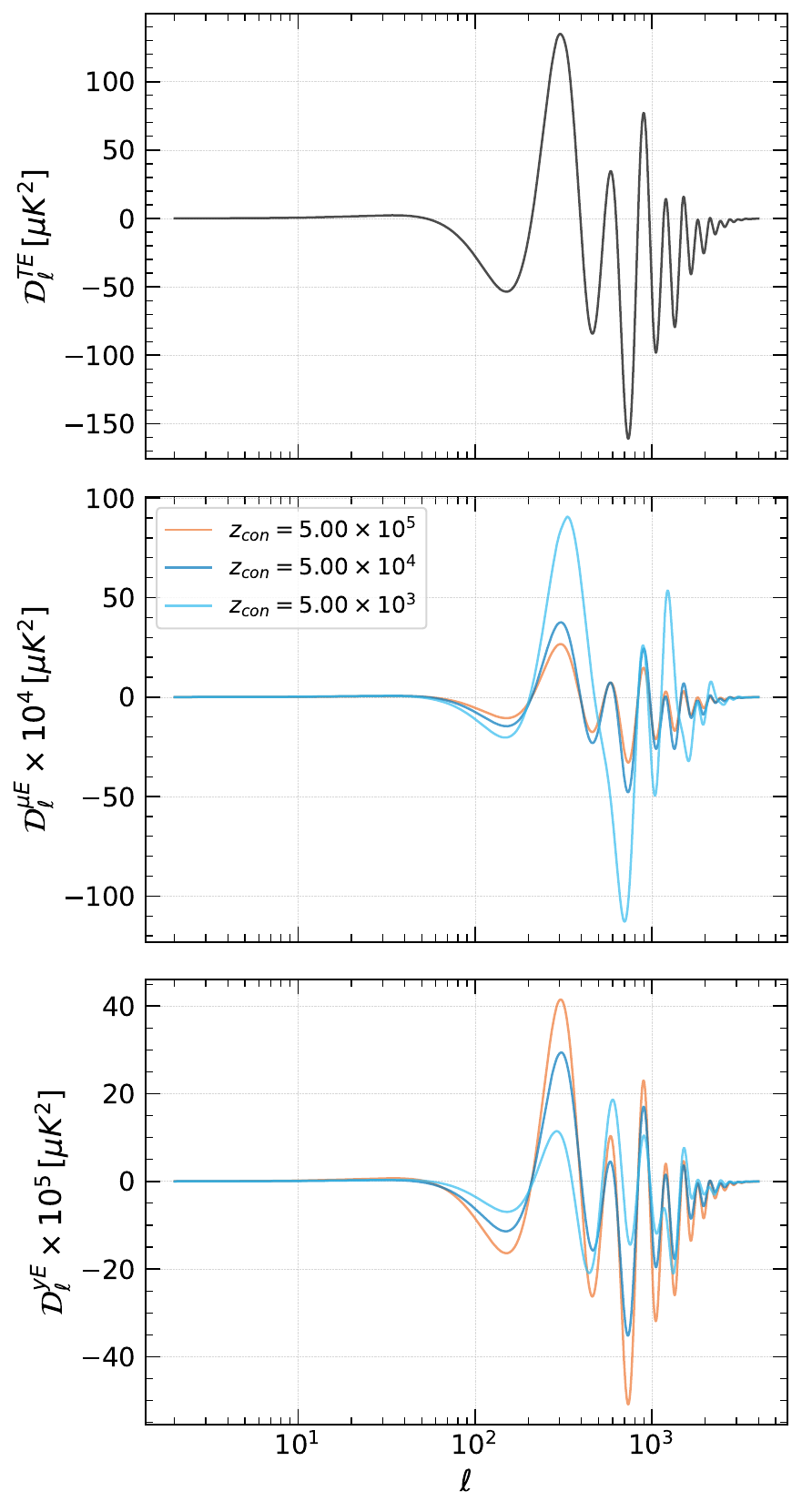}
    \caption{Cross-correlations of the CMB temperature and $E$-mode polarisation with the $\mu$ and $y$ spectral distortions, compared with the standard $TT$ and $TE$ power spectra. We selected a set of relevant conversion redshifts. As for the transfer functions and we fixed the energy loss to $\Drrtext = 10^{-5}$. }
    \label{fig:PS-TX-EX}
\end{figure}

\begin{figure}
\centering
\includegraphics[width=0.49\linewidth]{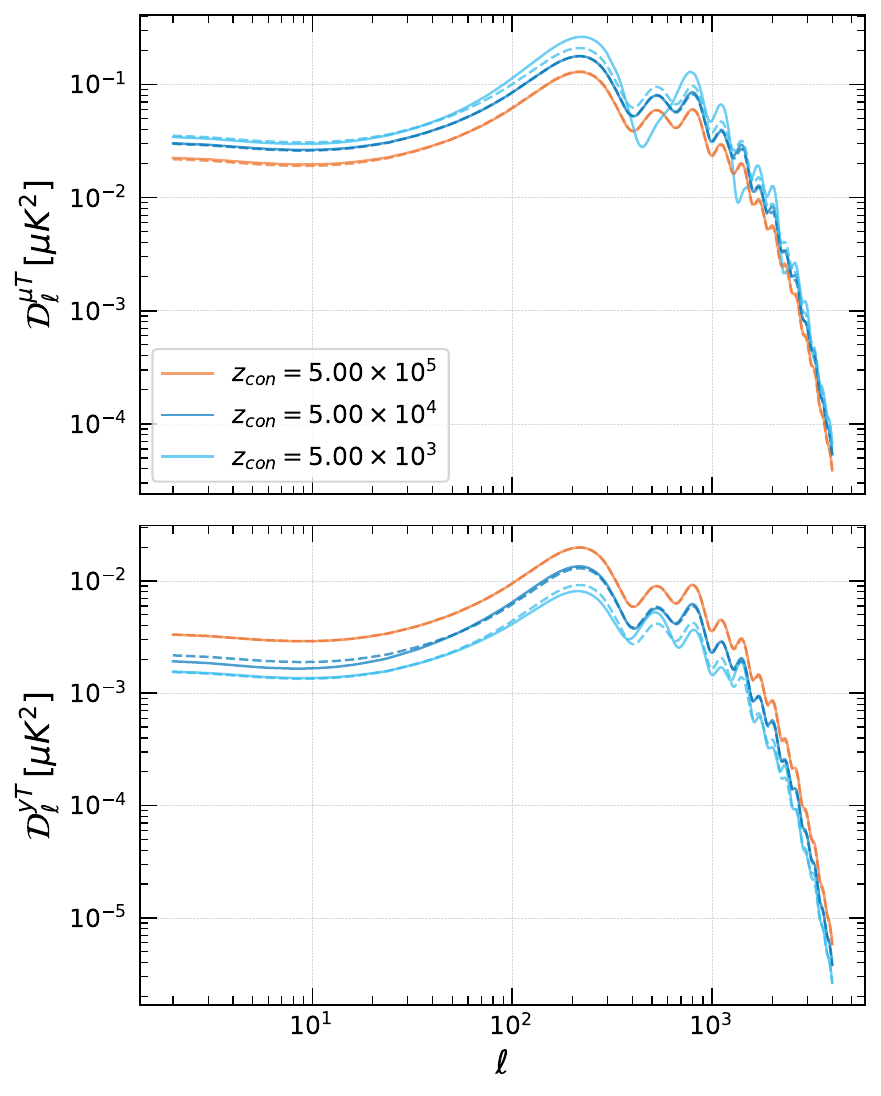}
\hspace{2mm}\includegraphics[width=0.49\linewidth]{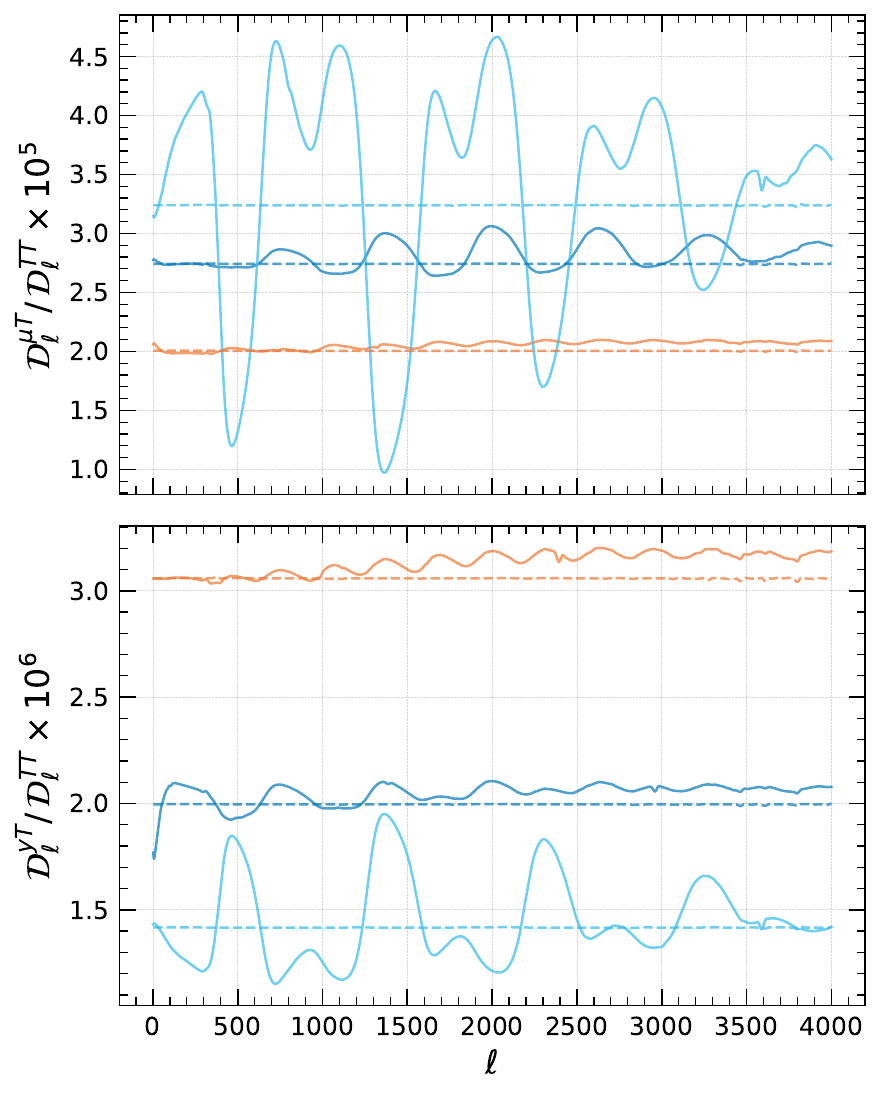}
\caption{On the left, the computed power spectra for $\mu T$ and $y T$ are compared to the stationary solution (dashed line), highlighting noticeable differences that increase for later conversion redshifts. On the right, we present their ratio with respect to the standard $TT$ power spectrum. The full solution presents a complex pattern close to recombination, which becomes progressively flatter going back in time. On the other hand, the approximate solution has no feature, being only a rescaling of the temperature spectrum by construction. For all conversions, the energy loss is fixed to $\Drrtext = 10^{-5}$.}
    \label{fig:PS-approx_TT}
\end{figure}

The power spectra are solved by \CT using the numerical setup introduced in section~4.1 of \cite{kite_spectro-spatial_2023-III}. In the system, the standard CMB temperature perturbations are solved separately from their corrections $\delta \Theta^{(1)}$, reproducing the \texttt{CLASS} results for $\mathcal{C}^{TT}_{\ell}$. All the results of this section are presented with the usual CMB normalisation 
$$\mathcal{D}_{\ell}^{XY} = \frac{\ell (\ell + 1)}{2 \pi} \mathcal{C}_{\ell}^{XY}.$$ 
As for the transfer functions, we fixed the relative energy loss at the $ 10^{-5}$ level and vary the dark photon mass (equivalent to varying the conversion redshift), focusing on the pre-recombination era only. We consider exactly the same cases as in the previous section, although in most of the figures $\zcon = 5 \times 10^6$ is omitted since it carries little distortion information due to extremely efficient thermalisation. We also checked that the results are almost independent of the computational basis size, so we adopted $N = 5$ for efficiency. Differently from the transfer functions, here we set $\ell_{\rm max} = 15$ because only for the largest scale modes we need to compute the transfer functions to late times, with truncation errors remaining small \citep{Ma1995, Seljak1996}.

We start by looking at Fig.~\ref{fig:PS-TX-EX}, where the standard power spectra for $TT$ and $TE$ (which do not vary with the conversion redshift) are compared to their cross-correlation with the distortion variables. As expected, the latter are orders of magnitude smaller than the former, with $\mu T$ and $\mu E$ dominating due to the decomposition of $D(x)$ observed in section~\ref{sec:decomposition}. Both correlations with the temperature and the $E$-modes present a rich time-dependent peak structure at small scales, fundamental to constraining the dark photon mass. On the other hand, we note that at low $\ell$ the signal becomes flatter and featureless, reproducing the well-known Sachs-Wolfe plateau. 

To highlight the time-dependent information a bit more, let us focus on the $\mu T$ and $y T$ signals. In the left panel of Fig.~\ref{fig:PS-approx_TT}, we show the full numerical $\mu T$ and $y T$ power spectra together with the ones obtained using the stationary solution $\vek{y}^{(1)}_\ell(\eta) \approx \vek{b}^{(0)}_0(\eta) \,\Theta^{(1)}_\ell(\eta)$ (dashed lines), which is essentially a rescaled version of the $TT$ spectrum as mentioned in section~\ref{sec:analytics}. The two are in good agreement, as are the transfer functions, especially for early conversions and at large angular scales. For lower conversion redshifts when electron scattering is less efficient, the approximation becomes less accurate, and larger differences are seen between the solutions. Finally, in the right panel of Fig.~\ref{fig:PS-approx_TT}, we show the ratio with respect to the fixed $TT$ spectrum, to better highlight the novel information. As regards the full solution, we can clearly observe a complex pattern which grows as the conversion redshift approaches recombination, while earlier conversions show almost no scale-dependent features, with only some oscillations.
We explored this aspect and believe it is caused by the fact that we are not including polarised effects in the spectral distortion hierarchy. Indeed, when we do not include polarization corrections in the computation of the power spectra based on the stationary solution $\vek{y}^{(1)}_\ell(\eta) \approx \vek{b}^{(0)}_0(\eta) \,\Theta^{(1)}_\ell(\eta)$, we found similar levels of residual wiggles. However, we leave a detailed investigation to future work, since a full treatment of polarized radiative transport lies beyond the scope of the present paper.\footnote{We also see some small numerical artifacts in the solutions whose origin we leave for future investigation.}  
%

\begin{figure}
    \centering
    \includegraphics[width=0.49\columnwidth]{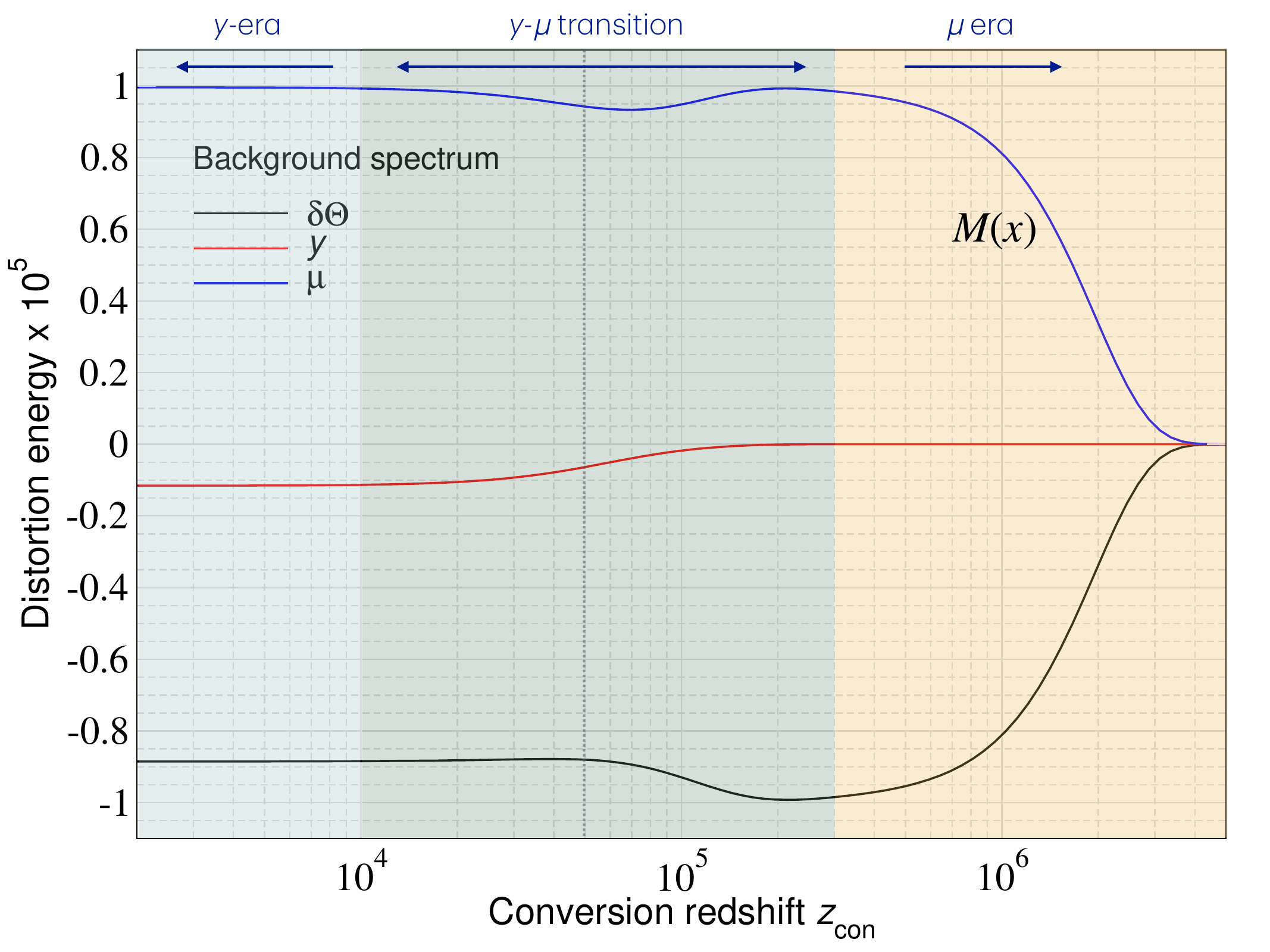}
    \hspace{0.2mm}
    \includegraphics[width=0.49\columnwidth]{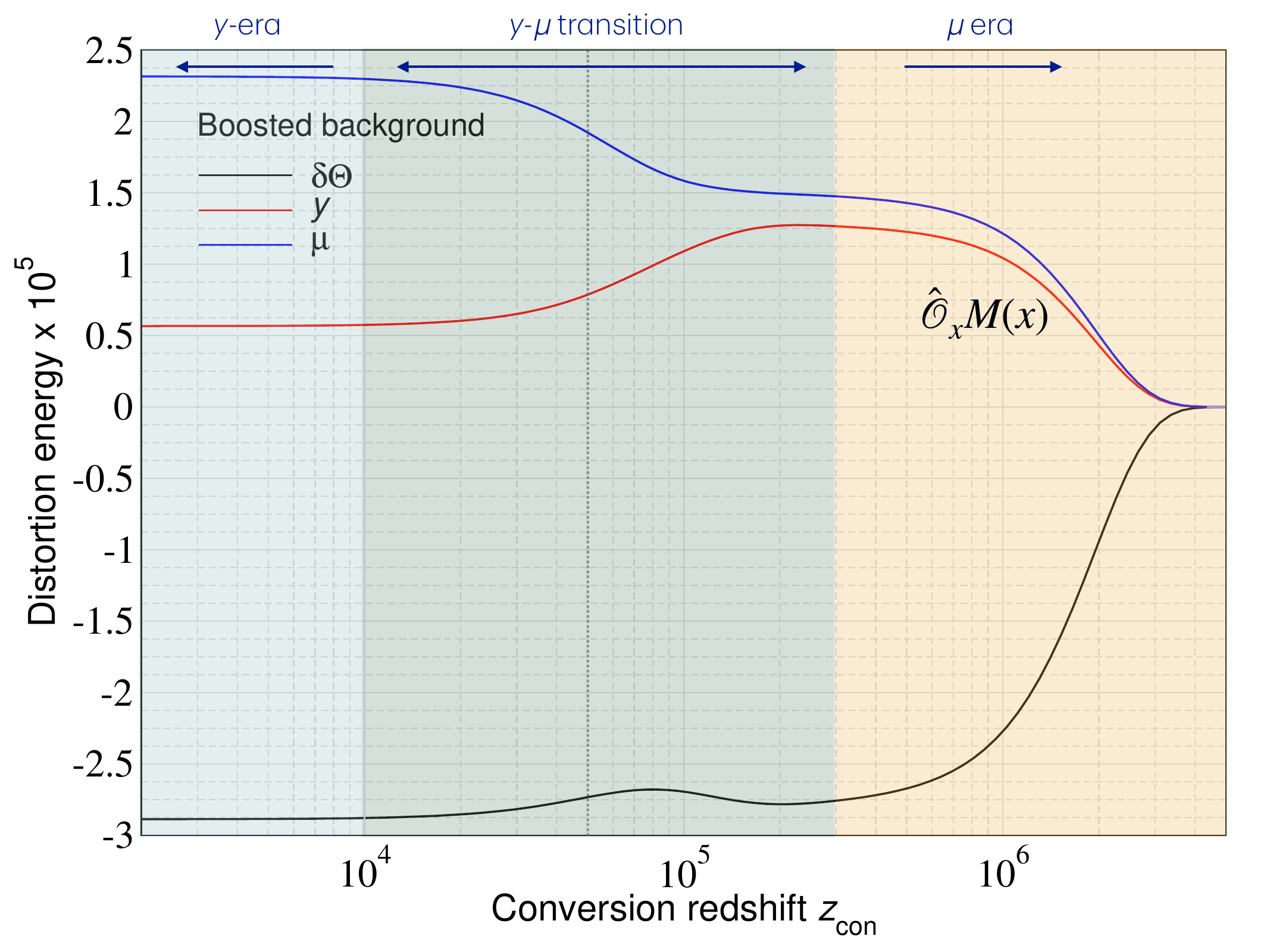}
    \\
    \caption{Decomposition of distortion signals in the observer basis and evaluated at $z=1100$ for varying conversion redshifts. 
We present all components in terms of their fractional energy contribution, $\Drrtext$, for a total distortion energy $\Drrtext\big|_{\rm d}=10^{-5}$.
    The left panel shows components of the background distortion, $\vek{y}^{(0)}_0$, while the right panel shows those for $\vek{b}_0^{(0)}=M_{\rm B} \vek{y}^{(0)}_0$. We also marked the main regimes of Compton scattering known from the thermalization Green's function \citep{Chluba2013Green}. The $y$-distortion component changes sign with respect to the background and also has a slightly higher amplitude at $\zcon>5 \times 10^4$ due to projections from the $y_{k>0}$. In the $\mu$-era at $\zcon > \pot{3}{5}$, where the background distortion is close to $M(x)$ (i.e., vanishing $y$ projection), the anisotropy distortion spectrum is close to that of $\boostO M(x)$ with roughly equal amounts of $\mu$ and $y$.
    \vspace{-0mm}}
    \label{fig:b0_illustration}
\end{figure}
We also highlight that for the considered cases the amplitudes of the $\mu T$ and $y T$ power spectra exhibit opposite trends with $\zcon$. Specifically, $y T$ seems to be increasing with $\zcon$ which is counter-intuitive given that at $\zcon > 50,000$ $\mu$-distortions are expected to be dominant.
In addition, the transfer functions for the $y$ distortion at the background level have a negative sign (e.g., Fig.~\ref{fig:tf_comparison}), and yet the $y T$ correlation is positive.
To understand these aspects, we have to consider the decomposition of $\vek{b}_0^{(0)}=M_{\rm B} \vek{y}^{(0)}_0$ at $z\simeq 1100$ in the observer basis, which is the one relevant to the power spectra [see Eq.~\eqref{eq:stationary_sol_ps}]. In Fig.~\ref{fig:b0_illustration} we illustrate the decomposition of $\vek{y}^{(0)}_0$ and $\vek{b}_0^{(0)}$ evaluated at $z=1100$ as a function of the conversion redshifts. Firstly, the background distortion indeed has $y<0$ even in the observer basis. We can also see that the total sum of the $\delta \Theta, y$ and $\mu$ components in terms of the energy density is extremely close to zero (at most a few percent), as happened in section ~\ref{sec:tf}, showing that these three components contain most of the information. 

Turning to $\vek{b}_0^{(0)}$, we observe that the contributions from $\mu$ and $y$ are now both positive. This is due to a combination of the effects of boosting and conversion into the observer basis. 
For instance, $\boostO M \approx 1.9 M+0.4 Y$ \citep{chluba_spectro-spatial_2023-II}, where computing the energy density fractions in $M$ and $Y$ shows that these essentially share to total energy in almost equal amounts, as can also be seen from the right panel in Fig.~\ref{fig:b0_illustration} at $\zcon \gtrsim 5 \times 10^4$.
In addition, the $\mu$-distortion projection rises at $\zcon \lesssim 5 \times 10^4$, while the $y$-projection decreases, again due to interplay between boosting and observer basis projection. This explains the overall trends seen in the right panel of Fig.~\ref{fig:PS-approx_TT}. It also implies that constraints that are derived using only $\mu T$ will not provide the full constraining power from distortion anisotropies, as the information in the $y T$ power spectrum is omitted. As we will see in section~\ref{sec:constraints}, the constraints indeed weaken for dark photon scenarios with conversion redshift $\zcon > 5 \times 10^4$.

Finally, the right panel of Fig.~\ref{fig:PS-approx_TT} together with Fig.~\ref{fig:b0_illustration} show an additional important point: at large angular scales and for early conversions ($\zcon>5 \times 10^4$) the spectrum of the anisotropies is always very close to $\boostO M$ in the observer basis, with $\simeq 50\%$ of the energy being in $y$. This means that from the observational point of view it would be best to constrain dark photon models analyzing the CMB data using this spectral template instead of $M$, thereby ensuring that nearly all the spectro-spatial information is accounted for. For later conversions, the loss of information when using the spectrum $M$ for the analysis is only $\simeq 20\%$. This indicates that the choice of the spectral template can make a difference and also depends on the scenario that is being constrained.

\begin{figure}
    \centering
    \includegraphics[width=\columnwidth]{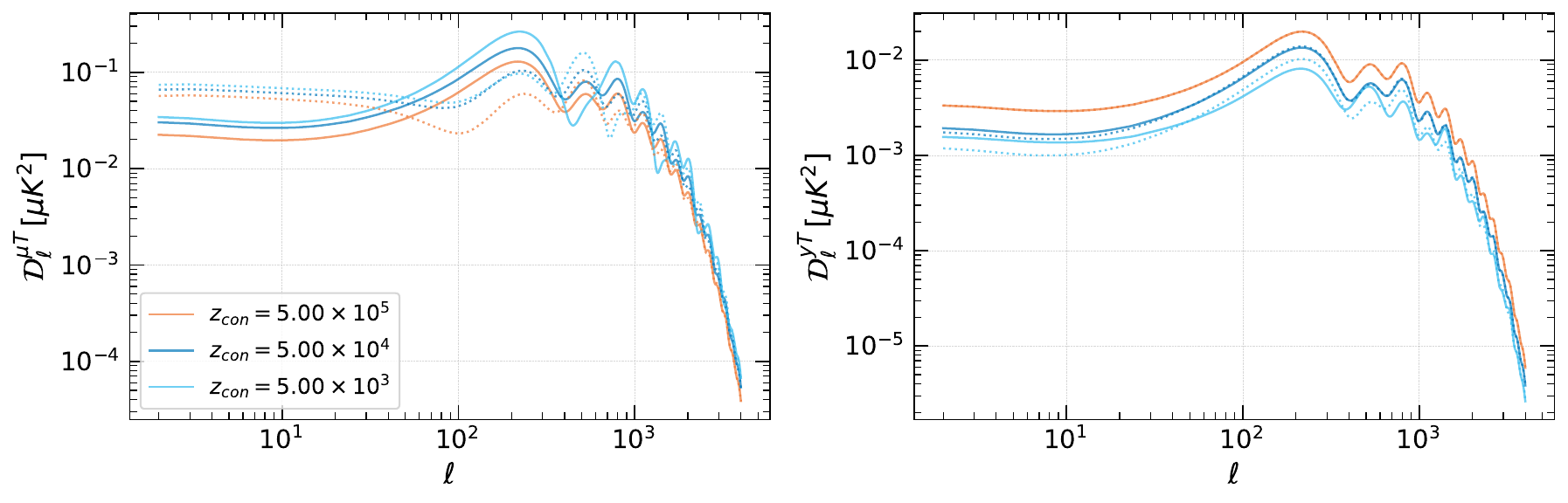}
    \vspace{-0mm}
    \caption{Distortion cross power spectra for three different conversion redshifts, showing the difference with (solid line) and without (dotted line) the anisotropic source term $\id \mathcal{S}^{(1)} / \id z$. Significant differences are visible for the $\mu T$ case at large scales.
    \vspace{-0mm}}
    \label{fig:PS-noS1}
\end{figure}

\begin{figure}
    \centering
    \includegraphics[width=0.49\linewidth]{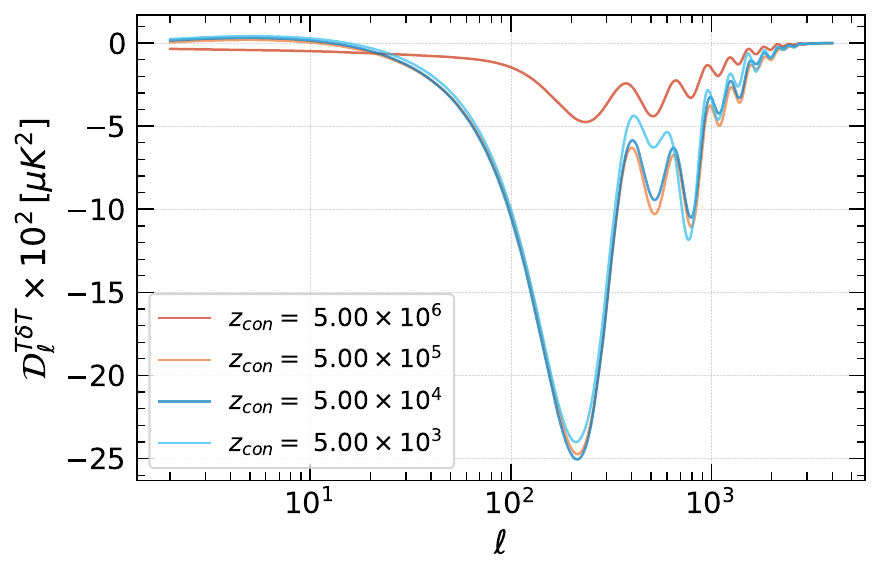}
    \hspace{1mm}
    \includegraphics[width=0.49\linewidth]{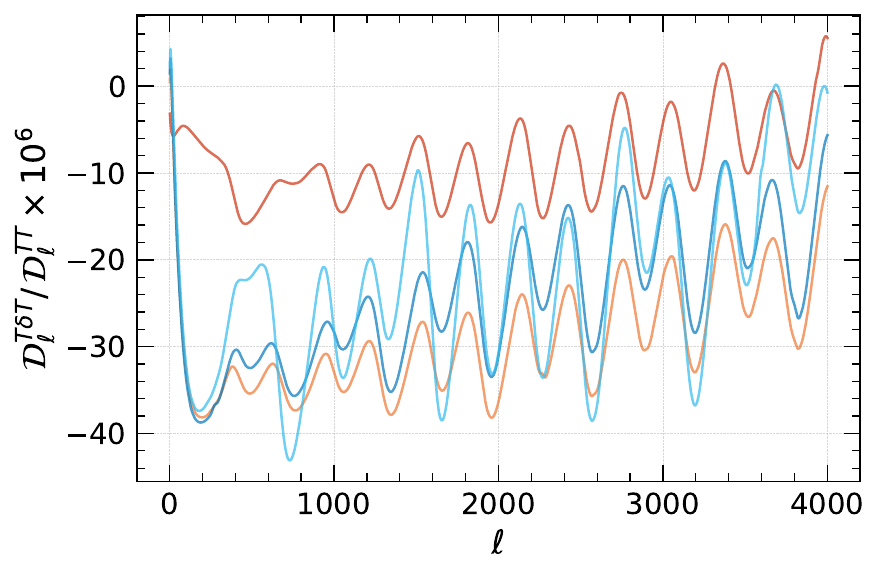}
    \vspace{-0mm}
    \caption{On the left, the cross-correlations of the CMB temperature with its corrections is shown. On the right, one can observe the ratio with respect to the TT spectrum. Differently from previous figures, the conversion happening at $5 \times 10^6$ is included, since it also carry important information. The pattern of the solution is different from the $\mu$ and $y$ case, exhibiting a characteristic iso-curvature phase shift. 
    \vspace{-0mm}}
    \label{fig:PS-TTcorr}
\end{figure}

\subsubsection{Effect of the anisotropic source term}
In analogy to what has been done in section \ref{sec:tf-noS1}, we explore how the effect of anisotropic sourcing affects the signal power spectra. An illustrative comparison is shown in Fig.~\ref{fig:PS-noS1}.
For both the $\mu T$ and $y T$ power spectra noticeable differences are present when anisotropic sourcing is omitted, especially at large angular scales. This is most obvious for the $\mu$ distortion at all considered conversion redshifts, as opposed to the $y$ distortion case where for early conversions there is no visible difference. This reflects the results found for the transfer functions and it is related both to the spectral decomposition of the source term and the efficiency of the thermalisation.

\subsubsection{Cross-correlation with the temperature corrections}
\label{sec:delta_Theta_terms}
As a final point of the power spectrum analysis, we study the cross correlation of the standard CMB temperature anisotropies with respect to its corrections ($\delta T=T_0 \delta \Theta$). As shown in Fig.~\ref{fig:PS-TTcorr}, this cross correlation exhibits a behavior that differs from the correlation with distortions. In this context, it is not useful to compare with the stationary solution since the differences are expected to be too large, as also highlighted at the end of section \ref{sec:tf}. 

One fundamental difference is that now even conversions taking place at $\zcon = 5 \times 10^6$ provide information that can in principle lead to constraints. This is visible in both panels of Fig.~\ref{fig:PS-TTcorr}, where the corresponding line is comparable in magnitude to the cases previously considered.  Importantly, it is clear that photon to dark photon conversions change the composition of the cosmic fluid, thus causing an {\it iso-curvature type perturbation}, which can in principle be used to constrain scenarios with large dark photon mass ($\mdp>10^{-4}\,{\rm eV}$) when no spectral distortion signatures are present. This is especially visible for the early conversion scenarios shown in Fig.~\ref{fig:PS-TTcorr}, where one can observe a characteristic phase shift with respect to the adiabatic perturbations that dominate $C_\ell^{TT}$.

However, to precisely treat the effect of dark photon conversion on the temperature perturbations we would also have to consider how the dark photon perturbations propagate. On average, we can neglect any change to the background expansion as photons and dark photons are both relativistic and hence no corrections are expected. For the perturbation, we could simply consider the dark photon anisotropy sources as {\it effective} sources in the neutrino equations (i.e., assuming the dark photons simply free stream). Because the direct changes to the temperature variables do not propagate back to the distortion variables at the considered perturbation order and are thus not separately distinguishable. We anticipate that even if we can in principal probe higher masses, the corresponding constraints from iso-curvature perturbations will be weaker than those obtained with other probes, such that we decided to leave a more detailed treatment to future work. 

\begin{figure}[t]
    \centering
    \includegraphics[width=0.49\columnwidth]{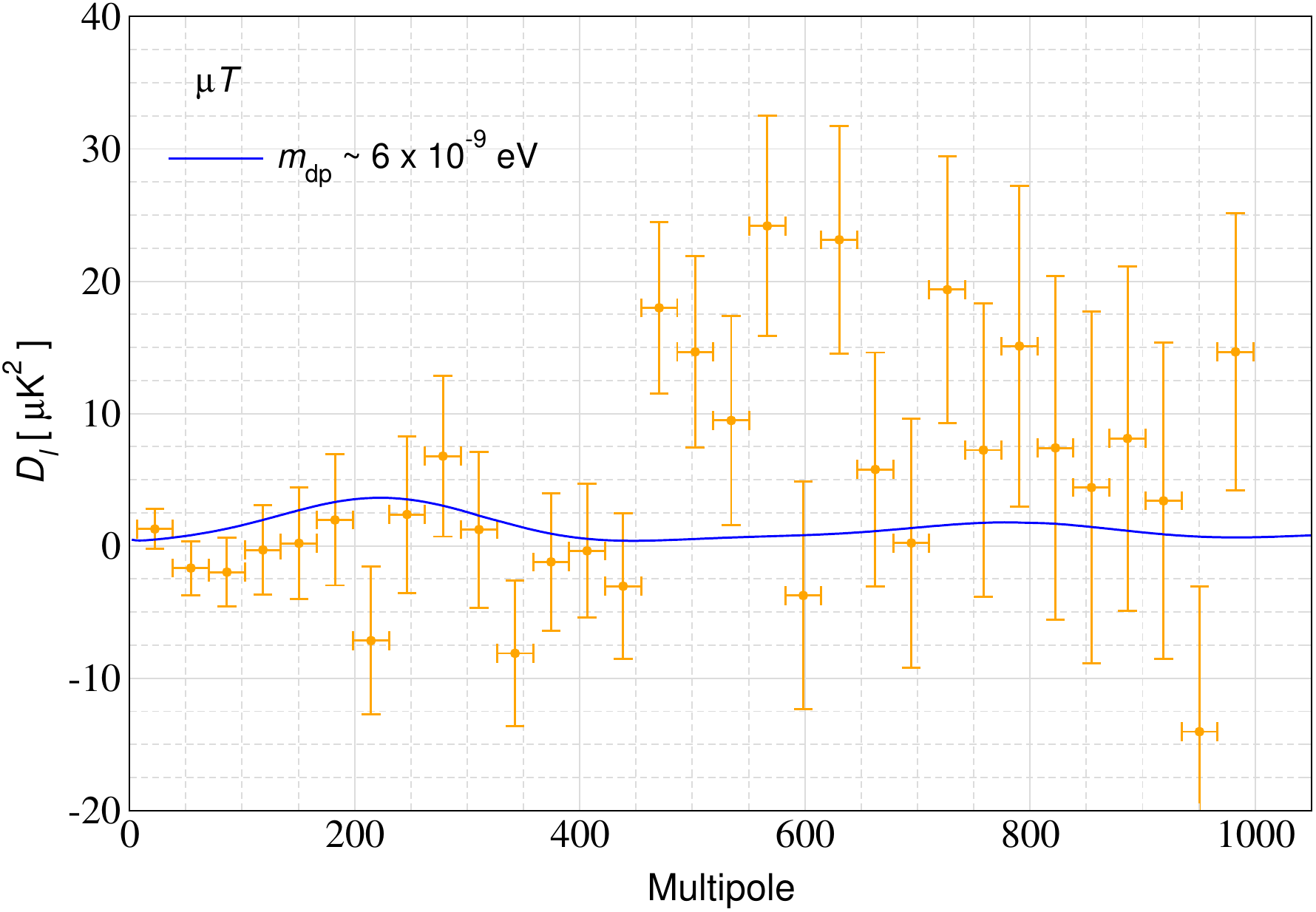}
    \hspace{1mm}
    \includegraphics[width=0.49\columnwidth]{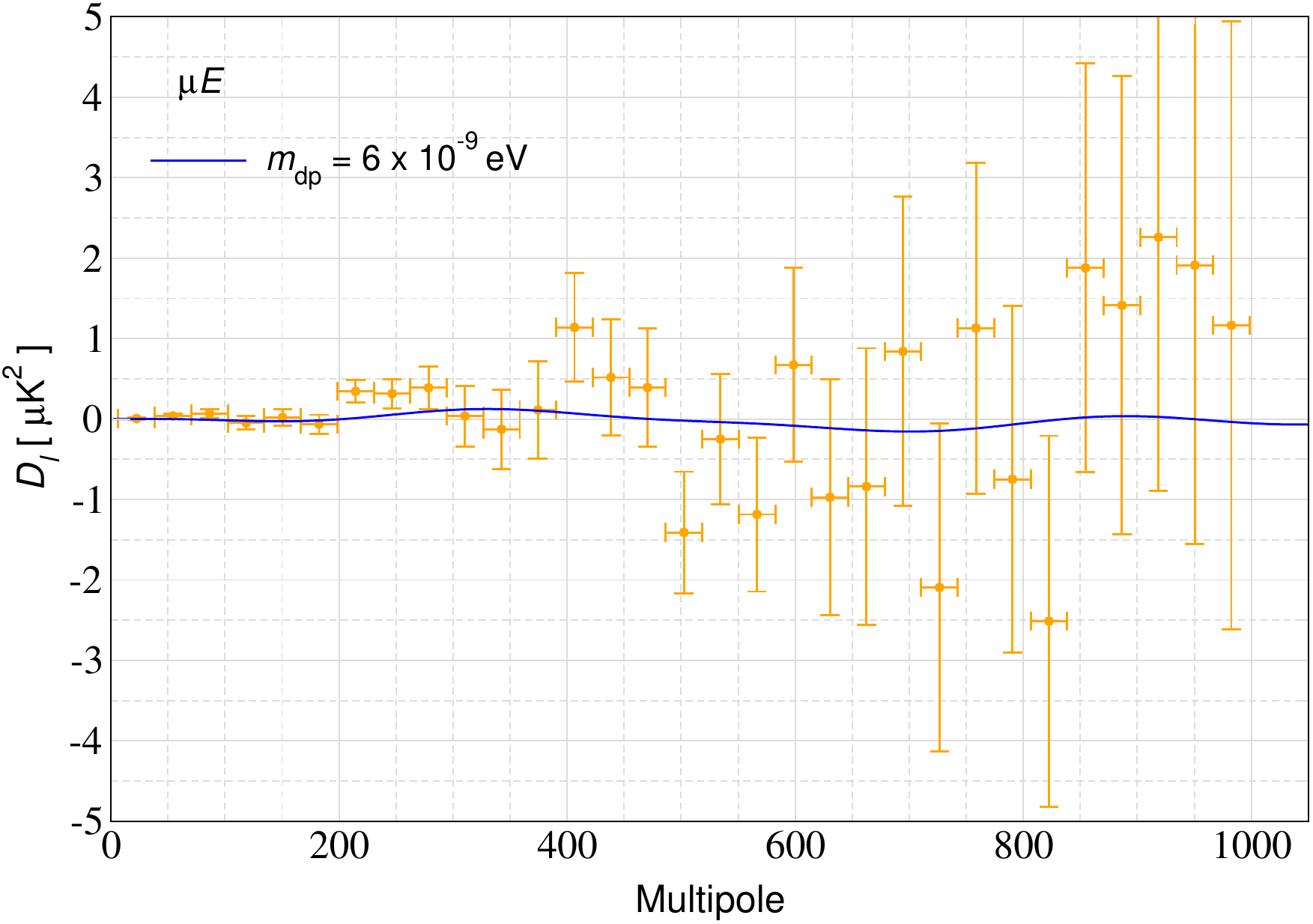}
    \caption{Distortion cross power spectrum measurements from \Planck \citep{Rotti2022muT}. We show the binned data points and error bars for the $\mu T$ and $\mu E$ cross power spectra together with a dark photon model (dark photon mass $\mdp\simeq \pot{6}{-9}\,{\rm eV}$, conversion redshift $\zcon = 5000$ and $\epsilon \simeq \pot{8}{-8}$) that is consistent with the data at 95\% c.l.}
    \label{fig:constraints_PS}
\end{figure}

\section{Dark photon constraints using \Planck data}
\label{sec:constraints}
Now that we have the cross-correlation spectra for various conversion redshifts, we can use existing limits on $\mu T$ and $\mu E$ obtained using \Planck data \citep{Rotti2022muT} to constrain the photon conversion process. In principle, the $y T$ signal also contains some valuable information, but for the considered dark photon models, the overall contribution is subdominant and in addition is contaminated by ISW effects from clusters of galaxies \citep[see discussion in][]{kite_spectro-spatial_2023-III}, such that we do not consider it here. 

Here, we estimate the constraints on dark photon scenarios with conversions at $\zcon\gtrsim 200$. For $\zcon\geq 2000$ (i.e., $\mdp\gtrsim 10^{-9}\,{\rm eV}$), our line-of-sight treatment should be highly accurate and the assumption of single conversions is also well justified. However, at $\zcon\lesssim  2000$ our current setup becomes less accurate and in addition one could encounter multiple conversions in a frequency-dependent manner \citep{Cyr2024Axions}. The main reason for loss of precision is because the direct photon source terms start becoming noticeable,  since the optical depth suppression (i.e., isotropization) becomes less complete at $\zcon \lesssim 1400$ and the direct visibility of sources increases. We can ease the treatment by increasing the width of the Gaussian that models the rapid conversion to $10\%$. At late times, when thermalization process are already negligible, this does not affect the results much but improve numerical convergence. 
However, a more careful analysis will be in order in the future.

\begin{figure}
    \centering
    \includegraphics[width=0.98\columnwidth]{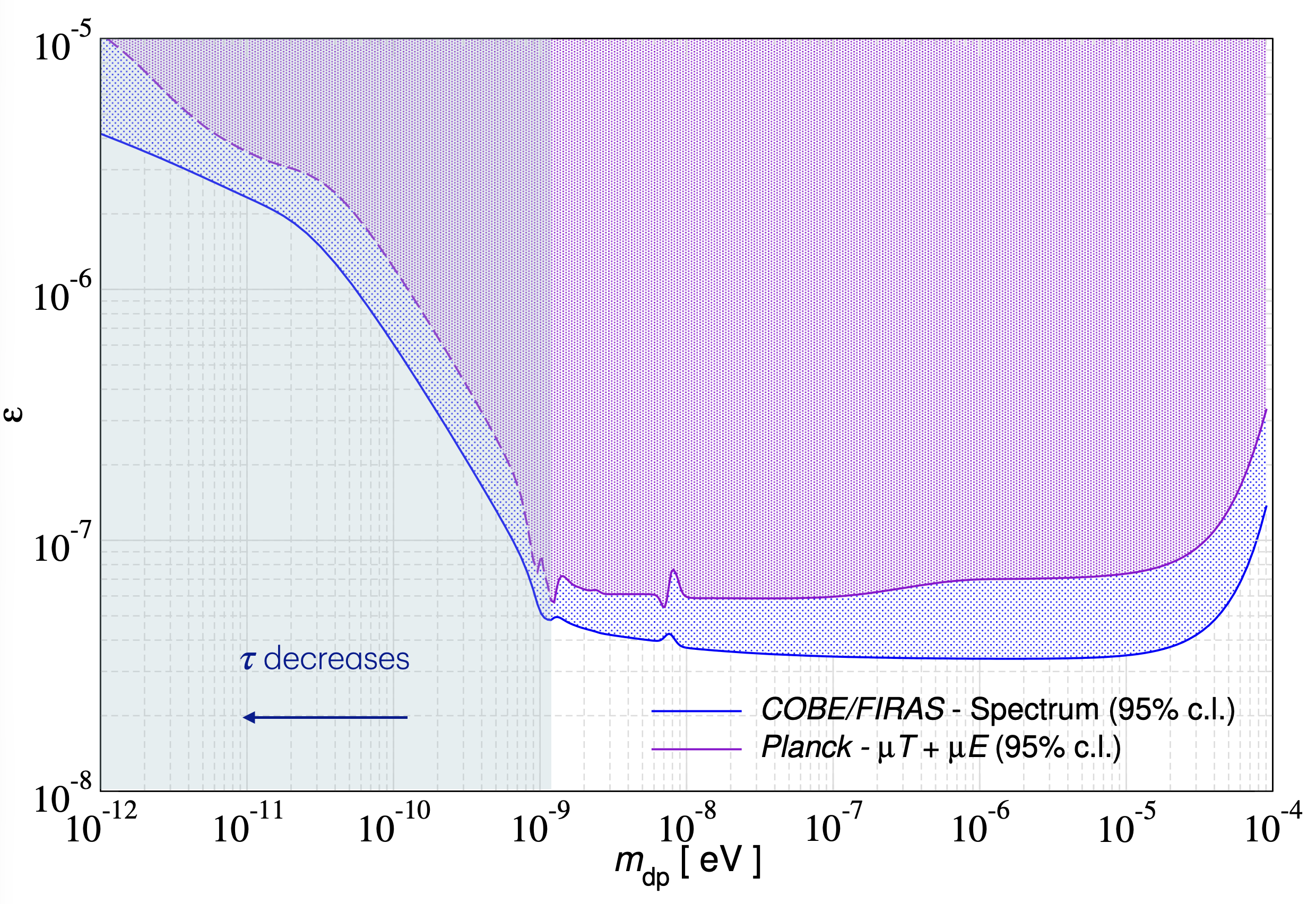}
    \caption{Constraints on the dark photon parameter space. The results obtained here from the analysis of \Planck $\mu T$ and $\mu E$ data are compared to those from the analysis of the CMB monopole spectrum \citep{Chluba2024DP}. The region of decreasing Thomson optical depth is highlighted, indicating the domain where the direct photon source terms become important and our current computations of the $\mu T$ power spectra are expected to be less accurate. The features in the constraints stem from changes in the conversion efficiency during the cosmological recombination eras of helium and hydrogen.
}
    \label{fig:constraints_eps}
\end{figure}

To derive the constraints, we use a simple $\chi^2$-test assuming that the data covariance matrix is diagonal. Given the model power spectra $C^{\mu X}_\ell(\Drrtext=10^{-5})$ for $\Drrtext=10^{-5}$, cases with larger or lower energy conversion can be obtained by simple scaling.
This gives the log-likelihood function\footnote{We evaluate the model power spectra at the central values of the data points only. We also neglect the possible covariance of the data points, deferring a more rigorous analysis to the future.}
\bealf{
\ln \mathcal{L}\approx-\sum_{X=T,E}\,\sum_{\ell} \frac{1}{2}\,\left(\frac{\alpha C^{\mu X}_\ell(\Drrtext=10^{-5})-C^{X T, \rm data}_\ell}{\sigma^{X T, \rm data}_{\ell}}\right)^2,
}
where $C^{X T, \rm data}_\ell$ are the \Planck measurements with errors $\sigma^{X T, \rm data}_{\ell}$ \citep[see orange points of Fig.~12 and 13 in][]{Rotti2022muT}. We also introduced the energy release parameter $\alpha$, for which the energy release constraints in units of $\Drrtext=10^{-5}$ is estimated by using the second derivative of the log-likelihood with respect to $\alpha$. Since the signal amplitude is proportional to $\Drrtext( \propto \gammacon)$, we can the use the dependence of $\gammacon$ on the dark photon mass, Eq.~\eqref{eq:gammacon}, to constraint the kinetic mixing parameter, $\epsilon$. 

An example of the signal power spectra together with the data points are shown in Fig.~\ref{fig:constraints_PS}, while the constraints on $\epsilon$ are shown in Fig.~\ref{fig:constraints_eps}. The estimated anisotropy limits obtained with \Planck are marginally weaker than those obtained directly from direct CMB distortion measurements performed by \COBEF \citep{Fixsen1996}. However, the methods used here are completely independent. 
The limits weaken slightly for conversions at $\zcon\gtrsim 50,000$. This is because the $\mu T$ signal power spectra drop in amplitude with more information remaining in the $y T$ power spectra which we did not include in our analysis here (see discussion surrounding figure~\ref{fig:PS-approx_TT} and \ref{fig:b0_illustration}). In contrast, for the \COBEF constraint from the monopole spectrum, both the $\mu$ and $y$ terms contribute to the constraint at all conversion redshifts, leaving the constraint featureless around $z\simeq 50,000$. We also note that the small feature around $\mdp \simeq 10^{-8}\,{\rm eV}$ stems from the effect of the first recombination of helium on the conversion parameter. Similarly, the features around $\mdp \simeq 10^{-9}\,{\rm eV}$ are caused by the second recombination of helium followed by hydrogen recombination.

We found that the constraints are dominated by measurements at large angular scales $\ell \lesssim 400$. In addition, polarization data currently only tightens the limits marginally.
The constraints could already be improved by adding the latest CMB data from ACT \citep{ACTDR62025} and SPT \citep{SPT2026}. In addition, future data from \Litebird in combination with SKA, should lead significant improvements \citep{Remazeilles2018muT, kite_spectro-spatial_2023-III, Zegeye2023S4, Zegeye2024}, possibly even surpassing the limits from \COBEF for certain scenarios. 
Overall, this highlights the unique potential of combining CMB distortion and CMB anisotropy measurements to constrain beyond the standard model physics, in the future possibly even allowing us to break model parameter degeneracies that the individual probes cannot deliver.

 \section{Conclusions}
 \label{sec:conclusions}
Going beyond the CMB spectral distortion monopole and starting to look at the anisotropies can offer new interesting insights on various astrophysical and cosmological processes. Moreover, it allows one to set novel constraints using currently existing data from \Planck, ACT and SPT, since an absolute calibration of the monopole is not needed. 
In this work, we focus on the conversion of photons into dark photons in the pre-recombination era. With respect to previous works \citep[e.g.,][]{Chluba2024}, here we account for anisotropic conversions and perturbations of local quantities (e.g. $\delta X_{\rm e}$ and $\delta N_{\rm H}$) in the source term, Eq.~\eqref{eq:dS1_rad}. Thanks to an approximate system of equations known as the Frequency Hierarchy [which we summarize in its most updated form in Eq.~\eqref{eq:evol_1_final_hierarchy}] the anisotropic thermalisation problem becomes numerically tractable using {\tt CosmoTherm}, providing details about the transfer functions that can be seen in figures~\ref{fig:tf6}--\ref{fig:tf3}. By making use of the generalized line-of-sight approach [see section~\ref{sec:line-of-sight}], we computed the cross-correlation power spectra for various parameters (figures ~\ref{fig:PS-TX-EX} and \ref{fig:PS-approx_TT}). These results show rich spectro-spatial information which varies with the conversion redshift, and highlights the importance of including the anisotropic source term in our treatment. As one can observe in Fig.~\ref{fig:tf_comparison} and ~\ref{fig:PS-noS1}, the absence of $\id \mathcal{S}^{(1)} / \id z$ leads to significant differences, particularly for the $\mu$ term.  We also compare the results with the analytical solution introduced in \cite{Chluba2026TC} and quoted in Eq.~\eqref{eq:stationary_sol}. The approximate and full solutions present good agreement at most times, apart from a correction to the solution that decays away quickly after the conversion. This approximation also allows us to explain several interesting theoretical aspects related to the amplitude of the power spectra, presented in Fig.~\ref{fig:b0_illustration}, showing that for $\zcon > 50,000$ the $\mu$ spectrum is better represented by $\boostO \Mspec(x)$. However, details of the spectro-spatial information are not captures for conversions at $\zcon \lesssim 10^4$ such that a detailed calculation is required.

Our calculations also show that the corrections to the standard CMB temperature perturbations ($\delta \Theta$) behave slightly differently in the presence of a conversion than the other distortion variables. As one can see in Fig.~\ref{fig:tf6} and \ref{fig:PS-TTcorr}, even at very high redshifts ($z \gtrsim 2\times 10^6$) the perturbation can be non-negligible, when on the other hand $\mu$ and $y$ are completely washed out due to the efficient thermalization processes. In principle, this allows one to put (weaker) constraints for very early conversions. This is simply because the conversion process leads to a variation in the local composition of the cosmic fluid, seemingly producing an iso-curvature type contribution. In this work, we show the effects of $\delta \Theta$ only to illustrate this point, since a proper treatment would require us to track dark photon perturbations in the hierarchy after conversion. We leave a full perturbative treatment of the converted dark photon population for future work. 

Finally, we perform a simplistic likelihood analysis where we use \Planck $\mu T$ and $\mu E$ data \citep{Rotti2022muT} to place constraints on the kinetic mixing parameter $\epsilon$ and the dark photon mass $\mdp$. 
In Fig.~\ref{fig:constraints_eps}, we show the $95\%$ c.l. curve together with the one obtained previously from \COBEF using the distortion monopole \cite{Chluba2024}. These two completely independent methods provide very similar limits, with the ones obtained by \Planck being only slightly weaker. 
The resulting signal from $y T$ correlations was not considered since it is usually slightly smaller and contaminated by ISW effects from cluster of galaxies \citep[see discussion in][]{kite_spectro-spatial_2023-III}. However, for conversions in the $\mu$-era this weakens our constraints.  We also note that the constraints obtained with our setup for $\zcon \lesssim 2000$ are less accurate, since it requires the integration of sharp step functions. 

We expect that future analysis using ACT and SPT data will further improve the limits obtained here, likely leading to bounds competitive with \COBEF. Alternatively, one could get more constraining power by combining both the monopole and anisotropic likelihoods (i.e., using a monopole+$\mu T$ + $\mu E$ likelihood). 
This could even allow us to break existing parameter degeneracies, given that the {\it epoch-dependent information} is encoded in different ways.  
Finally, analyzing the CMB data\footnote{The extraction of $\mu$ maps uses the so-called constrained internal linear combination method to extract the signals assuming the spectrum of the target signal is given by $M(x)$ \citep[e.g.,][]{Remazeilles2018muT, Rotti2022muT}.} for early conversions using $\boostO M(x)$ instead of just $M(x)$ should be more optimal and hence improve the high mass limits.  

To summarize, in this work we have demonstrated the value of using anisotropic CMB spectral distortions as a probe of new physics, providing information complementary to the constraints usually derived using the sky-averaged distortion signal. Our treatment could be similarly adopted to studying alternative particle physics processes, such as axion conversions and decaying particles. 
The analytic and numerical tools developed in the past few years provide a powerful formalism that opens the door to a new and interesting field of research. 
We also stress that the observational frontier of CMB spectral distortion science is advancing steadily with upcoming experiments such as TMS \citep{Jose2020TMS}, COSMO \citep{Masi2021} and BISOU \citep{BISOU} and the development of innovative analysis methods \citep[e.g.,][]{Rotti2021SD,
Bianchini2022, Sabyr2025y}. In combination, this will allow us to maximize the science return of CMB spectral distortion studies in preparation for future CMB spectrometer mission such as {\it FOSSIL}.

\section*{Acknowledgments}
This work was supported by the UKSA grant: LiteBIRD UK ST/Y005945/1. The authors would like to thank Colin Hill, Fiona McCarthy, Matthew Johnson and Will Coulton for helpful comments on the manuscript. SE is thankful to Dean's Doctoral Scholarship awarded by the University of Manchester. BC is grateful for support from an NSERC Banting Fellowship, as well as the Simons Foundation (Grant Number 929255).

{\small
\bibliographystyle{plain}
\bibliography{Lit-all,references}

@ARTICLE{Chluba2026TC,
       author = {{Chluba}, Jens and {Evangelista}, Sara and {Cyr}, Byrce},
        title = "{Analytic approximations for the distortion transfer functions}",
      journal = {in prep.},
         year = 2026,
        month = aug
}

@ARTICLE{SPT2026,
       author = {{Camphuis}, E. and {Quan}, W. and {Balkenhol}, L. and {Khalife}, A.~R. and {Ge}, F. and {Guidi}, F. and {Huang}, N. and {Lynch}, G.~P. and {Omori}, Y. and {Trendafilova}, C. and {Anderson}, A.~J. and {Ansarinejad}, B. and {Archipley}, M. and {Barry}, P.~S. and {Benabed}, K. and {Bender}, A.~N. and {Benson}, B.~A. and {Bianchini}, F. and {Bleem}, L.~E. and {Bouchet}, F.~R. and {Bryant}, L. and {Campitiello}, M.~G. and {Carlstrom}, J.~E. and {Chang}, C.~L. and {Chaubal}, P. and {Chichura}, P.~M. and {Chokshi}, A. and {Chou}, T.-L. and {Coerver}, A. and {Crawford}, T.~M. and {Daley}, C. and {de Haan}, T. and {Dibert}, K.~R. and {Dobbs}, M.~A. and {Doohan}, M. and {Doussot}, A. and {Dutcher}, D. and {Everett}, W. and {Feng}, C. and {Ferguson}, K.~R. and {Fichman}, K. and {Foster}, A. and {Galli}, S. and {Gambrel}, A.~E. and {Gardner}, R.~W. and {Goeckner-Wald}, N. and {Gualtieri}, R. and {Guns}, S. and {Halverson}, N.~W. and {Hivon}, E. and {Holder}, G.~P. and {Holzapfel}, W.~L. and {Hood}, J.~C. and {Hryciuk}, A. and {K{\'e}ruzor{\'e}}, F. and {Knox}, L. and {Korman}, M. and {Kornoelje}, K. and {Kuo}, C.-L. and {Levy}, K. and {Lowitz}, A.~E. and {Lu}, C. and {Maniyar}, A. and {Martsen}, E.~S. and {Menanteau}, F. and {Millea}, M. and {Montgomery}, J. and {Nakato}, Y. and {Natoli}, T. and {Noble}, G.~I. and {Ouellette}, A. and {Pan}, Z. and {Paschos}, P. and {Phadke}, K.~A. and {Pollak}, A.~W. and {Prabhu}, K. and {Raghunathan}, S. and {Rahimi}, M. and {Rahlin}, A. and {Reichardt}, C.~L. and {Rouble}, M. and {Ruhl}, J.~E. and {Schiappucci}, E. and {Simpson}, A. and {Sobrin}, J.~A. and {Stark}, A.~A. and {Stephen}, J. and {Tandoi}, C. and {Thorne}, B. and {Umilta}, C. and {Vieira}, J.~D. and {Vitrier}, A. and {Wan}, Y. and {Whitehorn}, N. and {Wu}, W.~L.~K. and {Young}, M.~R. and {Zebrowski}, J.~A. and {SPT-3G Collaboration}},
        title = "{SPT-3G D1: CMB temperature and polarization power spectra and cosmology from 2019 and 2020 observations of the SPT-3G main field}",
      journal = {\prd},
         year = 2026,
        month = apr,
       volume = {113},
       number = {8},
          eid = {083504},
        pages = {083504},
          doi = {10.1103/7wt3-9v2y},
archivePrefix = {arXiv},
       eprint = {2506.20707},
 primaryClass = {astro-ph.CO},
       adsurl = {https://ui.adsabs.harvard.edu/abs/2026PhRvD.113h3504C}
}

@ARTICLE{ACTDR62025,
       author = {{Louis}, Thibaut and {La Posta}, Adrien and {Atkins}, Zachary and {Jense}, Hidde T. and {Abril-Cabezas}, Irene and {Addison}, Graeme E. and {Ade}, Peter A.~R. and {Aiola}, Simone and {Alford}, Tommy and {Alonso}, David and {Amiri}, Mandana and {An}, Rui and {Austermann}, Jason E. and {Barbavara}, Eleonora and {Battaglia}, Nicholas and {Battistelli}, Elia Stefano and {Beall}, James A. and {Bean}, Rachel and {Beheshti}, Ali and {Beringue}, Benjamin and {Bhandarkar}, Tanay and {Biermann}, Emily and {Bolliet}, Boris and {Bond}, J. Richard and {Calabrese}, Erminia and {Capalbo}, Valentina and {Carrero}, Felipe and {Chen}, Shi-Fan and {Chesmore}, Grace and {Cho}, Hsiao-mei and {Choi}, Steve K. and {Clark}, Susan E. and {Cothard}, Nicholas F. and {Coughlin}, Kevin and {Coulton}, William and {Crichton}, Devin and {Crowley}, Kevin T. and {Darwish}, Omar and {Devlin}, Mark J. and {Dicker}, Simon and {Duell}, Cody J. and {Duff}, Shannon M. and {Duivenvoorden}, Adriaan J. and {Dunkley}, Jo and {Dunner}, Rolando and {Embil Villagra}, Carmen and {Fankhanel}, Max and {Farren}, Gerrit S. and {Ferraro}, Simone and {Foster}, Allen and {Freundt}, Rodrigo and {Fuzia}, Brittany and {Gallardo}, Patricio A. and {Garrido}, Xavier and {Gerbino}, Martina and {Giardiello}, Serena and {Gill}, Ajay and {Givans}, Jahmour and {Gluscevic}, Vera and {Goldstein}, Samuel and {Golec}, Joseph E. and {Gong}, Yulin and {Guan}, Yilun and {Halpern}, Mark and {Harrison}, Ian and {Hasselfield}, Matthew and {Healy}, Erin and {Henderson}, Shawn and {Hensley}, Brandon and {Herv{\'\i}as-Caimapo}, Carlos and {Hill}, J. Colin and {Hilton}, Gene C. and {Hilton}, Matt and {Hincks}, Adam D. and {Hlo{\v{z}}ek}, Ren{\'e}e and {Ho}, Shuay-Pwu Patty and {Hood}, John and {Hornecker}, Erika and {Huber}, Zachary B. and {Hubmayr}, Johannes and {Huffenberger}, Kevin M. and {Hughes}, John P. and {Ikape}, Margaret and {Irwin}, Kent and {Isopi}, Giovanni and {Joshi}, Neha and {Keller}, Ben and {Kim}, Joshua and {Knowles}, Kenda and {Koopman}, Brian J. and {Kosowsky}, Arthur and {Kramer}, Darby and {Kusiak}, Aleksandra and {Lagu{\"e}}, Alex and {Lakey}, Victoria and {Lee}, Eunseong and {Li}, Yaqiong and {Li}, Zack and {Limon}, Michele and {Lokken}, Martine and {Lungu}, Marius and {MacCrann}, Niall and {MacInnis}, Amanda and {Madhavacheril}, Mathew S. and {Maldonado}, Diego and {Maldonado}, Felipe and {Mallaby-Kay}, Maya and {Marques}, Gabriela A. and {van Marrewijk}, Joshiwa and {McCarthy}, Fiona and {McMahon}, Jeff and {Mehta}, Yogesh and {Menanteau}, Felipe and {Moodley}, Kavilan and {Morris}, Thomas W. and {Mroczkowski}, Tony and {Naess}, Sigurd and {Namikawa}, Toshiya and {Nati}, Federico and {Nerval}, Simran K. and {Newburgh}, Laura and {Nicola}, Andrina and {Niemack}, Michael D. and {Nolta}, Michael R. and {Orlowski-Scherer}, John and {Pagano}, Luca and {Page}, Lyman A. and {Pandey}, Shivam and {Partridge}, Bruce and {Perez Sarmiento}, Karen and {Prince}, Heather and {Puddu}, Roberto and {Qu}, Frank J. and {Ragavan}, Damien C. and {Ried Guachalla}, Bernardita and {Rogers}, Keir K. and {Rojas}, Felipe and {Sakuma}, Tai and {Schaan}, Emmanuel and {Schmitt}, Benjamin L. and {Sehgal}, Neelima and {Shaikh}, Shabbir and {Sherwin}, Blake D. and {Sierra}, Carlos and {Sievers}, Jon and {Sif{\'o}n}, Crist{\'o}bal and {Simon}, Sara and {Sonka}, Rita and {Spergel}, David N. and {Staggs}, Suzanne T. and {Storer}, Emilie and {Surrao}, Kristen and {Switzer}, Eric R. and {Tampier}, Niklas and {Thornton}, Robert and {Trac}, Hy and {Tucker}, Carole and {Ullom}, Joel and {Vale}, Leila R. and {Van Engelen}, Alexander and {Van Lanen}, Jeff and {Vargas}, Cristian and {Vavagiakis}, Eve M. and {Wagoner}, Kasey and {Wang}, Yuhan and {Wenzl}, Lukas and {Wollack}, Edward J. and {Zheng}, Kaiwen and {The Atacama Cosmology Telescope collaboration}},
        title = "{The Atacama Cosmology Telescope: DR6 power spectra, likelihoods and {\ensuremath{\Lambda}}CDM parameters}",
      journal = {\jcap},
         year = 2025,
        month = nov,
       volume = {2025},
       number = {11},
          eid = {062},
        pages = {062},
          doi = {10.1088/1475-7516/2025/11/062},
archivePrefix = {arXiv},
       eprint = {2503.14452},
 primaryClass = {astro-ph.CO},
       adsurl = {https://ui.adsabs.harvard.edu/abs/2025JCAP...11..062L}
}

@ARTICLE{Chluba2026,
       author = {{Chluba}, Jens and {Evangelista}, Sara and {Daman}, Tom and {Vasil}, Geoff},
        title = "{Improved frequency hierarchy treatment for anisotropic spectral distortions}",
      journal = {arXiv e-prints},
         year = 2026,
        month = feb,
          eid = {arXiv:2602.14963},
        pages = {arXiv:2602.14963},
          doi = {10.48550/arXiv.2602.14963},
archivePrefix = {arXiv},
       eprint = {2602.14963},
 primaryClass = {astro-ph.CO},
       adsurl = {https://ui.adsabs.harvard.edu/abs/2026arXiv260214963C}
}

@ARTICLE{Evangelista2025,
       author = {{Evangelista}, Sara and {Chluba}, Jens and {Pace}, Francesco},
        title = "{The late-time heating Green's function and improvements to distortion frequency hierarchy treatment}",
      journal = {\mnras},
         year = 2025,
        month = may,
       volume = {539},
       number = {2},
        pages = {1640-1650},
          doi = {10.1093/mnras/staf594},
archivePrefix = {arXiv},
       eprint = {2501.12822},
 primaryClass = {astro-ph.CO},
       adsurl = {https://ui.adsabs.harvard.edu/abs/2025MNRAS.539.1640E}
}

@ARTICLE{Seljak1996,
       author = {{Seljak}, Uros and {Zaldarriaga}, Matias},
        title = "{A Line-of-Sight Integration Approach to Cosmic Microwave Background Anisotropies}",
      journal = {\apj},
         year = 1996,
        month = oct,
       volume = {469},
        pages = {437},
          doi = {10.1086/177793},
archivePrefix = {arXiv},
       eprint = {astro-ph/9603033},
 primaryClass = {astro-ph},
       adsurl = {https://ui.adsabs.harvard.edu/abs/1996ApJ...469..437S}
}

@ARTICLE{Remazeilles2018muT,
       author = {{Remazeilles}, M. and {Chluba}, J.},
        title = "{Extracting foreground-obscured {\ensuremath{\mu}}-distortion anisotropies to constrain primordial non-Gaussianity}",
      journal = {\mnras},
         year = 2018,
        month = jul,
       volume = {478},
       number = {1},
        pages = {807-824},
          doi = {10.1093/mnras/sty1034},
archivePrefix = {arXiv},
       eprint = {1802.10101},
 primaryClass = {astro-ph.CO},
       adsurl = {https://ui.adsabs.harvard.edu/abs/2018MNRAS.478..807R}
}

@article{Berlin2022,
    author = "Berlin, Asher and Dror, Jeff A. and Gan, Xucheng and Ruderman, Joshua T.",
    title = "{Millicharged relics reveal massless dark photons}",
    eprint = "2211.05139",
    archivePrefix = "arXiv",
    primaryClass = "hep-ph",
    reportNumber = "FERMILAB-PUB-21-724-SQMS-T",
    doi = "10.1007/JHEP05(2023)046",
    journal = "JHEP",
    volume = "05",
    pages = "046",
    year = "2023"
}

@ARTICLE{Rotti2022muT,
       author = {{Rotti}, Aditya and {Ravenni}, Andrea and {Chluba}, Jens},
        title = "{Non-Gaussianity constraints with anisotropic {\ensuremath{\mu}} distortion measurements from Planck}",
      journal = {\mnras},
         year = 2022,
        month = oct,
       volume = {515},
       number = {4},
        pages = {5847-5868},
          doi = {10.1093/mnras/stac2082},
archivePrefix = {arXiv},
       eprint = {2205.15971},
 primaryClass = {astro-ph.CO},
       adsurl = {https://ui.adsabs.harvard.edu/abs/2022MNRAS.515.5847R}
}

@ARTICLE{Rotti2021SD,
       author = {{Rotti}, Aditya and {Chluba}, Jens},
        title = "{Combining ILC and moment expansion techniques for extracting average-sky signals and CMB anisotropies}",
      journal = {\mnras},
         year = 2021,
        month = jan,
       volume = {500},
       number = {1},
        pages = {976-985},
          doi = {10.1093/mnras/staa3292},
archivePrefix = {arXiv},
       eprint = {2006.02458},
 primaryClass = {astro-ph.CO},
       adsurl = {https://ui.adsabs.harvard.edu/abs/2021MNRAS.500..976R}
}

@ARTICLE{Zegeye2024,
       author = {{Zegeye}, David and {Crawford}, Thomas and {Chluba}, Jens and {Remazeilles}, Mathieu and {Grainge}, Keith},
        title = "{Square Kilometer Array as a cosmic microwave background experiment}",
      journal = {\prd},
         year = 2025,
        month = mar,
       volume = {111},
       number = {6},
          eid = {063517},
        pages = {063517},
          doi = {10.1103/PhysRevD.111.063517},
archivePrefix = {arXiv},
       eprint = {2406.04326},
 primaryClass = {astro-ph.CO},
       adsurl = {https://ui.adsabs.harvard.edu/abs/2025PhRvD.111f3517Z}
}

@ARTICLE{Chluba2024DP,
       author = {{Chluba}, Jens and {Cyr}, Bryce and {Johnson}, Matthew C.},
        title = "{Revisiting dark photon constraints from CMB spectral distortions}",
      journal = {\mnras},
         year = 2024,
        month = dec,
       volume = {535},
       number = {2},
        pages = {1874-1887},
          doi = {10.1093/mnras/stae2464},
archivePrefix = {arXiv},
       eprint = {2409.12115},
 primaryClass = {astro-ph.CO},
       adsurl = {https://ui.adsabs.harvard.edu/abs/2024MNRAS.535.1874C}
}

@article{Mirizzi:2009iz,
    author = "Mirizzi, Alessandro and Redondo, Javier and Sigl, Gunter",
    title = "{Microwave Background Constraints on Mixing of Photons with Hidden Photons}",
    eprint = "0901.0014",
    archivePrefix = "arXiv",
    primaryClass = "hep-ph",
    reportNumber = "DESY-08-205, MPP-2008-169",
    doi = "10.1088/1475-7516/2009/03/026",
    journal = "JCAP",
    volume = "03",
    pages = "026",
    year = "2009"
}

@ARTICLE{Bianchini2022,
       author = {{Bianchini}, F. and {Fabbian}, G.},
        title = "{CMB spectral distortions revisited: A new take on {\ensuremath{\mu}} distortions and primordial non-Gaussianities from FIRAS data}",
      journal = {\prd},
         year = 2022,
        month = sep,
       volume = {106},
       number = {6},
          eid = {063527},
        pages = {063527},
          doi = {10.1103/PhysRevD.106.063527},
archivePrefix = {arXiv},
       eprint = {2206.02762},
 primaryClass = {astro-ph.CO},
       adsurl = {https://ui.adsabs.harvard.edu/abs/2022PhRvD.106f3527B}
}

@INPROCEEDINGS{Jose2020TMS,
       author = {{Rubi{\~n}o Mart{\'\i}n}, Jos{\'e} Alberto and {Alonso Arias}, Paz and {Hoyland}, Roger J. and {Aguiar-Gonz{\'a}lez}, Marta and {De Miguel-Hern{\'a}ndez}, Javier and {G{\'e}nova-Santos}, Ricardo T. and {Gomez-Re{\~n}asco}, Maria F. and {Guidi}, Federica and {Fern{\'a}ndez-Izquierdo}, Patricia and {Fern{\'a}ndez-Torreiro}, Mateo and {Fuerte-Rodriguez}, Pablo A. and {Hernandez-Monteagudo}, Carlos and {L{\'o}pez-Caraballo}, Carlos H. and {Perez-de-Taoro}, Angeles and {Peel}, Michael W. and {Rebolo}, Rafael and {Zamora-Jimenez}, Antonio and {Gonz{\'a}lez-Carretero}, Eduardo D. and {Colodro-Conde}, Carlos and {P{\'e}rez-Lemus}, Cristina and {Toledo-Moreo}, Rafael and {P{\'e}rez-Liz{\'a}n}, David and {Cuttaia}, Francesco and {Terenzi}, Luca and {Franceschet}, Cristian and {Realini}, Sabrina and {Chluba}, Jens and {Murga-Llano}, Gaizka and {Sanquirce-Garcia}, Ruben},
        title = "{The Tenerife Microwave Spectrometer (TMS) experiment: studying the absolute spectrum of the sky emission in the 10-20GHz range}",
    booktitle = {Society of Photo-Optical Instrumentation Engineers (SPIE) Conference Series},
         year = 2020,
       series = {Society of Photo-Optical Instrumentation Engineers (SPIE) Conference Series},
       volume = {11453},
        month = dec,
          eid = {114530T},
        pages = {114530T},
          doi = {10.1117/12.2561309},
       adsurl = {https://ui.adsabs.harvard.edu/abs/2020SPIE11453E..0TR}
}

@ARTICLE{Masi2021,
       author = {{Masi}, S. and {Battistelli}, E. and {de Bernardis}, P. and {Coppolecchia}, A. and {Columbro}, F. and {D'Alessandro}, G. and {De Petris}, M. and {Lamagna}, L. and {Marchitelli}, E. and {Mele}, L. and {Paiella}, A. and {Piacentini}, F. and {Pisano}, G. and {Bersanelli}, M. and {Franceschet}, C. and {Manzan}, E. and {Mennella}, D. and {Realini}, S. and {Cibella}, S. and {Martini}, F. and {Pettinari}, G. and {Coppi}, G. and {Gervasi}, M. and {Limonta}, A. and {Zannoni}, M. and {Piccirillo}, L. and {Tucker}, C.},
        title = "{The COSmic Monopole Observer (COSMO)}",
      journal = {arXiv e-prints},
         year = 2021,
        month = oct,
          eid = {arXiv:2110.12254},
        pages = {arXiv:2110.12254},
          doi = {10.48550/arXiv.2110.12254},
archivePrefix = {arXiv},
       eprint = {2110.12254},
 primaryClass = {astro-ph.IM},
       adsurl = {https://ui.adsabs.harvard.edu/abs/2021arXiv211012254M}
}

@ARTICLE{BISOU,
       author = {{Maffei}, B. and {Abitbol}, M.~H. and {Aghanim}, N. and {Aumont}, J. and {Battistelli}, E. and {Chluba}, J. and {Coulon}, X. and {De Bernardis}, P. and {Douspis}, M. and {Grain}, J. and {Gervasoni}, S. and {Hill}, J.~C. and {Kogut}, A. and {Masi}, S. and {Matsumura}, T. and {Sullivan}, C. O and {Pagano}, L. and {Pisano}, G. and {Remazeilles}, M. and {Ritacco}, A. and {Rotti}, A. and {Sauvage}, V. and {Savini}, G. and {Stever}, S.~L. and {Tartari}, A. and {Thiele}, L. and {Trappe}, N.},
        title = "{BISOU: a balloon project to measure the CMB spectral distortions}",
      journal = {arXiv e-prints},
         year = 2021,
        month = oct,
          eid = {arXiv:2111.00246},
        pages = {arXiv:2111.00246},
          doi = {10.48550/arXiv.2111.00246},
archivePrefix = {arXiv},
       eprint = {2111.00246},
 primaryClass = {astro-ph.IM},
       adsurl = {https://ui.adsabs.harvard.edu/abs/2021arXiv211100246M}
}

@INPROCEEDINGS{Kogut2016SPIE,
   author = {{Kogut}, A. and {Chluba}, J. and {Fixsen}, D.~J. and {Meyer}, S. and 
	{Spergel}, D.},
    title = "{The Primordial Inflation Explorer (PIXIE)}",
booktitle = {SPIE Conference Series},
     year = 2016,
   series = {Proc.SPIE},
   volume = 9904,
    month = jul,
      eid = {99040W},
    pages = {99040W},
      doi = {10.1117/12.2231090},
   adsurl = {http://adsabs.harvard.edu/abs/2016SPIE.9904E..0WK}
}

@ARTICLE{Chluba2015GreensII,
       author = {{Chluba}, Jens},
        title = "{Green's function of the cosmological thermalization problem - II. Effect of photon injection and constraints}",
      journal = {\mnras},
         year = 2015,
        month = dec,
       volume = {454},
       number = {4},
        pages = {4182-4196},
          doi = {10.1093/mnras/stv2243},
archivePrefix = {arXiv},
       eprint = {1506.06582},
 primaryClass = {astro-ph.CO},
       adsurl = {https://ui.adsabs.harvard.edu/abs/2015MNRAS.454.4182C}
}

@ARTICLE{CLASSCODE,
   author = {{Lesgourgues}, J.},
    title = "{The Cosmic Linear Anisotropy Solving System (CLASS) I: Overview}",
  journal = {ArXiv:1104.2932},
archivePrefix = "arXiv",
   eprint = {1104.2932},
 primaryClass = "astro-ph.IM",
     year = 2011,
    month = apr,
   adsurl = {http://adsabs.harvard.edu/abs/2011arXiv1104.2932L}
}

@ARTICLE{Hu1997,
   author = {{Hu}, W. and {White}, M.},
    title = "{CMB anisotropies: Total angular momentum method}",
  journal = {\prd},
   eprint = {astro-ph/9702170},
     year = 1997,
    month = jul,
   volume = 56,
    pages = {596-615},
      doi = {10.1103/PhysRevD.56.596},
   adsurl = {http://adsabs.harvard.edu/abs/1997PhRvD..56..596H}
}

@ARTICLE{Chluba2013PCA,
   author = {{Chluba}, J. and {Jeong}, D.},
    title = "{Teasing bits of information out of the CMB energy spectrum}",
  journal = {\mnras},
archivePrefix = "arXiv",
   eprint = {1306.5751},
 primaryClass = "astro-ph.CO",
     year = 2014,
    month = mar,
   volume = 438,
    pages = {2065-2082},
      doi = {10.1093/mnras/stt2327},
   adsurl = {http://adsabs.harvard.edu/abs/2014MNRAS.438.2065C}
}

@ARTICLE{Chluba2013Green,
   author = {{Chluba}, J.},
    title = "{Green's function of the cosmological thermalization problem}",
  journal = {\mnras},
archivePrefix = "arXiv",
   eprint = {1304.6120},
 primaryClass = "astro-ph.CO",
     year = 2013,
    month = sep,
   volume = 434,
    pages = {352-357},
      doi = {10.1093/mnras/stt1025},
   adsurl = {http://adsabs.harvard.edu/abs/2013MNRAS.434..352C}
}

@ARTICLE{Senatore2009,
   author = {{Senatore}, L. and {Tassev}, S. and {Zaldarriaga}, M.},
    title = "{Non-gaussianities from perturbing recombination}",
  journal = {\jcap},
archivePrefix = "arXiv",
   eprint = {0812.3658},
     year = 2009,
    month = sep,
   volume = 9,
    pages = {38},
      doi = {10.1088/1475-7516/2009/09/038},
   adsurl = {http://adsabs.harvard.edu/abs/2009JCAP...09..038S}
}

@ARTICLE{Khatri2009,
   author = {{Khatri}, R. and {Wandelt}, B.~D.},
    title = "{Crinkles in the last scattering surface: Non-Gaussianity from inhomogeneous recombination}",
  journal = {\prd},
archivePrefix = "arXiv",
   eprint = {0810.4370},
     year = 2009,
    month = jan,
   volume = 79,
   number = 2,
      eid = {023501},
    pages = {023501},
      doi = {10.1103/PhysRevD.79.023501},
   adsurl = {http://adsabs.harvard.edu/abs/2009PhRvD..79b3501K}
}

@ARTICLE{Ma1995,
   author = {{Ma}, C.-P. and {Bertschinger}, E.},
    title = "{Cosmological Perturbation Theory in the Synchronous and Conformal Newtonian Gauges}",
  journal = {\apj},
   eprint = {arXiv:astro-ph/9506072},
     year = 1995,
    month = dec,
   volume = 455,
    pages = {7-+},
      doi = {10.1086/176550},
   adsurl = {http://adsabs.harvard.edu/abs/1995ApJ...455....7M}
}

@ARTICLE{Chluba2011therm,
   author = {{Chluba}, J. and {Sunyaev}, R.~A.},
    title = "{The evolution of CMB spectral distortions in the early Universe}",
  journal = {\mnras},
archivePrefix = "arXiv",
   eprint = {1109.6552},
 primaryClass = "astro-ph.CO",
     year = 2012,
    month = jan,
   volume = 419,
    pages = {1294-1314},
      doi = {10.1111/j.1365-2966.2011.19786.x},
   adsurl = {http://adsabs.harvard.edu/abs/2012MNRAS.419.1294C}
}

@ARTICLE{Kogut2011PIXIE,
   author = {{Kogut}, A. and {Fixsen}, D.~J. and {Chuss}, D.~T. and {Dotson}, J. and 
	{Dwek}, E. and {Halpern}, M. and {Hinshaw}, G.~F. and {Meyer}, S.~M. and 
	{Moseley}, S.~H. and {Seiffert}, M.~D. and {Spergel}, D.~N. and 
	{Wollack}, E.~J.},
    title = "{The Primordial Inflation Explorer (PIXIE): a nulling polarimeter for cosmic microwave background observations}",
  journal = {\jcap},
archivePrefix = "arXiv",
   eprint = {1105.2044},
 primaryClass = "astro-ph.CO",
     year = 2011,
    month = jul,
   volume = 7,
    pages = {25-+},
      doi = {10.1088/1475-7516/2011/07/025},
   adsurl = {http://adsabs.harvard.edu/abs/2011JCAP...07..025K}
}

@ARTICLE{Fixsen1996,
   author = {{Fixsen}, D.~J. and {Cheng}, E.~S. and {Gales}, J.~M. and {Mather}, J.~C. and 
	{Shafer}, R.~A. and {Wright}, E.~L.},
    title = "{The Cosmic Microwave Background Spectrum from the Full COBE FIRAS Data Set}",
  journal = {\apj},
   eprint = {arXiv:astro-ph/9605054},
     year = 1996,
    month = dec,
   volume = 473,
    pages = {576-+},
      doi = {10.1086/178173},
   adsurl = {http://adsabs.harvard.edu/abs/1996ApJ...473..576F}
}

@ARTICLE{Mather1994,
   author = {{Mather}, J.~C. and {Cheng}, E.~S. and {Cottingham}, D.~A. and 
	{Eplee}, Jr., R.~E. and {Fixsen}, D.~J. and {Hewagama}, T. and 
	{Isaacman}, R.~B. and {Jensen}, K.~A. and {Meyer}, S.~S. and 
	{Noerdlinger}, P.~D. and {Read}, S.~M. and {Rosen}, L.~P.
		},
    title = "{Measurement of the cosmic microwave background spectrum by the COBE FIRAS instrument}",
  journal = {\apj},
     year = 1994,
    month = jan,
   volume = 420,
    pages = {439-444},
      doi = {10.1086/173574},
   adsurl = {http://adsabs.harvard.edu/abs/1994ApJ...420..439M}
}

@ARTICLE{Hu1993,
   author = {{Hu}, W. and {Silk}, J.},
    title = "{Thermalization and spectral distortions of the cosmic background radiation}",
  journal = {\prd},
     year = 1993,
    month = jul,
   volume = 48,
    pages = {485-502},
      doi = {10.1103/PhysRevD.48.485},
   adsurl = {http://adsabs.harvard.edu/abs/1993PhRvD..48..485H}
}

@ARTICLE{Burigana1991,
   author = {{Burigana}, C. and {Danese}, L. and {de Zotti}, G.},
    title = "{Formation and evolution of early distortions of the microwave background spectrum - A numerical study}",
  journal = {\aap},
     year = 1991,
    month = jun,
   volume = 246,
    pages = {49-58},
   adsurl = {http://adsabs.harvard.edu/abs/1991A%26A...246...49B}
}

@ARTICLE{Chluba2010b,
   author = {{Chluba}, J. and {Thomas}, R.~M.},
    title = "{Towards a complete treatment of the cosmological recombination problem}",
  journal = {\mnras},
archivePrefix = "arXiv",
   eprint = {1010.3631},
 primaryClass = "astro-ph.CO",
     year = 2011,
    month = apr,
   volume = 412,
    pages = {748-764},
      doi = {10.1111/j.1365-2966.2010.17940.x},
   adsurl = {http://adsabs.harvard.edu/abs/2011MNRAS.412..748C}
}

@ARTICLE{Sunyaev1970mu,
   author = {{Sunyaev}, R.~A. and {Zeldovich}, Y.~B.},
    title = "{The interaction of matter and radiation in the hot model of the Universe, II}",
  journal = {\apss},
     year = 1970,
    month = apr,
   volume = 7,
    pages = {20-30},
      doi = {10.1007/BF00653472},
   adsurl = {http://adsabs.harvard.edu/abs/1970Ap%26SS...7...20S}
}

@ARTICLE{Chluba2010,
   author = {{Chluba}, J. and {Vasil}, G.~M. and {Dursi}, L.~J.},
    title = "{Recombinations to the Rydberg states of hydrogen and their effect during the cosmological recombination epoch}",
  journal = {\mnras},
archivePrefix = "arXiv",
   eprint = {1003.4928},
 primaryClass = "astro-ph.CO",
     year = 2010,
    month = sep,
   volume = 407,
    pages = {599-612},
      doi = {10.1111/j.1365-2966.2010.16940.x},
   adsurl = {http://adsabs.harvard.edu/abs/2010MNRAS.407..599C}
}

@ARTICLE{Novosyadlyj2006,
   author = {{Novosyadlyj}, B.},
    title = "{Perturbations of ionization fractions at the cosmological recombination epoch}",
  journal = {\mnras},
     year = 2006,
    month = aug,
   volume = 370,
    pages = {1771-1782},
      doi = {10.1111/j.1365-2966.2006.10593.x},
   adsurl = {http://cdsads.u-strasbg.fr/cgi-bin/nph-bib_query?bibcode=2006MNRAS.370.1771N&db_key=AST}
}

@ARTICLE{CMBFAST,
   author = {{Seljak}, U. and {Zaldarriaga}, M.},
    title = "{A Line-of-Sight Integration Approach to Cosmic Microwave Background Anisotropies}",
  journal = {\apj},
   eprint = {arXiv:astro-ph/9603033},
     year = 1996,
    month = oct,
   volume = 469,
    pages = {437-+},
      doi = {10.1086/177793},
   adsurl = {http://adsabs.harvard.edu/abs/1996ApJ...469..437S}
}

@ARTICLE{CAMB,
   author = {{Lewis}, A. and {Challinor}, A. and {Lasenby}, A.},
    title = "{Efficient Computation of Cosmic Microwave Background Anisotropies in Closed Friedmann-Robertson-Walker Models}",
  journal = {\apj},
  eprinttype = "arxiv",
  eprint = {astro-ph/9911177},
     year = 2000,
    month = aug,
   volume = 538,
    pages = {473-476},
      doi = {10.1086/309179},
   adsurl = {http://adsabs.harvard.edu/abs/2000ApJ...538..473L}
}

@ARTICLE{Planck2018params,
       author = {{Planck Collaboration} and {Aghanim}, N. and {Akrami}, Y. and {Ashdown}, M. and {Aumont}, J. and {Baccigalupi}, C. and {Ballardini}, M. and {Banday}, A.~J. and {Barreiro}, R.~B. and {Bartolo}, N. and {Basak}, S. and {Battye}, R. and {Benabed}, K. and {Bernard}, J. -P. and {Bersanelli}, M. and {Bielewicz}, P. and {Bock}, J.~J. and {Bond}, J.~R. and {Borrill}, J. and {Bouchet}, F.~R. and {Boulanger}, F. and {Bucher}, M. and {Burigana}, C. and {Butler}, R.~C. and {Calabrese}, E. and {Cardoso}, J. -F. and {Carron}, J. and {Challinor}, A. and {Chiang}, H.~C. and {Chluba}, J. and {Colombo}, L.~P.~L. and {Combet}, C. and {Contreras}, D. and {Crill}, B.~P. and {Cuttaia}, F. and {de Bernardis}, P. and {de Zotti}, G. and {Delabrouille}, J. and {Delouis}, J. -M. and {Di Valentino}, E. and {Diego}, J.~M. and {Dor{\'e}}, O. and {Douspis}, M. and {Ducout}, A. and {Dupac}, X. and {Dusini}, S. and {Efstathiou}, G. and {Elsner}, F. and {En{\ss}lin}, T.~A. and {Eriksen}, H.~K. and {Fantaye}, Y. and {Farhang}, M. and {Fergusson}, J. and {Fernandez-Cobos}, R. and {Finelli}, F. and {Forastieri}, F. and {Frailis}, M. and {Fraisse}, A.~A. and {Franceschi}, E. and {Frolov}, A. and {Galeotta}, S. and {Galli}, S. and {Ganga}, K. and {G{\'e}nova-Santos}, R.~T. and {Gerbino}, M. and {Ghosh}, T. and {Gonz{\'a}lez-Nuevo}, J. and {G{\'o}rski}, K.~M. and {Gratton}, S. and {Gruppuso}, A. and {Gudmundsson}, J.~E. and {Hamann}, J. and {Handley}, W. and {Hansen}, F.~K. and {Herranz}, D. and {Hildebrandt}, S.~R. and {Hivon}, E. and {Huang}, Z. and {Jaffe}, A.~H. and {Jones}, W.~C. and {Karakci}, A. and {Keih{\"a}nen}, E. and {Keskitalo}, R. and {Kiiveri}, K. and {Kim}, J. and {Kisner}, T.~S. and {Knox}, L. and {Krachmalnicoff}, N. and {Kunz}, M. and {Kurki-Suonio}, H. and {Lagache}, G. and {Lamarre}, J. -M. and {Lasenby}, A. and {Lattanzi}, M. and {Lawrence}, C.~R. and {Le Jeune}, M. and {Lemos}, P. and {Lesgourgues}, J. and {Levrier}, F. and {Lewis}, A. and {Liguori}, M. and {Lilje}, P.~B. and {Lilley}, M. and {Lindholm}, V. and {L{\'o}pez-Caniego}, M. and {Lubin}, P.~M. and {Ma}, Y. -Z. and {Mac{\'\i}as-P{\'e}rez}, J.~F. and {Maggio}, G. and {Maino}, D. and {Mandolesi}, N. and {Mangilli}, A. and {Marcos-Caballero}, A. and {Maris}, M. and {Martin}, P.~G. and {Martinelli}, M. and {Mart{\'\i}nez-Gonz{\'a}lez}, E. and {Matarrese}, S. and {Mauri}, N. and {McEwen}, J.~D. and {Meinhold}, P.~R. and {Melchiorri}, A. and {Mennella}, A. and {Migliaccio}, M. and {Millea}, M. and {Mitra}, S. and {Miville-Desch{\^e}nes}, M. -A. and {Molinari}, D. and {Montier}, L. and {Morgante}, G. and {Moss}, A. and {Natoli}, P. and {N{\o}rgaard-Nielsen}, H.~U. and {Pagano}, L. and {Paoletti}, D. and {Partridge}, B. and {Patanchon}, G. and {Peiris}, H.~V. and {Perrotta}, F. and {Pettorino}, V. and {Piacentini}, F. and {Polastri}, L. and {Polenta}, G. and {Puget}, J. -L. and {Rachen}, J.~P. and {Reinecke}, M. and {Remazeilles}, M. and {Renzi}, A. and {Rocha}, G. and {Rosset}, C. and {Roudier}, G. and {Rubi{\~n}o-Mart{\'\i}n}, J.~A. and {Ruiz-Granados}, B. and {Salvati}, L. and {Sandri}, M. and {Savelainen}, M. and {Scott}, D. and {Shellard}, E.~P.~S. and {Sirignano}, C. and {Sirri}, G. and {Spencer}, L.~D. and {Sunyaev}, R. and {Suur-Uski}, A. -S. and {Tauber}, J.~A. and {Tavagnacco}, D. and {Tenti}, M. and {Toffolatti}, L. and {Tomasi}, M. and {Trombetti}, T. and {Valenziano}, L. and {Valiviita}, J. and {Van Tent}, B. and {Vibert}, L. and {Vielva}, P. and {Villa}, F. and {Vittorio}, N. and {Wandelt}, B.~D. and {Wehus}, I.~K. and {White}, M. and {White}, S.~D.~M. and {Zacchei}, A. and {Zonca}, A.},
        title = "{Planck 2018 results. VI. Cosmological parameters}",
      journal = {\aap},
         year = 2020,
        month = sep,
       volume = {641},
          eid = {A6},
        pages = {A6},
          doi = {10.1051/0004-6361/201833910},
archivePrefix = {arXiv},
       eprint = {1807.06209},
 primaryClass = {astro-ph.CO},
       adsurl = {https://ui.adsabs.harvard.edu/abs/2020A&A...641A...6P}
}

@article{chluba_spectro-spatial_2023-I,
	title = {Spectro-spatial evolution of the {CMB}. {Part} {I}. {Discretisation} of the thermalisation {Green}'s function},
	volume = {2023},
	issn = {1475-7516},
	url = {https://iopscience.iop.org/article/10.1088/1475-7516/2023/11/026},
	doi = {10.1088/1475-7516/2023/11/026},
	number = {11},
	urldate = {2024-12-10},
	journal = {Journal of Cosmology and Astroparticle Physics},
	author = {Chluba, Jens and Kite, Thomas and Ravenni, Andrea},
	month = nov,
	year = {2023},
	pages = {026},
}

@article{kite_spectro-spatial_2023-III,
	title = {Spectro-spatial evolution of the {CMB}. {Part} {III}. {Transfer} functions, power spectra and {Fisher} forecasts},
	volume = {2023},
	issn = {1475-7516},
	url = {https://iopscience.iop.org/article/10.1088/1475-7516/2023/11/028},
	doi = {10.1088/1475-7516/2023/11/028},
	number = {11},
	urldate = {2024-12-10},
	journal = {Journal of Cosmology and Astroparticle Physics},
	author = {Kite, Thomas and Ravenni, Andrea and Chluba, Jens},
	month = nov,
	year = {2023},
	pages = {028},
}

@article{chluba_spectro-spatial_2023-II,
	title = {Spectro-spatial evolution of the {CMB}. {Part} {II}. {Generalised} {Boltzmann} hierarchy},
	volume = {2023},
	issn = {1475-7516},
	url = {https://iopscience.iop.org/article/10.1088/1475-7516/2023/11/027},
	doi = {10.1088/1475-7516/2023/11/027},
	number = {11},
	urldate = {2024-12-10},
	journal = {Journal of Cosmology and Astroparticle Physics},
	author = {Chluba, Jens and Ravenni, Andrea and Kite, Thomas},
	month = nov,
	year = {2023},
	pages = {027},
}

@ARTICLE{McDermott2020DP,
       author = {{McDermott}, Samuel D. and {Witte}, Samuel J.},
        title = "{Cosmological evolution of light dark photon dark matter}",
      journal = {\prd},
         year = 2020,
        month = mar,
       volume = {101},
       number = {6},
          eid = {063030},
        pages = {063030},
          doi = {10.1103/PhysRevD.101.063030},
archivePrefix = {arXiv},
       eprint = {1911.05086},
 primaryClass = {hep-ph},
       adsurl = {https://ui.adsabs.harvard.edu/abs/2020PhRvD.101f3030M}
}

@ARTICLE{Sabyr2025y,
       author = {{Sabyr}, Alina and {Fabbian}, Giulio and {Hill}, J. Colin and {Bianchini}, Federico},
        title = "{A new constraint on the $y$-distortion with FIRAS: robustness of component separation methods}",
      journal = {arXiv e-prints},
         year = 2025,
        month = aug,
          eid = {arXiv:2508.04593},
        pages = {arXiv:2508.04593},
          doi = {10.48550/arXiv.2508.04593},
archivePrefix = {arXiv},
       eprint = {2508.04593},
 primaryClass = {astro-ph.CO},
       adsurl = {https://ui.adsabs.harvard.edu/abs/2025arXiv250804593S}
}

@ARTICLE{Kogut2023,
       author = {{Kogut}, A. and {Fixsen}, Dale and {Aghanim}, Nabila and {Chluba}, Jens and {Chuss}, David T. and {Delabrouille}, Jacques and {Hensley}, Brandon S. and {Hill}, J. Colin and {Maffei}, Bruno and {Pullen}, Anthony R. and {Rotti}, Additya and {Switzer}, Eric R. and {Woillack}, Edward J. and {Zelko}, Ioana},
        title = "{Systematic error mitigation for the PIXIE Fourier transform spectrometer}",
      journal = {\jcap},
         year = 2023,
        month = jul,
       volume = {2023},
       number = {7},
          eid = {057},
        pages = {057},
          doi = {10.1088/1475-7516/2023/07/057},
archivePrefix = {arXiv},
       eprint = {2304.00091},
 primaryClass = {astro-ph.CO},
       adsurl = {https://ui.adsabs.harvard.edu/abs/2023JCAP...07..057K}
}

@ARTICLE{Cyr2024Axions,
       author = {{Cyr}, Bryce and {Chluba}, Jens and {Gangrekalve Manoj}, Pranav Bharadwaj},
        title = "{Revisiting Constraints on Resonant Axion-Photon Conversions from CMB Spectral Distortions}",
      journal = {arXiv e-prints},
         year = 2024,
        month = nov,
          eid = {arXiv:2411.13701},
        pages = {arXiv:2411.13701},
          doi = {10.48550/arXiv.2411.13701},
archivePrefix = {arXiv},
       eprint = {2411.13701},
 primaryClass = {astro-ph.CO},
       adsurl = {https://ui.adsabs.harvard.edu/abs/2024arXiv241113701C}
}

@ARTICLE{Zegeye2023S4,
       author = {{Zegeye}, David and {Bianchini}, Federico and {Bond}, J. Richard and {Chluba}, Jens and {Crawford}, Thomas and {Fabbian}, Giulio and {Gluscevic}, Vera and {Grin}, Daniel and {Hill}, J. Colin and {Meerburg}, P. Daniel and {Orlando}, Giorgio and {Partridge}, Bruce and {Reichardt}, Christian L. and {Remazeilles}, Mathieu and {Scott}, Douglas and {Wollack}, Edward J. and {CMB-S4 Collaboration}},
        title = "{CMB-S4 forecasts for constraints on f$_{NL}$ through {\ensuremath{\mu}} -distortion anisotropy}",
      journal = {\prd},
         year = 2023,
        month = nov,
       volume = {108},
       number = {10},
          eid = {103536},
        pages = {103536},
          doi = {10.1103/PhysRevD.108.103536},
archivePrefix = {arXiv},
       eprint = {2303.00916},
 primaryClass = {astro-ph.CO},
       adsurl = {https://ui.adsabs.harvard.edu/abs/2023PhRvD.108j3536Z}
}

@misc{Chluba2024,
    title = {Revisiting {Dark} {Photon} {Constraints} from {CMB} {Spectral} {Distortions}},
    url = {http://arxiv.org/abs/2409.12115},
    doi = {10.48550/arXiv.2409.12115},
    language = {en},
    urldate = {2025-11-18},
    publisher = {arXiv},
    author = {Chluba, Jens and Cyr, Bryce and Johnson, Matthew C.},
    month = sep,
    year = {2024},
    note = {arXiv:2409.12115 [astro-ph]},
}

@article{Chluba2025encylopedia,
    author = "Chluba, J.",
    title = "{The Cosmic Microwave Background: Spectral Distortions}",
    eprint = "2502.05188",
    archivePrefix = "arXiv",
    primaryClass = "astro-ph.CO",
    month = "1",
    year = "2025"
}

@article{Okun1982,
    author  = "Okun, L. B.",
    title   = "{Limits of electrodynamics: paraphotons?}",
    journal = "Sov. Phys. JETP",
    volume  = "56",
    pages   = "502",
    year    = "1982",
    note    = "[Zh. Eksp. Teor. Fiz. \textbf{83}, 892 (1982)]"
}

@article{Holdom1985,
    author  = "Holdom, Bob",
    title   = "{Two U(1)'s and Epsilon Charge Shifts}",
    journal = "Phys. Lett. B",
    volume  = "166",
    pages   = "196--198",
    year    = "1986",
    doi     = "10.1016/0370-2693(86)91377-8"
}

@article{Galison1983,
    author  = "Galison, Peter and Manohar, Aneesh",
    title   = "{Two Z's or Not Two Z's?}",
    journal = "Phys. Lett. B",
    volume  = "136",
    pages   = "279--283",
    year    = "1984",
    doi     = "10.1016/0370-2693(84)91161-4"
}

@article{Georgi1983,
    author = "Georgi, Howard and Ginsparg, Paul H. and Glashow, S. L.",
    title = "{Photon Oscillations and the Cosmic Background Radiation}",
    reportNumber = "HUTP-83/A044",
    doi = "10.1038/306765a0",
    journal = "Nature",
    volume = "306",
    pages = "765--766",
    year = "1983"
}

@article{Davidson2000,
    author        = "Davidson, Sacha and Hannestad, Steen and Raffelt, Georg",
    title         = "{Updated bounds on millicharged particles}",
    journal       = "JHEP",
    volume        = "05",
    pages         = "003",
    year          = "2000",
    eprint        = "hep-ph/0001179",
    archivePrefix = "arXiv",
    doi           = "10.1088/1126-6708/2000/05/003"
}

@article{Davidson1993,
    author        = "Davidson, Sacha and Peskin, Michael E.",
    title         = "{Astrophysical bounds on millicharged particles in models with a paraphoton}",
    journal       = "Phys. Rev. D",
    volume        = "49",
    pages         = "2114--2117",
    year          = "1994",
    eprint        = "hep-ph/9310288",
    archivePrefix = "arXiv",
    doi           = "10.1103/PhysRevD.49.2114"
}

@article{Fitzpatrick2021,
  author        = {Fitzpatrick, Patrick J. and Liu, Hongwan and Slatyer, Tracy R. and Tsai, Yu-Dai},
  title         = {{New Thermal Relic Targets for Inelastic Vector-Portal Dark Matter}},
  journal       = {Phys. Rev. D},
  volume        = {106},
  number        = {8},
  pages         = {083507},
  year          = {2022},
  doi           = {10.1103/PhysRevD.106.083507},
  eprint        = {2105.05255},
  archivePrefix = {arXiv},
  primaryClass  = {hep-ph}
}

@article{Berlin2023,
  author        = {Berlin, Asher and Krnjaic, Gordan and Pinetti, Elena},
  title         = {{Reviving MeV-GeV Indirect Detection with Inelastic Dark Matter}},
  journal       = {Phys. Rev. D},
  volume        = {110},
  number        = {3},
  pages         = {035015},
  year          = {2024},
  doi           = {10.1103/PhysRevD.110.035015},
  eprint        = {2311.00032},
  archivePrefix = {arXiv},
  primaryClass  = {hep-ph}
}

@article{Roy2026,
  author        = {Roy, Abhishek and Sanyal, Prasenjit and Scopel, Stefano},
  title         = {{Dark Photon mediated Inelastic Dark Matter in Cosmology, Astrophysics and Colliders}},
  year          = {2026},
  eprint        = {2602.18051},
  archivePrefix = {arXiv},
  primaryClass  = {hep-ph}
}

@article{TuckerSmith2001,
  author        = {Tucker-Smith, David and Weiner, Neal},
  title         = {{Inelastic Dark Matter}},
  journal       = {Phys. Rev. D},
  volume        = {64},
  pages         = {043502},
  year          = {2001},
  doi           = {10.1103/PhysRevD.64.043502},
  eprint        = {hep-ph/0101138},
  archivePrefix = {arXiv}
}

@article{Ackerman2008,
  author        = {Ackerman, Lotty and Buckley, Matthew R. and Carroll, Sean M. and Kamionkowski, Marc},
  title         = {{Dark Matter and Dark Radiation}},
  journal       = {Phys. Rev. D},
  volume        = {79},
  pages         = {023519},
  year          = {2009},
  doi           = {10.1103/PhysRevD.79.023519},
  eprint        = {0810.5126},
  archivePrefix = {arXiv},
  primaryClass  = {hep-ph}
}

@article{Kaplan2009,
  author        = {Kaplan, David E. and Krnjaic, Gordan Z. and Rehermann, Keith R. and Wells, Christopher M.},
  title         = {{Atomic Dark Matter}},
  journal       = {JCAP},
  volume        = {05},
  pages         = {021},
  year          = {2010},
  doi           = {10.1088/1475-7516/2010/05/021},
  eprint        = {0909.0753},
  archivePrefix = {arXiv},
  primaryClass  = {hep-ph}
}

@article{CyrRacine2012,
  author        = {Cyr-Racine, Francis-Yan and Sigurdson, Kris},
  title         = {{The Cosmology of Atomic Dark Matter}},
  journal       = {Phys. Rev. D},
  volume        = {87},
  pages         = {103515},
  year          = {2013},
  doi           = {10.1103/PhysRevD.87.103515},
  eprint        = {1209.5752},
  archivePrefix = {arXiv},
  primaryClass  = {astro-ph.CO}
}

@article{Cline2012,
  author        = {Cline, James M. and Liu, Zuowei and Xue, Wei},
  title         = {{Millicharged Atomic Dark Matter}},
  journal       = {Phys. Rev. D},
  volume        = {85},
  pages         = {101302},
  year          = {2012},
  doi           = {10.1103/PhysRevD.85.101302},
  eprint        = {1201.4858},
  archivePrefix = {arXiv},
  primaryClass  = {hep-ph}
}

@article{Bansal2022,
  author        = {Bansal, Saurabh and Barron, Jared and Curtin, David and Tsai, Yuhsin},
  title         = {{Precision Cosmological Constraints on Atomic Dark Matter}},
  journal       = {JHEP},
  volume        = {10},
  pages         = {095},
  year          = {2023},
  doi           = {10.1007/JHEP10(2023)095},
  eprint        = {2212.02487},
  archivePrefix = {arXiv},
  primaryClass  = {hep-ph}
}

@article{Adams2026,
    author = "Adams, Duncan K. and Barron, Jared and Cyr, Bryce and Zhang, Xiuyuan",
    title = "{CMB Spectral Distortions from Resonant Conversions in Atomic Dark Sectors}",
    eprint = "2602.13384",
    archivePrefix = "arXiv",
    primaryClass = "astro-ph.CO",
    reportNumber = "MIT-CTP/6003",
    month = "2",
    year = "2026"
}

@article{Caputo2021,
    author        = "Caputo, Andrea and Millar, Alexander J. and O'Hare, Ciaran A. J. and Vitagliano, Edoardo",
    title         = "{Dark photon limits: A handbook}",
    journal       = "Phys. Rev. D",
    volume        = "104",
    number        = "9",
    pages         = "095029",
    year          = "2021",
    eprint        = "2105.04565",
    archivePrefix = "arXiv",
    primaryClass  = "hep-ph",
    doi           = "10.1103/PhysRevD.104.095029"
}

@article{Caputo2025,
    author        = "Caputo, Andrea and Park, Jaeyoung and Yun, Seokhoon",
    title         = "{The Heavy Dark Photon Handbook: Cosmological and Astrophysical Bounds}",
    journal       = "Phys. Rev. D",
    year          = "2026",
    eprint        = "2511.15785",
    archivePrefix = "arXiv",
    primaryClass  = "hep-ph",
    doi           = "10.1103/m6zs-jxxp",
    note          = "in press"
}

@article{Caputo2026,
  author        = {Caputo, Andrea and Essig, Rouven},
  title         = {{The Dark Photon: a 2026 Perspective}},
  year          = {2026},
  eprint        = {2603.08430},
  archivePrefix = {arXiv},
  primaryClass  = {hep-ph},
  note          = {Contribution to Encyclopedia of Particle Physics}
}

@article{Antel2023,
    author        = "Antel, C. and others",
    title         = "{Feebly-interacting particles: FIPs 2022 Workshop Report}",
    journal       = "Eur. Phys. J. C",
    volume        = "83",
    number        = "12",
    pages         = "1122",
    year          = "2023",
    eprint        = "2305.01715",
    archivePrefix = "arXiv",
    primaryClass  = "hep-ph",
    doi           = "10.1140/epjc/s10052-023-12168-5"
}

@article{Hook2025,
    author = "Hook, Anson and Huang, Junwu and Shalaby, Mohamad",
    title = "{No cosmological constraints on dark photon dark matter from resonant conversion: Impact of nonlinear plasma dynamics}",
    eprint = "2510.13956",
    archivePrefix = "arXiv",
    primaryClass = "hep-ph",
    month = "10",
    year = "2025"
}

@article{Arsenadze2024,
    author = "Arsenadze, Giorgi and Caputo, Andrea and Gan, Xucheng and Liu, Hongwan and Ruderman, Joshua T.",
    title = "{Shaping dark photon spectral distortions}",
    eprint = "2409.12940",
    archivePrefix = "arXiv",
    primaryClass = "astro-ph.CO",
    reportNumber = "CERN-TH-2024-153, DESY-24-139, FERMILAB-PUB-24-0740-V",
    doi = "10.1007/JHEP03(2025)018",
    journal = "JHEP",
    volume = "03",
    pages = "018",
    year = "2025"
}

@article{Pirvu2023,
    author = "Pirvu, Dalila and Huang, Junwu and Johnson, Matthew C.",
    title = "{Patchy screening of the CMB from dark photons}",
    eprint = "2307.15124",
    archivePrefix = "arXiv",
    primaryClass = "hep-ph",
    doi = "10.1088/1475-7516/2024/01/019",
    journal = "JCAP",
    volume = "01",
    pages = "019",
    year = "2024"
}

@article{Krolewski2019,
    author = "Krolewski, Alex and Ferraro, Simone and Schlafly, Edward F. and White, Martin",
    title = "{unWISE tomography of Planck CMB lensing}",
    eprint = "1909.07412",
    archivePrefix = "arXiv",
    primaryClass = "astro-ph.CO",
    doi = "10.1088/1475-7516/2020/05/047",
    journal = "JCAP",
    volume = "05",
    pages = "047",
    year = "2020"
}

@article{Planck2018,
    author = "Aghanim, N. and others",
    collaboration = "Planck",
    title = "{Planck 2018 results. I. Overview and the cosmological legacy of Planck}",
    eprint = "1807.06205",
    archivePrefix = "arXiv",
    primaryClass = "astro-ph.CO",
    doi = "10.1051/0004-6361/201833880",
    journal = "Astron. Astrophys.",
    volume = "641",
    pages = "A1",
    year = "2020"
}

@article{Aramburo-Garcia2024,
    author = "Aramburo-Garcia, Andres and Bondarenko, Kyrylo and Boyarsky, Alexey and Kashko, Pavlo and Pradler, Josef and Sokolenko, Anastasia and Kugel, Roi and Schaller, Matthieu and Schaye, Joop",
    title = "{Dark photon constraints from CMB temperature anisotropies}",
    eprint = "2405.05104",
    archivePrefix = "arXiv",
    primaryClass = "astro-ph.CO",
    reportNumber = "UWThPh 2024-9, FERMILAB-PUB-24-0219-T, NORDITA 2024-013",
    doi = "10.1088/1475-7516/2024/11/049",
    journal = "JCAP",
    volume = "11",
    pages = "049",
    year = "2024"
}

@article{McCarthy2024,
    author = "McCarthy, Fiona and Pirvu, Dalila and Hill, J. Colin and Huang, Junwu and Johnson, Matthew C. and Rogers, Keir K.",
    title = "{Dark Photon Limits from Patchy Dark Screening of the Cosmic Microwave Background}",
    eprint = "2406.02546",
    archivePrefix = "arXiv",
    primaryClass = "hep-ph",
    doi = "10.1103/PhysRevLett.133.141003",
    journal = "Phys. Rev. Lett.",
    volume = "133",
    number = "14",
    pages = "141003",
    year = "2024"
}

@article{Arias2012,
    author = "Arias, Paola and Cadamuro, Davide and Goodsell, Mark and Jaeckel, Joerg and Redondo, Javier and Ringwald, Andreas",
    title = "{WISPy Cold Dark Matter}",
    eprint = "1201.5902",
    archivePrefix = "arXiv",
    primaryClass = "hep-ph",
    reportNumber = "DESY-11-226, MPP-2011-140, CERN-PH-TH-2011-323, IPPP-11-80, DCPT-11-160",
    doi = "10.1088/1475-7516/2012/06/013",
    journal = "JCAP",
    volume = "06",
    pages = "013",
    year = "2012"
}

@article{Caputo2020,
    author = "Caputo, Andrea and Liu, Hongwan and Mishra-Sharma, Siddharth and Ruderman, Joshua T.",
    title = "{Dark Photon Oscillations in Our Inhomogeneous Universe}",
    eprint = "2002.05165",
    archivePrefix = "arXiv",
    primaryClass = "astro-ph.CO",
    doi = "10.1103/PhysRevLett.125.221303",
    journal = "Phys. Rev. Lett.",
    volume = "125",
    number = "22",
    pages = "221303",
    year = "2020"
}
}

\newpage 
\appendix

\section{Relation between the energy release and the conversion parameter}
\label{sec:gammacon-to-Drr}
We have seen throughout the paper that for simulations, it is often more convenient to set the amount of energy converted and stored by the distortion instead of $\gammacon$ as a function of $\epsilon$ and $\mdp$. In this appendix, we aim to derive the relation between these two quantities as introduced in Eq.~\eqref{eq:gamamcon-to-Drr}. To compute the total effective energy release, it is necessary to take into account both the energy and entropy variations at the conversion using \cite{Chluba2015GreensII, Chluba2024DP}:
\begin{equation}
    \left. \Drr \right|_{\rm d} \simeq \left. \Drr \right|_{\rm con} - \frac{4}{3} \left. \frac{\Delta N_\gamma}{N_\gamma} \right|_{\rm con},
\end{equation}
on the left-hand side, we have the relevant energy and number density conversion contributions.
We therefore have to compute the energy and number integrals of the background source $\id \mathcal{S}^{(0)}(z, x)/\id z$ term as follows:
\bsub
\begin{align} 
\frac{\Delta \rho_\gamma}{\rho_\gamma} & = \frac{\gammacon}{G_3} \int x^3 S_{\rm d}(x) \id x = \frac{\gammacon}{G_3} \int x^3 \left[ - \frac{\nbb}{x}\right] \id x = - \frac{G_2}{G_3} \gammacon \\[2mm]
\frac{\Delta N_\gamma}{N_\gamma} & = \frac{\gammacon}{G_2} \int x^2 S_{\rm d}(x) \id x = \gammacon \int x^2 \left[ - \frac{\nbb}{x}\right] \id x = - \frac{G_1}{G_2} \gammacon 
\end{align}
\esub
where we used the definition of $S_{\rm d}(x)$ in Eq.~\eqref{eq:Dn_dp} and the moments of the Planckian distribution already defined above, $G_k = \int \frac{x^k}{\expf{x}-1} \id x$. We then find:
\begin{equation}
    \left. \Drr \right|_{\rm d} =  - \left[ \frac{G_2}{G_3} - \frac{4}{3} \frac{G_1}{G_2} \right]\gammacon  = 0.5421 \gammacon
\end{equation}
recovering the results of section \ref{sec:decomposition}.

\end{document}